\documentclass[10pt,a4paper]{article}
\usepackage[utf8]{inputenc}
\usepackage{amsmath}
\usepackage{amsfonts}
\usepackage{amssymb}
\usepackage{graphicx}
\usepackage{slashed}
\usepackage{enumitem}
\usepackage{subfigure,wrapfig}
\usepackage{caption}
\usepackage{titlesec}
\usepackage{blkarray}
\usepackage{multirow}
\usepackage{gensymb}
\usepackage[a4paper,bindingoffset=0.2in,
            left=0.5 in,right=0.5 in,top=0.5 in,bottom=0.5  in,
            footskip=.25in]{geometry}
\usepackage{physics}
\usepackage{dsfont}
\DeclareMathOperator{\Span}{Span}

\titleformat{\paragraph}
{\normalfont\normalsize\bfseries}{\theparagraph}{1em}{}
\titlespacing*{\paragraph}
{0pt}{3.25ex plus 1ex minus .2ex}{1.5ex plus .2ex}

\begin{document}

\title{\textbf{Revisiting Connes' Finite Spectral Distance on Non-commutative Spaces : Moyal Plane and Fuzzy Sphere}}
\author{Yendrembam Chaoba Devi\footnote{Corresponding author; S. N. Bose National Centre for Basic Sciences, JD Block, Sector III, Salt Lake City, Kolkata 700106, West Bengal, India; Email: ychaoba9@gmail.com} , Kaushlendra Kumar\footnote{Indian Institute of Science Education and Research, Mohanpur, Nadia 741246, West Bengal, India; Leibniz Universit\"at Hannover, Institut f\"ur Theoretische Physik, Appelstra\ss e 11A, 30167 Hannover, Germany; Email: kaushal.kumar224@gmail.com} , Biswajit Chakraborty$^{*}$\footnote{Email: biswajit@bose.res.in} , Frederik G. Scholtz\footnote{National Institute for Theoretical Physics (NITheP), Stellenbosch 7602, South Africa; Email: fgs@sun.ac.za}}
\date{}
\maketitle

\begin{abstract}
Beginning with a review of the existing literature on the computation of spectral distances on non-commutative spaces like Moyal plane and fuzzy sphere, adaptable to Hilbert-Schmidt operatorial formulation, we carry out a correction, revision and extension of the algorithm provided in \cite{FSBC} i.e. [Scholtz F G and Chakraborty B 2013 \emph{J. Phys. A: Math. Theor.} \textbf{46}, 085204] to compute the finite Connes' distance between normal states. The revised expression, which we provide here, involves the computation of the infimum of an expression which involves the ``transverse" $(\Delta \rho_\perp)$ component of the algebra element in addition to the ``longitudinal" component $\Delta \rho$ of \cite{FSBC}, identified with the difference of density matrices representing the states, whereas the expression given in \cite{FSBC} involves only $\Delta \rho$ and corresponds to the lower bound of the distance. This renders the revised formula less user-friendly, as the determination of the exact transverse component for which the infimum is reached remains a non-trivial task, but under rather generic conditions it turns out that the Connes' distance is proportional to the Hilbert-Schmidt norm of $\Delta \rho$, leading to considerable simplification.  In addition, we can determine an upper bound of the distance by emulating and adapting the approach of [P. Martinetti and L. Tomassini, \emph{Commun. Math. Phys.} \textbf{323} (2013) 107--141]. We then look for an optimal element for which the upper bound is reached. We are able to find one for the Moyal plane through the limit of a sequence obtained by finite dimensional projections of the representative of an element belonging to a multiplier algebra, onto the subspaces of the total Hilbert space, occurring in the spectral triple and spanned by the eigen-spinors of the respective Dirac operator. This is in contrast with the fuzzy sphere, where the upper bound, which is given by the geodesic of a commutative sphere is never reached for any finite $n$-representation of $SU(2)$. Indeed, for the case of maximal non-commutativity ($n = 1/2$), the finite distance is shown to coincide exactly with the above mentioned lower bound, with the transverse component playing no role. This, however starts changing from $n=1$ onwards and we try to improve the estimate of the finite distance and provide an almost exact result, using our revised algorithm. The contrasting features of these types of non-commutative spaces becomes quite transparent through the analysis, carried out in the eigen-spinor bases of the respective Dirac operators.
\end{abstract}


\section{Introduction}
It is quite plausible that the structure of space-time in the vicinity of Planck scale is described by a fuzzy ``quantum space-time". As shown by Doplicher \emph{et. al} \cite{Dop}, this quantum structure of space-time, where the space-time coordinates are operator-valued and satisfy a non-commuting coordinate algebra, can be one of the most plausible ways to prevent the gravitational collapse arising from the attempt to localise a space-time event within a Planck length scale. Since making a guess regarding the non-commutative structure of the coordinate algebra is difficult, one tries to postulate a simple structure of this form and study the geometry of the resulting non-commutative spaces. The simplest examples are the 2D Moyal plane ($\mathds{R}^2_*$):
\begin{equation}
[\hat{x}_1, \hat{x}_2] = i \theta,
\label{MP}
\end{equation} 
and fuzzy $\mathds{R}^3_*$, introduced by Madore in \cite{Mad}.
\begin{equation}
[\hat{x}_i, \hat{x}_j] = i \lambda \epsilon_{ijk} \hat{x}_k.
\label{FS}
\end{equation} 
Fuzzy sphere $\mathds{S}^2_*$ corresponds to a 2D subspace of $\mathds{R}^3_*$ with radius quantized as $r_n = \lambda \sqrt{n(n+1)}$ \cite{Jab}, \cite{Devi}. Because of the inherent uncertainty relations satisfied by these coordinates the usual concepts like points, lines etc. loose their meaning in these kinds of spaces. It thus becomes essential to use the mathematical formalism of Non-Commutative Geometry(NCG) as developed by Connes \cite{Con} to study the geometry of such spaces.

We would like to mention in this context that Connes himself, along with his other collaborators, have invested a lot of effort in formulating a completely new mathematical framework to describe the Standard model of particle physics, by invoking the so-called ``Almost Commutative Spaces" - built out of the usual commutative (Euclideanised) 4D curved space-time. This is expected to describe physics upto GUT scale ($\sim$ $10^{15}~GeV$). However, as we mentioned in the beginning, one perhaps has to take quantum space-times, i.e. spaces where the coordinates become operators satisfying a non-commutative algebra and for which (\ref{MP}) and (\ref{FS}) provide prototype examples, more seriously in the higher energy scales - like in the vicinity of Planck scales. (See for example \cite{van} for a review).  It is only recently that the above mentioned mathematical framework of NCG, \emph{a la} Connes, has been used to compute the spectral distances on ``quantized spaces" like the Moyal plane, fuzzy sphere etc. \cite{Mart},\cite{cag}-\cite{Var}. On the other hand, an algorithm was devised in \cite{FSBC} to compute infinitesimal distances using the Hilbert-Schmidt operatorial formulation of quantum mechanics \cite{SCGV}, \cite{Roh}. This Hilbert-Schmidt operatorial formulation has the advantage that it bypasses the use of any star product, like Moyal or Voros, and is therefore free from any ambiguities that can arise from there \cite{Liz}. Furthermore, it has an additional advantage that the above-mentioned algorithm/formula is adaptable to this Hilbert-Schmidt operatorial formulation and infinitesimally it has essentially the same structure as that of the induced metric from the Hilbert space inner product, obtained by Provost and Vallee \cite{Val} (See also \cite{Asht}), when expressed in terms of the density matrix \cite{FSBC} and yields for the Moyal plane the correct infinitesimal distance in the ``harmonic oscillator basis" and a flat metric in the coherent state basis, upto an overall numerical factor. However, the corresponding finite distance can not simply be obtained just by ``integrating" along the geodesics, as the very concept of geodesics in the conventional sense (i.e. like on a commutative differentiable manifold) may not exist at all. This motivates us to undertake the task of extending our algorithm of \cite{FSBC} so that one is able to compute finite distances as well. This new formula (see \eqref{rev_formula} below) is shown to involve the `transverse' component ($\Delta\rho_\perp$), in addition to the ``longitudinal" component ($\Delta\rho_{\parallel}\equiv \Delta\rho$) where $\Delta\rho$ is the difference between normal states represented by density matrices, as in \cite{FSBC}. This in turn shows that the formula of \cite{FSBC} actually corresponds only to the lower bound of the distance and not the actual distance. Indeed, on the way to our derivation of the generalised formula in section \ref{Sec5} we point out a flaw in the analysis in \cite{FSBC} where the same expression was shown to correspond to the upper bound as well. We point out in this context that the derivation in \cite{FSBC} was also commented on in \cite{FA}. It should, however, be mentioned that the counter example in \cite{FA} does not satisfy the condition, which requires that the optimal element should at least belong to the multiplier algebra, if not to the algebra itself (See section \ref{sec 2.5} below). Rather the error lies elsewhere, as we show in the sequel (see the line below \eqref{3.1.11} and at the end of the section \ref{sec5.2}). This error, however was not serious in examples studied previously to compute \emph{infinitesimal} distances as these distances coincided with the exact distance for discrete states and differed from the exact distance by a numerical factor only for the coherent states for both $\mathds{R}^2_*$ and $\mathds{S}^2_*$ \cite{FSBC, Devi}.

Returning to the revised formula \eqref{rev_formula}, we note that in the generic case there can be many choices of $\Delta\rho_\perp$ to a given $\Delta\rho$ (in fact an infinite number of them in $\mathds{R}^2_*$). Consequently the computation of the infimum involving $\Delta\rho_\perp$ occurring in the revised formula turns out to be quite non-trivial. One therefore has to try with different forms of $\Delta\rho_\perp$ and improve the estimate of the distance as best as one can. Alternatively we can emulate and adapt the approach of \cite{Mart} to our Hilbert-Schmidt operatorial formulation to obtain an upper bound to the distance and then look for an optimal element in the algebra saturating this upper bound. In case at least one such optimal element can be identified, this upper bound itself can obviously be recognised as the true distance. Otherwise one has to be content with the above-mentioned best possible estimate only. In fact this paper deals with an interplay of both the approaches, as they seem to complement each other in some sense. This brings out some stark differences between $\mathds{R}^2_*$ and $\mathds{S}^2_*$. Particularly for $\mathds{S}^2_*$ corresponding to any finite $n$-representation of $SU(2)$ one can not define a geodesic in the conventional sense; it reduces to commutative $\mathds{S}^2$ only in the limit $n \to \infty$. The distance turns out to be much smaller than the geodesic distance of $\mathds{S}^2$, the latter coinciding with the above-mentioned upper bound. Further in the case of maximal non-commutativity i.e. for $n = 1/2$, even the finite distance is shown to coincide exactly with the lower bound with $\Delta\rho_\perp$ playing no role and any pair of pure states is shown to be interpolated by a one-parameter family of mixed states, lying in the interior of $\mathds{S}^2_*$. The distance between any mixed state to the nearest pure state can be taken as a measure of the `mixedness'.

The whole analysis, particularly the ``ball condition'' is carried out in the eigen-spinor basis furnished by the respective Dirac operators. This is clearly a natural choice of basis, as the ball condition involves the Dirac operator. This definitely simplifies the computations considerably and even allows us to study the geometry of $\mathds{S}^2_*$ for $n = 1$, apart from reproducing many of the existing results in the literature, albeit with the help of \textit{Mathematica}. However the corresponding analysis for $n > 1$ remains quite intractable, even with \textit{Mathematica}.

The paper is organised as follows. In section \ref{sec2}, we provide a brief review of the Hilbert-Schmidt operatorial formulation of non-commutative quantum mechanics on the 2D-Moyal plane and fuzzy sphere and the associated spectral triples, introduced in \cite{FSBC, Devi} - required to study the geometrical aspects of them. We also introduce the corresponding Dirac operators and their eigen-spinors for both non-commutative spaces. To begin, we revisit the derivation of the formula given in \cite{FSBC} in section \ref{Sec5} and derive the corrected form. We then provide a computation of (finite) distances on the Moyal plane in the coherent and the  ``harmonic oscillator" basis in sections \ref{sec3} and \ref{sec4}, respectively. We then proceed onto the case of the fuzzy sphere in quite the same way adopted as in the Moyal plane in section \ref{sec6}. We note some fundamental differences from the case of the Moyal plane and adopt a different algorithm using the Dirac eigen-spinors and study the $n = 1/2$ and $n = 1$ representations of the fuzzy sphere algebra. We finally conclude in section \ref{sec_con}.

\section{Review of Hilbert-Schmidt operatorial formulation and the spectral triples for non-commutative spaces ($\mathds{R}^2_*$ and $\mathds{S}^2_*$)} \label{sec2}

\subsection{Moyal Plane ($\mathds{R}^2_*$)}

The Hilbert-Schmidt operatorial formulation of Non-Commutative Quantum Mechanics (NCQM) on the 2D Moyal plane \cite{SCGV,Roh}, described by the non-commutative Heisenberg algebra i.e. the co-ordinate algebra \eqref{MP} augmented by the following commutation relations involving linear momenta operators $\hat{p_i}$ satisfying (in units of $\hbar = 1$)
\begin{equation}
[\hat{x}_i, \hat{p}_j] = i \delta_{ij} ~~;~~ [\hat{p}_i, \hat{p}_j] = 0 \label{HA}
\end{equation}
begins by introducing an auxiliary Hilbert space $\mathcal{H}_c$ furnishing a representation of just the coordinate algebra (\ref{MP}). In this particular situation, since the algebra (\ref{MP}) is isomorphic to the algebra $[\hat{x}, \hat{p}] = i$ of a 1D harmonic oscillator, $\mathcal{H}_c$ can be constructed exactly in the same manner i.e.
\begin{equation}
\mathcal{H}_c = \Span \left\lbrace |n\rangle = \frac{(b^\dagger)^n}{\sqrt{n!}} |0\rangle \right\rbrace_{n=0}^\infty
\label{Hc}
\end{equation}
where $b = \frac{1}{\sqrt{2 \theta}} (\hat{x}_1 + i \hat{x}_2)$ and $b^{\dagger}$ are the respective lowering and raising operators satisfying $[b, b^\dagger] = 1$ and the "ground state" $|0\rangle$ satisfy $b|0\rangle = 0$. This $\mathcal{H}_c$, however, cannot furnish a representation of the linear momentum operators $\hat{p}_i$. As shown in \cite{SCGV, Roh}, we need to introduce the space $\mathcal{H}_q$ comprised of Hilbert-Schmidt (HS) operators acting on $\mathcal{H}_c$. These HS operators \cite{Conway} are composite operators $\Psi (\hat{x}_1, \hat{x}_2)$, built out of $\hat{x}_1$ and $\hat{x}_2$, having finite Hilbert-Schmidt (HS) norm $\|\cdot \|_{\text HS}$
\begin{equation} \label{E1}
\mathcal{H}_q = \bigg\lbrace \Psi (\hat{x}_1, \hat{x}_2) : \Psi (\hat{x}_1, \hat{x}_2) \in \mathcal{B}(\mathcal{H}_c);~ \| \Psi (\hat{x}_1, \hat{x}_2) \|_{\text HS} := \left( \Tr_c \left( \Psi (\hat{x}_1, \hat{x}_2)^\dagger \Psi (\hat{x}_1, \hat{x}_2)\right) \right)^{\frac{1}{2}} < \infty \bigg\rbrace .
\end{equation}
Here $\Tr_c$ denotes taking trace over $\mathcal{H}_c$ using any arbitrary orthonormal basis and $\mathcal{B}(\mathcal{H}_c)$ is the space of bounded operators on $\mathcal{H}_c$. This space has a natural inner product:
\begin{equation}\label{E2}
\left(\Psi_1 (\hat{x}_1, \hat{x}_2), \Psi_2 (\hat{x}_1, \hat{x}_2) \right) := \Tr_c \left( \Psi_1 (\hat{x}_1, \hat{x}_2)^\dagger \Psi_2 (\hat{x}_1, \hat{x}_2) \right) < \infty
\end{equation}
and the above HS-norm \eqref{E1} is related to this inner product. This $\mathcal{H}_q$ forms a 2-sided *-ideal of the Banach algebra $\mathcal{B}(\mathcal{H}_c)$ and forms a Hilbert space on its own, w.r.t the inner product \eqref{E2}. Generic states of $\mathcal{H}_c$ \eqref{Hc} and that of $\mathcal{H}_q$ \eqref{E1} are denoted by angular and round kets i.e. $|\cdot \rangle$ and $|\cdot)$ respectively. In fact, one can identify $\mathcal{H}_q$ as $\mathcal{H}_c \otimes \widetilde{\mathcal{H}}_c$ with $\widetilde{\mathcal{H}}_c$ being the dual of $\mathcal{H}_c$ \eqref{Hc} and this allows us to write a generic element of $\mathcal{H}_q$ as $|\psi, \phi) := |\psi\rangle\langle \phi | \in \mathcal{H}_q$ or their linear span. One may refer to $|\psi\rangle$ (resp. $\langle \phi |$) as the left (resp. right) hand sector. 

Further, the space of trace class operators form a dense subspace of $\mathcal{H}_q$, whose elements also have finite trace norm i.e.
\begin{equation}\label{tracen}
\| \Psi (\hat{x}_1, \hat{x}_2) \|_{tr} := \Tr_c \sqrt{\left( \Psi (\hat{x}_1, \hat{x}_2)^\dagger \Psi (\hat{x}_1, \hat{x}_2) \right)} < \infty.
\end{equation}
This is obviously a more stringent requirement than HS-norm as $\| \cdot \|_{\text HS} \leq \sqrt{\|\cdot \|_{op} ~ \|\cdot \|_{tr}}$ , where $\|\cdot \|_{op}$ is the operator norm and elements in $\mathcal{H}_q$ fulfilling (\ref{tracen}) form a proper subspace of $\mathcal{H}_q$. The completion of trace class operators w.r.t HS norm gives back $\mathcal{H}_q$. Further, the product of two HS operators gives a trace class operator and this fact is used to define the inner product \eqref{E2}. In what follows, we shall make use of only HS-norm $(\| \cdot \|_{HS})$ and operator norm i.e.  $\| \cdot \|_{op}$, which will occur only in the ball condition (see (\ref{corr4A1}) below).

A unitary representation of the non-commutative Heisenberg algebra (\ref{MP}) and (\ref{HA}) is obtained by the following actions:
\begin{equation}
\hat{X}_i | \psi ) = |\hat{x}_i \psi) ~~;~~ \hat{P}_i  | \psi )= \frac{1}{\theta} \epsilon_{ij} | [\hat{x}_j, \psi] )  \label{URep}.
\end{equation}
Note that we are using capital letters to distinguish them from the operators acting on $\mathcal{H}_c$. Apart from the ``harmonic oscillator" basis $| n \rangle$, introduced in \eqref{Hc}, satisfying $b^{\dagger}b | n \rangle = n | n \rangle$, one can also introduce normalized coherent states in terms of a dimensionless complex number $z = \frac{1}{\sqrt{2\theta}} (x_1 + ix_2)$ as
\begin{equation}
| z \rangle = U(z, \bar{z}) | 0 \rangle ~~;~~ U(z, \bar{z}) = e^{-\bar{z}b + zb^{\dagger}}  \label{CohSt},
\end{equation}
where $U(z, \bar{z})$ is a unitary operator furnishing a projective representation of the translation group. These states, where uncertainty relations are saturated as $\Delta\hat{x}_1\Delta\hat{x}_2 = \frac{\theta}{2}$ and $SO(2)$ - symmetric $\Delta\hat{x}_1 = \Delta\hat{x}_2 = \sqrt{\frac{\theta}{2}}$, provide an over-complete basis in $\mathcal{H}_c$. The corresponding non-orthogonal projection operator $\rho_z \equiv | z \rangle \langle z | \equiv | z) \in \mathcal{H}_q$ is an operator acting on $\mathcal{H}_c$ and is an eigenstate of $B$ (the representation of $\hat{b}$ in $\mathcal{H}_q$) : $B | z) = z | z )$. This represents a quantum state with maximal localization, where the position measure must now be interpreted in the context of a weak measurement (Positive Operator-Valued Measurement, POVM) rather than a strong measurement (Projective Valued Measurement, PVM).
 
These states $|z)$ can be visualized to be smeared version of sharply localized states of commutative quantum mechanics but centered around the mean point $z$ with minimal uncertainty and can be regarded as maximally localized.  As was shown in \cite{Basu} this quantum Hilbert space $\mathcal{H}_q$ has a built-in structure of an algebra in the sense that under the multiplication map $`m'$ the usual operator product of any arbitrary pair of elements $| \phi ), | \psi ) \in \mathcal{H}_q$ yields another element of $\mathcal{H}_q$:
\begin{eqnarray}
 m :  \mathcal{H}_q \otimes \mathcal{H}_q  &\to&  \mathcal{H}_q,  \label{multi-map}\\\nonumber
| \phi ) \otimes | \psi ) &\to& |\phi \psi ). 
\end{eqnarray}
It was further shown in \cite{Basu} that the representation i.e. the symbols of the above composite state $| \phi \psi )$ in the coherent state i.e. $( z | \phi \psi )$ is obtained by composing the respective representations of individual states i.e. $(z|\phi)$ and $(z|\psi)$ by using the Voros star product $\ast_V$:
\begin{equation}
(z | \phi \psi) = 4\pi^2 ( z | \phi ) \ast_V ( z | \psi ) ~~;~~~~ \ast_V = e^{\overleftarrow{\partial_{\bar{z}}} \overrightarrow{\partial_z}}.  \label{VSP}
\end{equation}
Furthermore, it was shown in \cite{Roh} that $|z) \in \mathcal{H}_q$ also provides an over-complete basis on $\mathcal{H}_q$ - the counterpart of $| z \rangle \in \mathcal{H}_c$, provided that they are composed using the above mentioned Voros star product \eqref{VSP} :
\begin{equation}
\mathds{1}_q = \int \frac{dz d\bar{z}}{\pi} |z) \ast_V (z|.
\end{equation}

As an aside, we mention here that analogous Moyal basis, the counterpart of $|z)$ above, can also be constructed, where the composition of individual symbols is performed through Moyal star product to yield the symbol of the composite operator \cite{Basu}. This is the counter part of \cite{Mart2}. However, since these Moyal basis do not conform to POVM, we do not pursue this case here any more.

\subsection{Fuzzy Sphere ($\mathds{S}^2_*$)}
Here, the position operators satisfy the $su(2)$ commutation relation  (\ref{FS}). This can be realized through a pair of oscillators using the Jordan-Schwinger map \cite{Grosse2, Wat} :
\begin{equation}
 \hat{x}_{i} =  \hat{\chi}^{\dagger}\sigma_{i}\hat{\chi} =  \hat{\chi}^{\dagger}_{\alpha}\sigma_{i}^{\alpha\beta}\hat{\chi}_{\beta},~~~\text{where $\sigma_i$ are the Pauli matrices}.\label{JSmap}
\end{equation}
Here, $\hat{\chi}_\alpha/\hat{\chi}_\alpha^\dagger$ satisfy the following commutation relations:
\begin{equation}
[\hat{\chi}_{\alpha},\hat{\chi}^{\dagger}_{\beta}] = \frac{1}{2}\lambda\delta_{\alpha\beta}~,~~[\hat{\chi}_{\alpha},\hat{\chi}_{\beta}] = 0 = [\hat{\chi}^{\dagger}_{\alpha},\hat{\chi}^{\dagger}_{\beta}]~;~~~~~~\alpha,\beta=1,2. \label{c10}
\end{equation}
Common eigenstates of $\hat{\vec{x}}^2$ and $\hat{x}_3$ can be constructed as follows:
\begin{equation}
 \lvert n, n_3 \rangle =  \sqrt{\frac{(2/\lambda)^{2n}}{(n+n_3)!(n-n_3)! }}\chi^{\dagger (n+n_3)}_{1}\chi^{\dagger(n-n_3)}_{2}\rvert 0\rangle,
\end{equation}
with eigenvalues $r_n^2=\lambda n(n+1)$ and $x_3=\lambda n_3$, respectively. We define the ladder operators $\hat{x}_{\pm}=\hat{x}_1\pm i\hat{x}_2$ which satisfy the commutation relations:
\begin{equation}
[\hat{x}_{3},\hat{x}_{\pm}]= \pm\lambda \hat{x}_{\pm}; ~~~~
[\hat{x}_{+},\hat{x}_{-}]=2\lambda \hat{x}_{3}.
\end{equation}
Hence the classical configuration space $\mathcal{F}_c$ of a fuzzy space of type (\ref{FS}) is given by  
\begin{eqnarray}
\mathcal{F}_c = Span\big\{\rvert n,n_3\rangle\big| \forall n\in \mathds{Z}/2, ~-n\leq n_3\leq n\big\}. \label{confspace}
\end{eqnarray}
 Each $n$ corresponds to the fixed sphere of radius $r_n=\lambda\sqrt{n(n+1)}$ such that for a fixed $n$ the corresponding Hilbert sub-space is a $(2n+1)$-dimensional sub-space
\begin{eqnarray}
\mathcal{F}_{n} = Span\{\rvert n,n_3\rangle ~~\rvert~~ \text{n is fixed}, -n \leq n_3\leq n\}. \label{F-n}
\end{eqnarray}

However, since the quantum Hilbert space, in which the physical states are represented, consists of those operators generated by coordinate operators only and since these commute with the Casimir the elements of the quantum Hilbert space must in addition commute with the Casimir, i.e., must be diagonal in $n$. Therefore, the quantum Hilbert space $\mathcal{H}_q$ of the fuzzy space of type (\ref{FS}) splits into the direct sum of sub-spaces as follows:
\begin{equation}
 \mathcal{H}_q=\{\Psi \in Span\{\rvert n,n_3\rangle\langle n,n'_3\rvert\}:~~\text{tr}_c(\Psi^\dag\Psi)<\infty\}=\bigoplus_{n}\mathcal{H}_n,
\end{equation}
where
\begin{equation}
 \mathcal{H}_n=\{\Psi \in Span\{\rvert n,n_3\rangle\langle n,n_3\rvert\}\equiv |n_3,n'_3):~~\text{tr}_c(\Psi^\dag\Psi)<\infty~~ \text{with fixed}~~ n\}, \label{H-n}
\end{equation}
represents the quantum Hilbert space of a fuzzy sphere with fixed radius $r_n=\lambda\sqrt{n(n+1)}$.

\subsubsection{Perelomov's SU(2) coherent states}
The coherent states are the quantum states which saturate the Heisenberg's uncertainty relation. Each coherent state $|z\rangle$  is obtained by the action of a unitary operator $U(z)$ on the highest weight state $|n,n\rangle$, satisfying $\hat{x}_+|n,n\rangle =0$, and is specified by a complex number $z$. It is well-known that $\lVert|z\rangle-|z'\rangle\rVert\rightarrow 0 $ as $|z-z'|\rightarrow 0$ and hence implies that coherent states have the properties of classical states. Here we review the construction of generalized coherent states of the $SU(2)$ group \cite{Perelomov} such that the non-commutative analog of the homogeneous space of the fuzzy sphere can be constructed by using the Perelomov's $SU(2)$ coherent states \cite{Grosse2}. 

Note that the Heisenberg uncertainty relations for the fuzzy sphere is given by
\begin{equation}
\Delta\hat{x}_1\Delta\hat{x}_2\geq \frac{\lambda}{2}|\langle \hat{x}_3\rangle| \Longrightarrow \Delta\hat{x}_1\Delta\hat{x}_2=\frac{\lambda}{2}|\langle \hat{x}_3\rangle|=
~~~\frac{1}{2}\lambda^2 n \label{Hei.R-FS}
\end{equation}
for both bases build on $|n,n\rangle$ and $|n,-n\rangle$.  Thus, we can choose the highest weight state as $|n,n\rangle $ and the Perelomov $SU(2)$ coherent states can be obtained by the action of a representation $T(g)$ of $g\in SU(2)$ on $|n,n\rangle$. Moreover, any action of $\hat{x}_3$ on the state $|n,n\rangle$ does not change it so that the group element generated by $\hat{x}_3$ is the stability subgroup of $SU(2)$. This implies that the set of generalized coherent states of $SU(2)$ is topologically isomorphic to the coset space $SU(2)/U(1)\simeq \mathds{S}^2$. However, geometrically it reduces to $\mathds{S}^2$ only in the limit $n\rightarrow\infty$. We know that the operator $T(g), ~\forall g\in  SU(2)$ can be expressed in terms of Euler angles as $T(g)=e^{-i\frac{\varphi}{\lambda} \hat{x}_3}e^{-i\frac{\theta}{\lambda}\hat{x}_2}e^{-i\frac{\psi}{\lambda} \hat{x}_3}$ \cite{Perelomov}. For $\mathds{S}^2$, $\psi=0$ such that a generic coherent state specified by a point on $\mathds{S}^2$ is given by 
\begin{equation}
|z\rangle =e^{-i\frac{\varphi}{\lambda} \hat{x}_3}e^{-i\frac{\theta}{\lambda}\hat{x}_2}|n,n\rangle\xrightarrow{\varphi=0} e^{-i\frac{\theta}{\lambda} \hat{x}_2}|n,n\rangle=e^{\frac{\theta}{2\lambda}(\hat{x}_--\hat{x}_+)}|n,n\rangle.
\end{equation}
Here, $z$ is the stereographic variable of $\mathds{S}^2$ projected from the south pole to the complex planes $z=1$ and $z=-\tan\frac{\theta}{2}e^{i\varphi}$.

\subsection{Spectral Triple: Moyal Plane}

The fact that Moyal plane geometry can also be studied by introducing suitable spectral triple was shown quite some time back in \cite{Gay}. We however, present a review of basic features of such a construction using Hilbert-Schmidt operator formulation which was used also in \cite{FSBC}. The basic ingredients of studying the geometrical content, particularly Connes' distance on $\mathcal{H}_c$, are facilitated by specifying the spectral triple $(\mathcal{A}, \mathcal{H}, \mathcal{D})$ where the algebra $\mathcal{A}$ is identified with $\mathcal{H}_q$ as mentioned above (\ref{multi-map}) and the Hilbert space $\mathcal{H}$ is the module and identified with $\mathcal{H}_c \otimes \mathds{C}^2$ so that it represents the space of appropriate ``spinors" where a typical element $a \in \mathcal{A} = \mathcal{H}_q$ acts on $\Psi = \begin{pmatrix}
| \psi_1 \rangle\\
| \psi_2 \rangle\
\end{pmatrix}$ through the diagonal representation $\pi$ as
\begin{equation} \label{corr4A1}
\pi(a) \Psi = \pi(a) \begin{pmatrix}
| \psi_1 \rangle\\
| \psi_2 \rangle\
\end{pmatrix}= \begin{pmatrix}
a & 0 \\
0 & a \
\end{pmatrix} \begin{pmatrix}
| \psi_1 \rangle\\
| \psi_2 \rangle\
\end{pmatrix} = \begin{pmatrix}
a | \psi_1 \rangle\\
a | \psi_2 \rangle\
\end{pmatrix}.
\end{equation}
The Dirac operator, as explained in detail in \cite{Devi} (see also the Appendix A.1), is identified as
\begin{equation} \label{DirOp}
\mathcal{D} = \sqrt{\frac{2}{\theta}} \left( \begin{array}{cc}
0 & b^{\dagger}\\
b & 0\\
\end{array} \right),
\end{equation}
having a well-defined left action on $\mathcal{H}_c \otimes \mathds{C}^2$ as
\begin{equation} \label{corr4A2}
\mathcal{D} \Psi = \sqrt{\frac{2}{\theta}} \begin{pmatrix}
0 & b^\dagger \\
b & 0 \
\end{pmatrix} \begin{pmatrix}
|\psi_1\rangle\\
|\psi_2\rangle\
\end{pmatrix} = \sqrt{\frac{2}{\theta}} \begin{pmatrix}
b^\dagger |\psi_2\rangle \\
b |\psi_1\rangle\
\end{pmatrix}.
\end{equation}
The eigen-spinors of the Dirac operator for the Moyal plane are given by the following,
\begin{equation} \label{Dir_bas_M}
\begin{split}
| 0 \rangle \rangle := \begin{pmatrix}
\ket{0}\\
0\
\end{pmatrix} \in \mathcal{H}_c \otimes \mathds{C}^2 ~~;~~ | m \rangle \rangle_\pm := \frac{1}{\sqrt{2}} \begin{pmatrix}
\ket{m}\\
\pm \ket{m - 1}\
\end{pmatrix} \in \mathcal{H}_c \otimes \mathds{C}^2 ~~;~~ m = 1,2,3, \cdots
\end{split}
\end{equation}
with the eigenvalues $\lambda_m$ i.e. $\mathcal{D} | m \rangle \rangle_\pm = \lambda_m^\pm | m \rangle \rangle_\pm$, for any state $| m \rangle\rangle_\pm $ given by
\begin{equation*}
\lambda_0 = 0 ~~~~;~~~~ \lambda^\pm_m = \pm \sqrt{\frac{2m}{\theta}}.
\end{equation*}

Note that this orthonormal basis is obtained from the natural orthonormal and complete basis
\begin{equation} \label{diff_basis}
\left\lbrace | n, \uparrow \rangle \rangle = \ket{n} \otimes \begin{pmatrix}
1\\
0\
\end{pmatrix} = \begin{pmatrix}
\ket{n}\\
0\
\end{pmatrix} ~~;~~ | n, \downarrow \rangle \rangle = \ket{n} \otimes \begin{pmatrix}
0\\
1\
\end{pmatrix} = \begin{pmatrix}
0\\
\ket{n}\
\end{pmatrix} \right\rbrace
\end{equation}
for $\mathcal{H}_c \otimes \mathds{C}^2$ by first leaving out $\begin{pmatrix}
\ket{0}\\
0\
\end{pmatrix}\equiv |0\rangle\rangle$ separately and then pairing $| n, \uparrow \rangle\rangle$ and $| n-1, \downarrow \rangle\rangle$ as
\begin{equation}
| n \rangle\rangle_\pm = \frac{1}{\sqrt{2}} \left( | n, \uparrow \rangle\rangle \pm | n-1, \downarrow \rangle\rangle \right) = \frac{1}{\sqrt{2}} \begin{pmatrix}
\ket{n}\\
\pm \ket{n-1}\
\end{pmatrix}; ~~ n=1,2,3,\cdots
\end{equation}
Both of these (\ref{Dir_bas_M}, \ref{diff_basis}) furnish complete and orthonormal bases for $\mathcal{H}_c \otimes \mathds{C}^2$, so that the resolution of the identity, particularly for \eqref{Dir_bas_M} takes the form
\begin{equation} \label{res_id}
\mathds{1}_{\mathcal{H}_q \otimes M_2(\mathds{C})} =|0\rangle\rangle\langle\langle 0|+ \sum \limits_{m = 1}^{\infty} \Big( |m\rangle\rangle_+ \, _+\langle\langle m | + |m\rangle\rangle_- \, _-\langle\langle m | \Big)  ;~~ _\pm\langle\langle m | n \rangle\rangle_\pm = \delta_{mn};~~ _\pm\langle\langle m | n \rangle\rangle_\mp = 0
\end{equation}

\subsection{Spectral Triple : Fuzzy Sphere}  \label{section}

For the fuzzy sphere the spectral triple was already constructed in \cite{Devi}. We will be working with the same one here. The spectral triple consists of the algebra $\mathcal{A} \equiv \mathcal{H}_q \ni |m\rangle\langle n|$, the Hilbert space $\mathcal{H} \equiv \mathcal{H}_c \otimes \mathds{C}^2 \ni \begin{pmatrix}
|\psi_1\rangle \\ |\psi_2\rangle
\end{pmatrix}$, and the Dirac operator 
\begin{equation} \label{fuzz_dir}
\mathcal{D}\equiv \frac{1}{r_n} \hat{\vec{J}} \otimes \vec{\sigma} = \frac{1}{r_n}\begin{pmatrix}
\hat{J}_3 & \hat{J}_- \\ \hat{J}_+ & -\hat{J}_3
\end{pmatrix},
\end{equation}
where $\hat{J}_i$ is related to $\hat{x}_i$ by $\hat{J}_i=\frac{1}{\lambda}\hat{x}_i$ and $\hat{J}_{\pm} = \hat{J}_1 \pm \hat{J}_2$ are the usual ladder operators. We would like to mention in this context that although there are several prescriptions for the Dirac operator (see for example \cite{Grosse2, Wat}), we essentially follow that of \cite{Wat}. As mentioned earlier we will make use of the Dirac eigen-spinors $|n, n_3 \rangle\rangle_{\pm} \in \mathcal{H}_c \otimes \mathds{C}^2$, which for the fuzzy sphere are given by \cite{Var}
\begin{equation} \label{Dir_bas_F}
\begin{split}
|n, n_3 \rangle\rangle_+ & := \sqrt{\frac{n + n_3 + 1}{2n + 1}} \ket{n, n_3} \otimes \begin{pmatrix}
1\\
0\
\end{pmatrix} + \sqrt{\frac{n - n_3}{2n + 1}} \ket{n, n_3 + 1} \otimes \begin{pmatrix}
0\\
1\
\end{pmatrix},\\
|n, n_3^\prime \rangle\rangle_- & := - \sqrt{\frac{n - n_3}{2n + 1}} \ket{n, n_3} \otimes \begin{pmatrix}
1\\
0\
\end{pmatrix} + \sqrt{\frac{n + n_3 + 1}{2n + 1}} \ket{n, n_3 + 1} \otimes \begin{pmatrix}
0\\
1\
\end{pmatrix},
\end{split}
\end{equation}
with $-n-1 \leq n_3 \leq n$ and $-n \leq n_3^\prime \leq n-1$. The respective eigenvalues are given by
\begin{equation} \label{eig_fuzz}
\lambda^+_{n_3} = \frac{n}{r_n} ~~~~;~~~~ \lambda^-_{n_3^\prime} = -\frac{(n+1)}{r_n}
\end{equation}
for any $n_3$ or $n'_3$. Thus, the eigenvalues for a particular $n$ representation is independent of $n_3$ or $n'_3$ and is responsible for a $(2n+2)$-fold degeneracy in the positive eigenvalue sector and a $2n$-fold degeneracy in the negative eigenvalue sector. This can be understood from the tensor product structure of Dirac operator \eqref{fuzz_dir} and the Clebsch-Gordon decomposition of a tensor product of a pair of $SU(2)$ representations. For example, if $\hat{\vec{J}}$ in \eqref{fuzz_dir} corresponds to the $n = 1/2$ representation i.e. $\hat{\vec{J}} = \vec{\sigma}/2$, then it will split into the direct sum of $n = 1$ (triplet) and $n = 0$ (singlet) representations of three and one dimension, respectively.

\subsection{Spectral distance \emph{a la} Connes} \label{sec2.5} \label{sec 2.5}

States $\omega$ are positive linear functionals of norm 1 over $\mathcal{A}$. Pure states play a rather fundamental role and are defined as those functionals that cannot be written as a convex linear combination of two other functionals. The Connes’ spectral distance between two such states is then defined by
\begin{align}\label{ConDis}
\begin{split}
d(\omega, \omega^{\prime}) = \sup_{a \in B} \abs{\omega(a) - \omega^{\prime}(a)},\\
B = \left\lbrace a \in \mathcal{A} : \| [\mathcal{D}, \pi(a)] \|_{op} \leq 1 \right\rbrace,\\
\| a \|_{op} = \sup_{\phi \in \mathcal{H}} \frac{\| a \phi \|}{\| \phi \|}.\\
\end{split}
\end{align}
An algorithm, adaptable to this Hilbert-Schmidt operatorial formulation and subject to the following conditions :
\begin{itemize}
\item The states $\omega, \omega^{\prime}$ are normal states (see \cite{Rob} for the definition),
\item The optimal element $a_s$ for which the supremum value in $d(\omega, \omega')$ in \eqref{ConDis} is attained also should saturate the ball condition i.e. 
\begin{equation} \label{ball_sat}
d(\omega, \omega') = \abs{\omega(a_s) - \omega^{\prime}(a_s)};~ \| [\mathcal{D}, \pi(a_s)] \|_{op} = 1,
\end{equation}
and must also be an element of the algebra $\mathcal{A} = \mathcal{H}_q: a_s \in \mathcal{A}$ or at least be an element of the multiplier algebra\footnote{Multiplier algebra $M = M_{L} \cap M_{R}$ where $M_{L} = \left\lbrace T \mid \psi T \in \mathcal{H}_{q} ~\forall~ \psi \in \mathcal{H}_{q} \right\rbrace$ and $M_{R} = \left\lbrace T \mid T \psi \in \mathcal{H}_{q} ~\forall~ \psi \in \mathcal{H}_{q} \right\rbrace $}.
\item Let $V_0 = \left\lbrace a \in \mathcal{A} : \|[\mathcal{D}, \pi(a)]\|_{op} = 0 \right\rbrace$, then the states $\omega$, $\omega^{\prime}$ are such that $\omega(a) - \omega^{\prime}(a) = 0$ , $\forall~a \in V_0$,
\end{itemize}
was devised to compute Connes' spectral distance \cite{FSBC}. The present analysis will also assume the above conditions. Particularly, the third condition implies a certain irreducibility condition, as explained in Appendix A.3.  These conditions are actually quite mild and just imply that a generic state $\omega$ can be represented by a density matrix $\rho$. Furthermore, if the state $\omega$ is pure, then the corresponding density matrix $\rho$ too will be pure. More specifically, to illustrate through an example, let us consider the case of the Moyal plane $\mathds{R}^2_*$, which among many other possible states admit a special kind of states $|z\rangle \in \mathcal{H}_c$ \eqref{CohSt} or more precisely the associated density matrices $\rho_z = | z \rangle\langle z| \in \mathcal{H}_q$ representing self-adjoint projection operators ($\rho_z^2 = \rho_z = \rho_z^*$) and can be regarded as pure states in the $C^*$ algebraic framework. These $SO(2)$-symmetric maximally localized states of the Moyal plane $\mathds{R}^2_*$ can be thought of as deformed versions (of noncommutative origin with deformation parameter $\theta$) of sharply localized states given by the Dirac delta distribution $\delta^2_{\vec{a}}$ of commutative Moyal plane $\mathds{R}^2$ with which it has got a 1-1 correspondence. The action of $\delta^2_{\vec{a}}$  on a typical function $f(\vec{x}) \in C_0^{\infty}(\mathds{R}^2)$, the algebra of the spectral triple $\left( C_0^{\infty}(\mathds{R}^2), L^2(\mathds{R}^2)\otimes \mathds{C}^2, -i\slashed{\partial}  \right)$ describing the geometry of $\mathds{R}^2$-plane yields, by definition, $f(\vec{a})$ i.e. it evaluates the function at $\vec{x} = \vec{a}$.  Indeed in the limit $\theta \rightarrow 0$ the former goes to latter. Apart from these states, built out of coherent states, there are also numerous other pure states like the so called harmonic oscillator states $\rho_n = | n \rangle\langle n| \in \mathcal{H}_q $ - the density matrices built out of the vectors $|n \rangle \in \mathcal{H}_c$ in \eqref{Hc}. One can also introduce the squeezed coherent states, where $\Delta \hat{x}_1 \not= \Delta \hat{x}_2$ but the product is still minimal: $\Delta \hat{x}_1 \Delta \hat{x}_2 = \frac{\theta}{2}$. We shall, however, not be concerned about these squeezed coherent states in this paper.

Now coming back to algebra $\mathcal{A}$, which we have identified with $\mathcal{H}_q$ can indeed be identified with an involutive algebra, which is a dense sub-algebra of a $C^*$-algebra, where the hermitian conjugation ($\dagger$) plays the role of involution operator. Since both $\rho$ and $a$ are elements of $\mathcal{A} = \mathcal{H}_q$, the Connes' distance \eqref{ConDis} between a pair of states, now represented by density matrices $\rho$ and $\rho^\prime$ (See appendix A.3), can be recast in terms of the inner product \eqref{C3} as
\begin{equation} \label{dis}
d(\rho, \rho^\prime) = \sup_{a \in B} |(\Delta \rho, a)| ~;~~ \Delta \rho = \rho - \rho^\prime \in \mathcal{A} = \mathcal{H}_q.
\end{equation}
In this context, it is worthwhile to recall another important role of the third condition in the present analysis. If this condition were to be violated then we can find an $a_0 \in V_0$ such that $\abs{\omega^\prime(a_0) - \omega(a_0)} \neq 0$. However, since the ball condition \eqref{ConDis} places no constraint on $\|a_0\|_{tr}$, it is clear that no upper bound exists for this distance function and the resulting distance diverges. When this condition holds, the spectral distance is always finite. This point was illustrated through the example of the $\mathds{C}P^1$ model in \cite{Devi}. 

This example, which was first considered in \cite{Mart3}, is described by the following spectral triple :
\begin{equation}
\mathcal{A} = M_2(\mathds{C}), ~~ \mathcal{H} = \mathds{C}^2 ~~\text{and}~~ \mathcal{D} = \begin{pmatrix}
D_1 & 0\\
0 & D_2\
\end{pmatrix}. \label{cp1}
\end{equation}
Note that the Dirac operator has been written here in diagonal form with $D_1$, $D_2$ being its eigenvalues. It was shown there that the space of pure states corresponds to $\mathds{S}^2 \cong \mathds{C}P^1$. Now with $D_1 = D_2$, the Dirac operator will become proportional to the identity operator, so that $\| [\mathcal{D}, \pi(a)] \|_{\text{op}} = 0 ~\forall~a \in M_2(\mathds{C})$ holds trivially and no constraint is imposed by the ball condition on any $a \in M_2(\mathds{C})$, giving rise to infinite distance between any pair of pure states. Even setting $D_1 \neq D_2$ subsequently, one finds that only parts of the algebra and more specifically the off-diagonal elements $a_{12}$ of $a = \begin{pmatrix}
a_{11} & a_{12}\\
a^*_{12} & a_{22}\
\end{pmatrix}$ gets constrained as $\abs{a_{12}} \leq \frac{1}{\abs{D_1 - D_2}}$, whereas no constraint is imposed on the real diagonal elements $a_{11}$ and $a_{22}$. Consequently, the distance between any pair of states, belonging to different latitudes diverges and distance between those belonging to the same latitude can only be finite and calculated (See \eqref{cp1_dist} below). We would like to mention in this context that any pair of states belonging to same latitude be related by the unitary matrix $U_3(\phi) = exp(\frac{i}{2}\phi\sigma_3) = \begin{pmatrix}
e^{i\frac{\phi}{2}} & 0 \\ 0 & e^{-i\frac{\phi}{2}}
\end{pmatrix} \in SU(2)$, which being a diagonal matrix commutes with the Dirac operator $\mathcal{D}$ \eqref{cp1}. This, in turn implies that the commutator undergoes no change under this $U_3(\phi)$ action : $[U_3(\phi) \mathcal{D} U_3(\phi), a] = [\mathcal{D}, a]$ $\forall ~a \in M_2(\mathds{C})$ and therefore the pair of states $(\omega_1 \circ \alpha_{U_3(\phi)}, \omega_1 \circ \alpha_{U_3(\phi)})$ obtained from the corresponding pair of states $(\omega_1, \omega_2)$ belonging to the same latitude through an inner-automorphism $\alpha_{U_3(\phi)}$  of the algebra $M_2(\mathds{C})$ \eqref{cp1} defined as $\alpha_{U_3(\phi)} (a) = U_3(\phi) a U^{-1}_3(\phi)$; $a \in M_2(\mathds{C})$, too belongs to the same latitude and is an isometric transformation:
\begin{equation} \label{iso_trans}
d(\omega_1 \circ \alpha_{U_3(\phi)}, \omega_1 \circ \alpha_{U_3(\phi)}) = d(\omega_1, \omega_2).
\end{equation}
In contrast, any other unitary transformation like $U_1 (\theta ) = exp(\frac{i}{2}\theta\sigma_1)$, which do not commute with Dirac operator \eqref{cp1} and relates states belonging to different latitudes will not be an isometric transformation (see lemma 2.3 in \cite{R1}) any more. However, the point we would like to emphasize is that these states, if can be made to belong to the rotated equator corresponding to an equivalent spectral triple, then the original distance can be restored (See Appendix A.2).

In the next section we take up the issue of this same computation, but for \emph{finite} distances, by devising an appropriate and a more general algorithm that also corrects the error in \cite{FSBC}.  As mentioned earlier, knowledge of the infinitesimal distance is not enough to compute the finite distance between a pair of finitely separated points by integrating along the geodesic connecting these points, if the concerned non-commutative space does not allow one to define a geodesic in the conventional sense. In fact, as we shall see later, although the  Moyal plane $\mathds{R}^2_*$ allows one to identify the straight line as the geodesic just like the commutative $\mathds{R}^2$, the great circle in the case of fuzzy sphere $\mathds{S}^2_*$ can not be identified in this manner. 


\section{Towards an algorithm to compute finite distances} \label{Sec5}

As mentioned earlier, we consider the particular case where the states in \eqref{ConDis} are normal and bounded so that they are representable by density matrices. With this the Connes distance function \eqref{dis} becomes
\begin{equation}
d(\omega , \omega^{\prime}) =\sup_{a \in B^\prime}\abs{\text{tr}(\Delta\rho \, a)} = \sup_{a \in B^\prime } \abs{(\Delta\rho , a)},
\label{3.1.3}
\end{equation}		
where, $\rho_{\omega}$ is the density matrix associated with the state $\omega$ so that
\begin{equation}
\omega(a)=\text{tr}(\rho_{\omega}a) , ~~  \Delta\rho = \rho_{\omega} - \rho_{\omega^\prime}  
\label{3.1.4}
\end{equation}
and
\begin{equation}
(\Delta\rho , a)=\text{tr}\big((\Delta\rho)^{\dagger}a\big) = \text{tr}(\Delta\rho~a).
\label{3.1.5}
\end{equation}
We also introduce the following subsets:
\begin{equation} 
V^{\perp}_{o} =  \{ a \in \mathcal{A} :  \|\left[\mathcal{D},\pi(a)\right]\|_{\text{op}} \ne 0\} ,
\label{3.1.6}
\end{equation} 
\begin{equation} 
B^\prime = \{ a \in \mathcal{A} : 0 < \|\left[\mathcal{D},\pi(a)\right]\|_{\text{op}} \le 1\},
\label{3.1.7} 
\end{equation}
\begin{equation} 
W =  \{ a \in \mathcal{A} :  (\Delta\rho , a) = 0\}.  
\label{3.1.8}
\end{equation}  
Note that $B'$ is a dense subset of $B$: $B' \subset B$, so the supremum evaluated over $B$ will be identical with the one evaluated over $B'$.

First we find the lower bound. For this, let us make an assumption that the optimal element $a = a_{s} \in \mathcal{A} = \mathcal{H}_{q}$  is aligned in a direction which is either parallel or anti-parallel to $\Delta\rho$. Equivalently $a_{s}$ can be taken to be proportional to $\Delta\rho$. We thus consider the one-parameter family of algebra elements 
\begin{equation*}
\Lambda= \left\lbrace a \in \mathcal{A} : a= \lambda \Delta\rho ,  0 \le \lambda \le \frac{1}{ ||\left[\mathcal{D},\pi(a)\right]||_{op}} \right\rbrace \subset B^\prime.
\end{equation*}
Taking the extremal element $a = \frac{\Delta\rho}{\|\left[\mathcal{D},\pi(\Delta\rho)\right]\|_{\text{op}}}$ yields the lower bound as
\begin{equation}
d(\rho , \rho\prime) \geq \frac{\text{tr}\big((\Delta\rho)^2\big)}{\|\left[\mathcal{D},\pi(\Delta\rho)\right]\|_{\text{op}}} .
\label{3.1.9}
\end{equation}
Clearly this lower bound will correspond to the true distance iff the optimal element $a_s$ for which the supremum is attained in \eqref{ConDis} turns out to be proportional to $\Delta \rho$. The same lower bound can be obtained alternatively in a more rigorous manner by noting that the HS norm of any element within the ball $B^{\prime}$ \eqref{3.1.7} is bounded above.  To see this, consider an element $a \in B^{\prime}$ s.t. $\|\left[\mathcal{D},\pi(a)\right]\|_{\text{op}} \le 1$. Now writing $a = \|a\|_{\text HS} \hat{a}$ in terms of the ``unit vector" $\hat{a}$ satisfying $\|\hat{a}\|_{\text HS} = 1$, allows us to extract $\|\hat{a}\|_{\text HS}$ out of this inequality to write $\|a\|_{\text HS} \|\left[\mathcal{D},\pi(\hat{a})\right]\|_{\text{op}} \le 1$ yielding
\begin{equation} \label{low_bound}
\|a\|_{\text HS} \le \frac{1}{\|\left[\mathcal{D},\pi(\hat{a})\right]\|_{\text{op}}}.
\end{equation}
Therefore the supremum of $\|a\|_{\text HS}, a \in B^{\prime} $ is bounded above by
\begin{equation}
\sup\limits_{a \in B'} \|a\|_{\text HS} \leq \frac{1}{s} ~,~~ s = \inf_{a \in B^\prime}\|\left[\mathcal{D},\pi(\hat{a})\right]\|_{\text{op}}.
\label{3.1.10}
\end{equation}
It should be noted in this context that the inequality will be saturated iff `$a$' saturates the ball condition: $\|[\mathcal{D}, \pi(a)]\|_{op} = 1$. We can now decompose  $\hat{a}$ in a ``longitudinal" ($\widehat{\Delta\rho}$) and ``transverse" ($\widehat{\Delta\rho}_{\perp}$) component as $\hat{a} = \cos\theta \widehat{\Delta\rho} + \sin\theta \widehat{\Delta\rho}_{\perp}$, where $\|\widehat{\Delta\rho}\|_{HS} = \| \widehat{\Delta\rho}_{\perp} \|_{HS} = 1$ and $\widehat{\Delta\rho}_{\perp} \in W$, is taken to be orthogonal to $\widehat{\Delta\rho}$ i.e. $(\widehat{\Delta\rho}_\perp,\widehat{\Delta\rho})=\Tr(\widehat{\Delta\rho}_\perp\widehat{\Delta\rho})=0$ and corresponds to a unit vector in the plane formed by $\hat{a}$ and $\widehat{\Delta\rho}$. It is clear from \eqref{3.1.3}, that we can choose $\theta$ to be an acute angle i.e $0 \le \theta < \frac{\pi}{2}, $ ensuring that both $\sin \theta$, $\cos\theta$ are positive. Note that here  $\cos\theta \ne 0$ since we want $(d\rho, \hat{a}) \ne 0$ otherwise $d(\omega,\omega')$ in \eqref{3.1.3} would collapse to zero. We can then re-write $\|\left[\mathcal{D},\pi(\hat{a})\right]\|_{\text{op}}$, as in \cite{FSBC} and invoke the triangular inequality to get
\begin{equation}
\|\left[\mathcal{D},\pi(\hat{a})\right]\|_{\text{op}}  = \|\left[ \mathcal{D},\cos\theta \, \pi(\widehat{\Delta\rho}) + \sin\theta \, \pi(\widehat{\Delta\rho}_{\perp}) \right]\|_{\text{op}} \le \abs{cos\theta} \| \left[ \mathcal{D},\pi(\widehat{\Delta\rho}) \right] \|_{\text{op}}  + \abs{\sin\theta} \| \left[ \mathcal{D},\pi(\widehat{\Delta\rho}_{\perp}) \right] \|_{\text{op}}.
\label{3.1.10.1}
\end{equation}
This, however, just  means that the infimum of the L.H.S is bounded from above as
\begin{equation}
s = \inf_{ \theta \in [0 , \pi / 2) }\|\left[\mathcal{D},\pi(\hat{a})\right]\|_{\text{op}} \le  \inf_{\theta \in [0 , \pi / 2)}  \abs{cos\theta}\|[\mathcal{D},\pi(\widehat{\Delta\rho})]\|_{\text{op}} + \abs{\sin\theta} \|[\mathcal{D},\pi(\widehat{\Delta\rho}_{\perp})]\|_{\text{op}}.  \label{3.1.11}
\end{equation}
Here, of course, the mod over $\sin \theta$ and $\cos \theta$ are quite redundant, as $\theta$ is an acute angle as we have mentioned above. 

In \cite{FSBC}, $\| [\mathcal{D}, \pi(\widehat{\Delta\rho})] \|_{op}$ was erroneously identified as the upper bound of $s$, enabling us to identify, again erroneously, the RHS of \eqref{3.1.9}  as the upper bound. Recall in this context that by definition of the infimum, it should rather correspond to the highest lower bound, say $C$, satisfying
\begin{equation}
C \le  \| \cos\theta [\mathcal{D}, \pi(\widehat{\Delta\rho})] + \sin\theta [\mathcal{D}, \pi(\widehat{\Delta\rho}_{\perp}) ] \|_{\text{op}}  ~\forall~~ \Delta\rho_{\perp} \in W ,~\theta \in [0 , \pi / 2).
\end{equation}
Since the determination of a general formula to obtain $C$ is difficult, we have to be contented with just writing
\begin{equation}
s = \inf_{a \in B^\prime} \| \left[ \mathcal{D}, \pi(\hat{a}) \right] \|_{\text{op}} = \inf_{\substack{\Delta\rho}_{\perp} \in W \\ \theta \in [0 , \pi / 2)}  \| \cos\theta [ \mathcal{D}, \pi(\widehat{\Delta\rho})] + \sin\theta [\mathcal{D}, \pi(\widehat{\Delta\rho}_{\perp}) ] \|_{\text{op}}
\label{3.1.12}
\end{equation}
and find $s$ case by case. 

Returning to our derivation, we begin by rewriting the infinitesimal Connes distance function \eqref{3.1.3},\eqref{3.1.4} as
\begin{equation} \label{newB1}
\begin{split}
d(\rho,\rho^\prime) & = \sup_{a \in B^\prime } \abs{(\Delta\rho , a)} = \sup_{a \in B^\prime}  \|a\|_{\text HS} \abs{(\Delta\rho,\hat{a})} ~;~~ \Delta\rho = \rho - \rho^\prime\\
 & = \sup_{a \in B^\prime}  \|a\|_\text{HS} \abs{\left(\Delta\rho, \cos\theta \widehat{\Delta\rho} + \sin\theta \widehat{\Delta\rho}_{\perp} \right)} = \tilde{N} \| \Delta\rho \|_{\text HS},
\end{split}
\end{equation}
where $\tilde{N}$ is given by
\begin{equation} \label{newB2}
\tilde{N} = \sup_{a \in B^\prime} \left( \|a\|_\text{HS} |\cos \theta| \right).
\end{equation}
We now make use of (\ref{3.1.10}, \ref{3.1.12}) to recast $\tilde{N}$ as
\begin{equation}
\tilde{N} \leq  \frac{1}{\inf\limits_{\substack{\Delta\rho_{\perp} \in W \\ \theta \in [0 , \pi / 2)}}  \| [\mathcal{D}, \pi(\widehat{\Delta\rho})] + \tan\theta  [\mathcal{D}, \pi(\widehat{\Delta\rho}_{\perp})] \|_\text{op} }
\label{3.1.13}
\end{equation}
with $\tan \theta$ varying within the range $0 \leq \tan \theta < \infty$. We would like to point out at this stage that the optimal element $a_s \in \mathcal{A}$ for which the supremum in \eqref{newB1} is reached has to necessarily saturate the ball condition as well i.e. $\| [\mathcal{D}, \pi (a_s)]\|_{op} = 1.$ To see this, suppose $\| [\mathcal{D}, \pi (a_s)]\|_{op}  = \lambda < 1.$ Then we can easily see that $\left( \frac{a_s}{\lambda} \right)$ will saturate the ball condition and then $d(\rho, \rho')$ obtained by the inner product $\big|\left(\Delta \rho, \frac{a_s}{\lambda}\right)\big|$ will attain even a higher value: $\frac{1}{\lambda} |(\Delta \rho, a_s)| > |(\Delta \rho, a_s)|$. We therefore see that the optimal element can be effectively searched among the algebra elements $a_s \in \mathcal{A}$, which are positive self-adjoint elements \cite{Mart3} and importantly, saturate the ball condition. This implies, in turn, that the inequalities in (\ref{low_bound},\ref{3.1.13}) will actually correspond to equalities. With this, we can write 
\begin{equation}
\widetilde{N} = \frac{1}{\inf\limits_{\Delta\rho \in W,~ \theta \in [0, \frac{\pi}{2}]} \| [\mathcal{D}, \pi(\widehat{\Delta\rho})] + \tan\theta[\mathcal{D}, \pi(\widehat{\Delta\rho}_\perp)] \|_{op}}.
\end{equation}
Note that here too we can invoke the triangle-inequality to write
\begin{equation}
\| [\mathcal{D}, \pi(\widehat{\Delta\rho})] +  \tan\theta [\mathcal{D}, \pi(\widehat{\Delta\rho}_{\perp})] \|_\text{op}  \\
  \leq \| [\mathcal{D}, \pi(\widehat{\Delta\rho})] \|_\text{op} +  \abs{\tan\theta} \|[\mathcal{D}, \pi(\widehat{\Delta\rho}_{\perp}) ] \|_\text{op} \label{norm-ineq}
\end{equation}
with the R.H.S. being a monotonically increasing function of $\theta$. This implies that its value at $\theta = 0$ yields the upper bound of the L.H.S. Correspondingly, this yields back the lower bound \eqref{3.1.9}. Finally, we can eliminate the unit vectors $\widehat{\Delta\rho}$ and $\widehat{\Delta\rho}_{\perp}$ in the ``longitudinal" and ``transverse" directions respectively to re-write the formula (\ref{newB1}-\ref{3.1.13}) in terms of the original vectors $\Delta\rho$ and $\Delta\rho_\perp$ themselves, by multiplying both numerator and denominator by $\| \Delta\rho \|_\text{HS}$ to get 
\begin{equation} \label{rev_formula}
d(\rho, \rho+\Delta\rho) = N \| \Delta\rho \|^2_\text{HS} =\tilde{N}\| \Delta\rho \|_\text{HS},
\end{equation}
where
\begin{equation} \label{N_finite}
N = \frac{1}{\inf\limits_{\substack{\Delta\rho}_{\perp} \in W,~\kappa \in [0 , \infty)} \| [\mathcal{D}, \pi(\Delta\rho)] + \kappa [\mathcal{D}, \pi(\Delta\rho_{\perp})] \|_\text{op}}~;~~~\kappa=\frac{\| \Delta\rho \|_\text{HS}}{\| \Delta\rho_\perp \|_\text{HS}}\tan\theta.
\end{equation}
From \eqref{3.1.3} it is clear that the Connes' distance function can only depend on $\Delta\rho$, i.e. 
\begin{equation}
d(\rho,\rho+\Delta\rho)=d(\Delta\rho).
\end{equation}
It is also elementary to see that for any unitary transformation $U$
\begin{equation}
d(U\Delta\rho U^\dagger)=d(\Delta\rho).
\end{equation}
This also implies, from \eqref{rev_formula}, that 
\begin{equation}
\label{N-trans}
\tilde N(U\Delta\rho U^\dagger)=\tilde N(\Delta\rho), \quad N(U\Delta\rho U^\dagger)= N(\Delta\rho).
\end{equation}
In general we note from \eqref{3.1.13} and \eqref{N_finite} that $N$ and $\tilde N$ depend on the 'direction' of $\Delta\rho$, in the sense that even if $\|\Delta\rho^\prime\|_\text{HS}=\|\Delta\rho\|_\text{HS}$, $N(\Delta\rho^\prime)\ne N(\Delta\rho)$.  However, if $\|\Delta\rho^\prime\|_\text{HS}=\|\Delta\rho\|_\text{HS}$ implies $\Delta\rho^\prime=U\Delta\rho U^\dagger$, this dependence disappears using \eqref{N-trans}, $\tilde N$ is a constant as $\widehat{\Delta\rho}^\prime$ and $\widehat{\Delta\rho}$ both have norm one and $N=\frac{\tilde N}{\|\Delta\rho\|_\text{HS}}$.  This is the case for the coherent state basis in the Moyal plane, where equality of the HS norms implies that $\Delta\rho^\prime$ and $\Delta\rho$ differ by a rotation of the form $R=e^{i\phi b^\dagger b}$.  This explains why the Connes' distance on the Moyal plane is proportional to the HS norm, which is simply the Euclidean distance, infinitesimally and for finite distances.  We corroborate this result in the next section through a more explicit calculation.  

More generally, it is not difficult to verify that $\|\Delta\rho^\prime\|_\text{HS}=\|\Delta\rho\|_\text{HS}$ implies $\Delta\rho^\prime=U\Delta\rho U^\dagger$ if $\Delta\rho$ and $\Delta\rho^\prime$ are the difference of two orthogonal pure states, in which case $\tilde N$ is again a constant.  This, and more general scenarios under which this holds will be explored elsewhere.
 
However, this clearly cannot hold in general, but when this is the case, it readily follows that $\tilde{N}$, as determined from (\ref{3.1.13}) indeed corresponds to a numerical constant and upto this constant \eqref{newB2} the metric $g_{ij}$ that can be read off from the infinitesimal distance (\ref{newB1}), is indeed given by the Provost-Vallee form i.e. $g_{ij} \varpropto (\partial_i \rho , \partial_j \rho)$ \cite{FSBC, Val} in the coherent state basis in particular. In the case of the Moyal plane, this readily yields a flat metric, as in \cite{FSBC}, so that the straight lines are expected to play the role of geodesics. Indeed this fact was used implicitly in the parametrization \eqref{eqA1} below. On the other hand, for the case of the fuzzy sphere, although the metric is that of commutative sphere - upto an overall numerical factor, it turns out that the finite distance is quite different from the geodesic (great circle). Indeed, for the $n = 1/2$ representation of $SU(2)$ i.e. for the case of maximal non-commutativity, the distance turns out to be half of the chordal distance \cite{Var} and there does not exist any geodesic in the conventional sense, preventing one to integrate the infinitesimal distance to compute finite distance and the commutative result is obtained only in the $n \to \infty$ limit. As we shall see in the sequel that essentially similar results are obtained by applying our formula \eqref{rev_formula}. We begin with the computation of the finite distance between coherent states on the Moyal plane in the next section.

\section{Distance between finitely separated coherent states in $\mathds{R}^2_*$} \label{sec3}

The purpose of this section is to determine the spectral distance, \emph{a la} Connes, between an arbitrary pair of finitely separated coherent states $\ \ket{z} , \ket{z'} \in \mathcal{H}_{c}$, or more precisely between $\rho_{z} = \dyad{z}{z}$ and $\rho_{z'} = \dyad{z'}{z'} \in \mathcal{A} = \mathcal{H}_{q}$. Although a formal algorithm (\ref{rev_formula}) and (\ref{N_finite}) was devised for this purpose in the preceding section, this is not very user-friendly, as one needs an infinite number of variables to parametrise a general $\Delta \rho_\perp$, so that the computation of infimum in \eqref{N_finite} by varying all these parameters is virtually an impossible task. On the other hand, the lower bound (\ref{3.1.9}) can be easily computed as was done in \cite{FSBC} (at most up to a numerical constant). The strategy we therefore adopt is to emulate \cite{Mart} to obtain the corresponding upper bound and then look for an optimal element $a_s$ for which the saturation condition holds. If there exists at least one $a_s$ (note that this may not be unique!) then we can identify the upper bound to be the true distance. It may also happen in some situations that both upper and lower bounds coincide. In this case also, their common value can be identified as the distance. Otherwise, one has to play with different choices of $\Delta\rho_\perp$ in (\ref{N_finite}) to find the best possible estimate, as the upper bound cannot be identified as the true distance. We shall encounter a variety of such situations in the rest of the paper, which will help us to study and contrast various non-commutative spaces through the examples of $\mathds{R}^2_*$ and $\mathds{S}^2_*$. The present section deals with $\mathds{R}^2_*$.

We begin by considering the action of the state $\omega_z$, (associated to the fuzzy point $z$ in the Moyal plane, in the spirit of Gelfand and Naimark), on a generic algebra element $a \in \mathcal{H}_q$ as,
\begin{equation}
\begin{split}
\omega_{z}(a) = \text{tr}(\rho_{z} a) & = \text{tr}\left( U(z, \bar{z})\dyad{0}{0}U^{\dagger}(z , \bar{z}) a \right)\\
 & = \bra{0}\left(U^{\dagger}(z , \bar{z}) a  U(z, \bar{z})\right)\ket{0}.
\end{split}
\label{1.4}
\end{equation}
Note that we have made use of \eqref{CohSt} here. This means that the algebra element $a \in \mathcal{A} = \mathcal{H}_{q}$ gets translated by the adjoint action of $U(z, \bar{z})$ thereby furnishing a proper representation of the translation group. Without loss of generality we therefore only have to compute the distance between the pair of states $\omega_z$ and $\omega_0$ (taken to correspond to the ``origin" $z = 0$) as can easily be seen by invoking the transformational property of the Dirac operator $\mathcal{D}$ under translation \eqref{Dirac-z}, (as explained in Appendix A) and can be written, by using \eqref{ConDis}, as
\begin{equation}
d(\omega_{z} , \;\omega_{0}) = \sup_{a \in B} \abs{ \bra{0}(U^{\dagger}(z , \overline{z}) a  U(z, \overline{z}))\ket{0} - \bra{0} a\ket{0}}. \:\:
\label{1.5}
\end{equation}
Intuitively, $d(\omega_{z},\omega_{0})$ is the maximum change in the  expectation values of the $a \in B$ and the translated algebra element $U^{\dagger}(z , \overline{z}) a  U(z, \overline{z})$ in the same state $\ket{0} \in \mathcal{H}_{c}$. This is somewhat reminiscent of the transition from the Schr\"odinger to Heisenberg picture, where the operators are subjected to the unitary evolution in time through an adjoint action of the unitary operator, while the states are held frozen in time.

To proceed further, let us introduce a one-parameter family of density matrices $\rho_{zt} = \dyad{zt}{zt}$ with $t \in [0 ,1]$ taken to be, for convenience, a real affine parameter along the straight line connecting the origin $z = 0$ and the point $z$ in the complex plane. We can then introduce the function
\begin{equation}\label{eqA1}
W(t)= \omega_{zt}(a)=\text{tr}(\rho_{zt} a).
\end{equation}
Consequently, we have the following inequality,
\begin{equation}
\abs{\omega_z(a) - \omega_0(a)} = \abs{\int_{0}^{1} \derivative{W(t)}{t} dt }  \leq  \int_{0}^{1} \abs{\derivative{W(t)}{t} } dt.
\label{1.6}
\end{equation}
Since only a hermitian element $\left(a=a^{\dagger} \in \mathcal{A}\right)$ can give the supremum in the Connes' distance function \cite{Mart3}, an upper bound for $\abs{\derivative{W(t)}{t}}$ by making use of the Cauchy-Schwarz inequality can readily be obtained as
\begin{equation} \label{upper_bound}
\abs{\derivative{W(t)}{t} } = \abs{\bar{z} \omega_{zt}([b, a]) + z \omega_{zt}([b , a]^{\dagger})} \leq \sqrt{2}\abs{z} \sqrt{ \abs{\omega_{zt}([b, a]) }^{2} +   \abs{\omega_{zt}([b , a]^{\dagger}) }^{2}} \leq \sqrt{2}\abs{z} \sqrt{ ||[b, a]||_{op}^{2} +   ||[b,a]^{\dagger}||_{op}^{2}}.
\end{equation}
The ``ball" condition \eqref{ConDis} (With the Dirac operator $\mathcal{D}$ \eqref{DirOp}) reduces to a simpler form:
\begin{equation}
\| [b, a] \|_{op} = \|[b^{\dagger},a]\|_{op} \leq \sqrt{\frac{\theta}{2}}  ~~\textrm{for} ~a \in B.
\label{1.12}
\end{equation}
From \eqref{upper_bound} and \eqref{1.12}, one can therefore write
\begin{equation}
\abs{\derivative{W(t)}{t}} \leq \sqrt{2\theta}\abs{z}. 
\label{1.13}
\end{equation}
Hence from eq (\ref{ConDis}), (\ref{1.6} and (\ref{1.13}) we have the following upper bound for the Connes' distance:
\begin{equation}
d(\omega_{z}, \omega_{0}) \le \sqrt{2 \theta}\abs{z}.
\label{1.14}
\end{equation}
An optimal element saturating this upper bound can now be easily identified as
\begin{equation}
a_{s} = \sqrt{\frac{\theta}{2}} \left( b  e ^{-i \alpha} + b^{\dagger} e ^{i \alpha}\right),
\label{1.15.1}
\end{equation}
where $z =\abs{z} e^{i\alpha}$. However a simple inspection shows that although $a_s \in B$:
\begin{equation}\label{eqA4}
\|[\mathcal{D},\pi(a_{s})]\|_{op} =  1,
\end{equation}
it fails to be a HS operator. In fact it is not even a compact operator. Consequently $a_{s} \notin \mathcal{H}_{q} = \mathcal{A}$, but can be thought of as belonging to the multiplier algebra. A resolution of this problem was provided in \cite{Mart}. We briefly describe their approach here, as we propose an alternative approach in the next sub-section.

Here one weakens the strong requirement that $a_s\in\mathcal{H}_q=\mathcal{A}$ and rather looks for a sequence $\left\lbrace a_{n} \right\rbrace$, $a_{n} \in B$ and also $a_n \in \mathcal{A}=\mathcal{H}_{q}$, by inserting a suitable operator-valued "Gaussian" factor and thereby rendering it a HS operator:
\begin{equation}\label{eqA6}
a_{n} = \sqrt{\frac{\theta}{2}} \left( b  e ^{-i \alpha} (e^{-\lambda_{n} b^{\dagger}b} )+  (e^{-\lambda_{n}  b^{\dagger}b }) b^{\dagger}  e ^{i \alpha} \right).
\end{equation}
Then the following proposition was proved in \cite{Mart} (proposition 3.5):
\paragraph*{Proposition:} Let $z = \abs{z} e^{\ i\alpha}$ be a fixed translation and $\lambda > 0$. Define $a =  \sqrt{\frac{\theta}{2}} \left( b^{'} + b^{'\dagger} \right)$,  where $b^{\prime} = b  e^{-i \alpha} \left( e^{-\lambda b^{\dagger}b} \right)$. Then there exists a $\gamma > 0$ s.t. $a  \in B$ (Lipschitz ball) for any $\lambda \le \gamma$.
Using this proposition, any generic element of the sequence \eqref{eqA6} can be written in terms of above $b^\prime$ as
\begin{equation*}
a = \sqrt{\frac{\theta}{2}} ( b' + b^{'\dagger} ) \in B ~~~~\mathrm{with}~~\lambda \le \gamma.
\end{equation*}
Now, it can be easily shown that
\begin{equation*}
\omega_{0}(a) = 0 ~~\mathrm{and}~~ \omega_{z}(a)= \sqrt{2 \theta} \ \abs{z} \exp(-\abs{z}^{2}(1- e^{-\lambda})).
\end{equation*}
Therefore, 
\begin{equation} \label{1.15.2}
d(\omega_z, \omega_0) = \lim\limits_{n\rightarrow\infty} \abs{\omega_{z}(a)- \omega_{0}(a)}=\lim_{\lambda \rightarrow 0}  \abs{\omega_{z}(a)- \omega_{0}(a)} =\lim_{\lambda \rightarrow 0}  \ \sqrt{2 \theta} \ \abs{z} \exp(-\abs{z}^{2}(1- e^{-\lambda})) =  \sqrt{2 \theta} \, \abs{z} .
\end{equation}

\subsection{Infinitesimal distance and optimal element}

It is clear, on the other hand, from the translational symmetry of the Dirac operator $\mathcal{D}$ \eqref{DirOp} (see appendix A) that it is quite adequate to look for the optimal element $a_s$ at the level of the infinitesimal distance itself. The anticipated advantage is that $a_s$ \eqref{1.15.1} when projected to a finite dimensional sub-space will be automatically HS. Besides, it will help us to put the computation presented in \cite{FSBC} in the context of the present analysis and to zoom in on the source of the mismatch between the result of \cite{FSBC} and \cite{cag}, by a factor of $\sqrt{3}$. We therefore turn our attention towards the computation of the Connes distance between infinitesimally separated coherent states $\rho_z = | z )$ and $\rho_{z+dz}$. Clearly, this can just be read off from \eqref{1.15.2} to yield, by invoking translational symmetry (see Appendix A.1),
\begin{equation}\label{eqA7}
d(\rho_0,\rho_{dz})=d(\rho_z, \rho_{z + dz}) = \sqrt{2\theta} \abs{dz}
\end{equation}
and the apparent optimal element for the infinitesimal case is easily obtained by projecting $a_s$ (\ref{1.15.1}) into the 2D subspace $\Span\{|0\rangle,|1\rangle\}$ to get
\begin{equation}\label{1.16}
a^{eff}_{s}  = P_2 a_s P_2 = \sqrt{\frac{\theta}{2}} \frac{d\rho}{\abs{dz}}~; ~~~P_2 \equiv (\ket{0}\bra{0}+\ket{1}\bra{1}) \in \mathcal{H}_q
\end{equation}
where
\begin{equation}\label{a1}
d\rho = \rho_{dz} - \rho_0 = \dyad{dz}{dz} - \dyad{0}{0} = d\bar{z} \dyad{0}{1} + dz \dyad{1}{0}
\end{equation}
upto $\mathcal{O}(dz^2, d\bar{z}^2)$. Thus $a_s^{eff}$ \eqref{1.16} is proportional to $d\rho$ and the lower bound \eqref{3.1.9} is expected to yield the correct distance. But this $P_2 \in \mathcal{H}_q$ although seems to be quite adequate to yield the infinitesimal distance through the inner product \eqref{dis}, the effective optimal element $a_s^{eff}$ \eqref{1.16} is thrown out of the ball $B$ by this projector. This stems from the fact that $P_2$ or for that matter the representation $\pi(P_N) = \begin{pmatrix}
P_N & 0\\
0 & P_N\
\end{pmatrix} \in \mathcal{H}_q \otimes M_2(\mathbb{C})$ for the more general projection operator $P_{N+1} = \sum\limits_{n=0}^{N} |n \rangle \langle n | \in \mathcal{H}_q$ is \emph{not} built out of the eigen-spinors \eqref{Dir_bas_M} of the Dirac operators \eqref{corr4A2}. Rather $\pi\left(P_{N+1}\right)$ can be naturally associated to the basis \eqref{res_id}. Consequently the subspace $\pi\left(P_{N+1} a_s P_{N+1}\right)$ is not invariant under the adjoint action of the Dirac operator \eqref{DirOp}; there is a protruding part. This results in a linearly divergent operator norm for the commutator $\left[ \mathcal{D}, \pi\left(P_{N+1} a_s P_{N+1}\right) \right]$ :
\begin{equation*}
\| \left[ \mathcal{D}, \pi\left(P_{N+1} a_s P_{N+1}\right) \right] \|_{op} = \sqrt{N(N+1) + 1} ~ > 1 ~~;~~ \mathrm{for}~ N \geq 1
\end{equation*}
For $N=1$ in particular, this yields
\begin{equation*}
\| \left[ \mathcal{D}, \pi\left(P_{N+1} a_s P_{N+1}\right) \right] \|_{op} = \sqrt{3}
\end{equation*}
and accounts for the mis-match of a factor $\sqrt{3}$ between the results of \cite{FSBC} and \cite{cag}.

We are now therefore motivated to present an alternative to above ``Gaussian" sequence approach of (\ref{eqA6})-(\ref{1.15.2}) \cite{Mart} by constructing a sequence of projected $\pi(a_s)$ in $\mathcal{H}_q \otimes M_2(\mathds{C})$, rather than projected $a_s$ in $\mathcal{H}_q$ as in \eqref{1.16}, using a projector which is appropriate for the eigen-spinor basis of the Dirac operator \eqref{Dir_bas_M}.

We begin with the diagonal representation of $a\in\mathcal{A}=\mathcal{H}_q$, i.e. $\pi(a) = \begin{pmatrix}
a & 0\\
0 & a\
\end{pmatrix} \in \mathcal{H}_q \otimes M_2(\mathds{C})$. In particular for $a = d\rho$ \eqref{a1}, we have
\begin{equation} \label{pi_d_rho}
\pi(d\rho) \equiv \begin{pmatrix}
d\rho & 0\\
0 & d\rho\
\end{pmatrix} = \begin{pmatrix}
0 & \frac{d\bar{z}}{\sqrt{2}} & \frac{d\bar{z}}{\sqrt{2}} & 0 & 0\\
\frac{dz}{\sqrt{2}} & 0 & 0 & \frac{d\bar{z}}{2} & -\frac{d\bar{z}}{2}\\
\frac{dz}{\sqrt{2}} & 0 & 0 & -\frac{d\bar{z}}{2} & \frac{d\bar{z}}{2}\\
0 & \frac{dz}{2} & -\frac{dz}{2} & 0 & 0\\
0 & -\frac{dz}{2} & \frac{dz}{2} & 0 & 0\
\end{pmatrix} \in \mathcal{H}_q \otimes M_2(\mathds{C}),
\end{equation}
where the columns and rows are labelled by $| 0 \rangle\rangle, | 1 \rangle\rangle_+, | 1 \rangle\rangle_-, | 2 \rangle\rangle_+, | 2 \rangle\rangle_-$ of (\ref{Dir_bas_M}), respectively. Note that it has vanishing entries in the remaining rows/columns, indexed by $|n\rangle\rangle_{\pm}$, with $n\ge 3$.

Proceeding with the same proposed optimal element \eqref{1.15.1}, we see that it is quite adequate to project it on the same above $5$D subspace spanned by the eigen-spinors $| 0 \rangle\rangle, | 1 \rangle\rangle_\pm, | 2 \rangle\rangle_\pm$.
\begin{equation}
\pi(a_s) \to \mathds{P}_2 \pi(a_s) \mathds{P}_2 ~~\mathrm{with}~~ \mathds{P}_2 = | 0 \rangle\rangle \langle\langle 0 |  +  | 1\rangle\rangle_+ \, _+\langle\langle 1| + | 1\rangle\rangle_- \, _-\langle\langle 1| + | 2\rangle\rangle_+ \, _+\langle\langle 2| + | 2\rangle\rangle_- \, _-\langle\langle 2|.
\end{equation}
On computation this yields
\begin{equation}\label{proj-a_s}
\mathds{P}_2 \pi(a_s) \mathds{P}_2 = \sqrt{\frac{\theta}{2}} \begin{pmatrix}
0 & \frac{1}{\sqrt{2}} & \frac{1}{\sqrt{2}} & 0 & 0\\
\frac{1}{\sqrt{2}} & 0 & 0 & \frac{\sqrt{2} + 1}{2} & \frac{\sqrt{2} - 1}{2}\\
\frac{1}{\sqrt{2}} & 0 & 0 & \frac{\sqrt{2} - 1}{2} & \frac{\sqrt{2} + 1}{2}\\
0 & \frac{\sqrt{2} + 1}{2} & \frac{\sqrt{2} - 1}{2} & 0 & 0\\
0 & \frac{\sqrt{2} - 1}{2} & \frac{\sqrt{2} + 1}{2} & 0 & 0\
\end{pmatrix}.
\end{equation}
Consequently,
\begin{equation}
[\mathcal{D}, \mathds{P}_2 \pi(a_s) \mathds{P}_2] \equiv \begin{pmatrix}
0 & -\frac{1}{\sqrt{2}} & \frac{1}{\sqrt{2}} & 0 & 0\\
\frac{1}{\sqrt{2}} & 0 & 0 & -\frac{1}{2} & \frac{1}{2}\\
-\frac{1}{\sqrt{2}} & 0 & 0 & -\frac{1}{2} & \frac{1}{2}\\
0 & \frac{1}{2} & \frac{1}{2} & 0 & 0\\
0 & -\frac{1}{2} & -\frac{1}{2} & 0 & 0\
\end{pmatrix}
\end{equation}
and finally we have
\begin{equation}
[\mathcal{D}, \mathds{P}_2 \pi(a_s) \mathds{P}_2]^\dagger [\mathcal{D}, \mathds{P}_2 \pi(a_s) \mathds{P}_2] \equiv \left( \begin{tabular}{c|c}
$\mathds{1}_{3 \times 3}$ & $0_{3 \times 2}$\\
\hline
$0_{2 \times 3}$ & $B_{2 \times 2}$\
\end{tabular} \right),
\end{equation}
where
\begin{equation} \label{B_matrix}
B = \begin{pmatrix}
1/2 & -1/2\\
-1/2 & 1/2\
\end{pmatrix}
\end{equation}
is a $2 \times 2$ square matrix having eigenvalues $1$ and $0$ and $O_{m \times n}$ refers to a rectangular null matrix with $m$ rows and $n$ columns. Thus clearly we have in this case,
\begin{equation} \label{mix_ball}
\|[\mathcal{D}, \mathds{P}_2 \pi(a_s) \mathds{P}_2]\|_{op} = 1
\end{equation}
and,
\begin{equation} \label{innpro_moy}
\frac{1}{2} \abs{\langle\langle \pi(d\rho) \mid \mathds{P}_2 \pi(a_s) \mathds{P}_2 \rangle\rangle} = \sqrt{2\theta} \abs{dz},
\end{equation}
where $\langle\langle . \mid . \rangle\rangle$ denotes the inner product between a pair of elements $\mathcal{H}_q \otimes M_2(\mathds{C})$ given by
\begin{equation} \label{innpro_Hq}
\langle\langle A_1 \mid A_2 \rangle\rangle = Tr_{\mathcal{H}_c \otimes \mathds{C}^2} (A_1^\dagger A_2) ~~~~;~~ A_1, A_2 \in \mathcal{H}_q \otimes M_2(\mathds{C})
\end{equation}
and is the counter part of \eqref{C3}. Again, the subscript $\mathcal{H}_c \otimes \mathds{C}^2$ indicates that the trace has to computed over $\mathcal{H}_c \otimes \mathds{C}^2$. Further note that a factor of $1/2$ has been inserted in (\ref{innpro_moy}) in anticipation to relate the inner products \eqref{C3} and \eqref{innpro_Hq}, in case both $A_1$ and $A_2$ in \eqref{mix_ball} are representations of $a_1, a_2 \in \mathcal{H}_q$ such that $A_1 = \pi(a_1)$ and $A_2 = \pi(a_2)$. In that case, they can be related as, $(a_1, a_2) = \frac{1}{2} \langle\langle \pi(a_1) \mid \pi(a_2) \rangle\rangle$, as the factor $\frac{1}{2}$ compensates the double counting arising from the diagonal representation $\pi$. Of course here one can easily see that $\nexists$ any $a \in \mathcal{H}_q$ s.t. $\pi(a) = \mathds{P}_2 \pi(a_s) \mathds{P}_2$ and one can not simply relate \eqref{innpro_Hq} with any inner products $(.,.)$ of $\mathcal{H}_q$. Indeed, if it were to exist, we could have identified this `$a$', using (\ref{mix_ball}) and (\ref{innpro_moy}), to be the optimal element itself, which by definition has to belong to $\mathcal{H}_q=\mathcal{A}$, or at best to the multiplier algebra. In fact, this will be a persistent feature with any finite $(2N+1)$-dimensional projection $\mathds{P}_N \pi(a_s) \mathds{P}_N$ with
\begin{equation}\label{P-N}
\mathds{P}_N = | 0 \rangle\rangle \langle\langle 0 | + \sum\limits_{n = 1, \pm}^{N} | n \rangle\rangle_{\pm} \, _{\pm}\langle\langle n | = \begin{pmatrix}
P_N & 0 \\ 0 & P_{N-1}
\end{pmatrix}.
\end{equation}

One can note at this stage, however, that one can keep on increasing the rank of the projection operator $\mathds{P}_N$ indefinitely without affecting (\ref{mix_ball}) and (\ref{innpro_moy}) in the sense that the counter part of these equations still has the same form
\begin{equation} \label{proj_ball}
\| [\mathcal{D}, \mathds{P}_N \pi(a_s) \mathds{P}_N] \|_{op} = 1
\end{equation}
and
\begin{equation}
\frac{1}{2} \abs{\langle\langle \pi(d\rho) \mid \mathds{P}_N \pi(a_s) \mathds{P}_N \rangle\rangle} = \sqrt{2\theta} \abs{dz}
\end{equation}
are independent of $N$ if $N \geq 2$.

These equations again follow from the fact that
\begin{equation}
[\mathcal{D}, \mathds{P}_N \pi(a_s) \mathds{P}_N]^\dagger [\mathcal{D}, \mathds{P}_N \pi(a_s) \mathds{P}_N] = \left( \begin{array}{c|c}
\mathds{1}_{(2N-1) \times (2N-1)} & O_{(2N-1) \times 2}\\
\hline
O_{2 \times (2N-1)} & B\
\end{array} \right)
\end{equation}
with $B$ \eqref{B_matrix} again appearing as the lower block and $\pi(d\rho)$ \eqref{pi_d_rho} has "support" only on the first $5 \times 5$ block.

Finally, since in the limit $N \to \infty$, $\mathds{P}_N \to \mathds{1}$ by \eqref{res_id}, we have $\mathds{P}_N \pi(a_s) \mathds{P}_N \to \pi(a_s)$. One can thus interpret
\begin{equation}
\| [\mathcal{D}, \pi(a_s) \|_{op} \equiv \lim\limits_{N \to \infty} \| [\mathcal{D}, \mathds{P}_N \pi(a_s) \mathds{P}_N] \|_{op} = 1
\end{equation}
and
\begin{equation}
(d\rho, a_s) = \frac{1}{2} \abs{\langle\langle \pi(d\rho) \mid \pi(a_s) \rangle\rangle} \equiv \lim\limits_{N \to \infty} \frac{1}{2} \abs{\langle\langle \pi(d\rho) \mid \mathds{P}_N \pi(a_s) \mathds{P}_N \rangle\rangle} = \sqrt{2\theta} \, \abs{dz}.
\end{equation}
Thus, instead of inserting a Gaussian factor, as in \eqref{eqA6}, we have a sequence $\left\lbrace \mathds{P}_N \pi(a_s) \mathds{P}_N \right\rbrace$ of HS operators living in $\mathcal{H}_q \otimes M_2(\mathds{C})$ (note that $\mathcal{H}_q \otimes M_2(\mathds{C})$ can be regarded as Hilbert-Schmidt operators acting on $\mathcal{H}_c \otimes \mathds{C}^2$) and each of them satisfy the ball condition \eqref{proj_ball}. This is accomplished by projecting in the finite dimensional subspaces by $\mathds{P}_N$ \eqref{P-N}, spanned by Dirac eigen-spinors \eqref{res_id} rather than projecting just to $\mathcal{H}_q$ by $P_N$, where the ball condition gets violated and the operator norm diverges linearly. We would also like to point out in this context that $\pi(d\rho)$ (\ref{pi_d_rho}) and $\mathds{P}_2\pi(a_s)\mathds{P}_2$ (\ref{proj-a_s}) are not proportional anymore, unlike their counterparts (\ref{1.16}). However, here too we can easily split $\mathds{P}_2\pi(a_s)\mathds{P}_2$ or for that matter $\pi(a_s)$ itself in the limit $N\rightarrow\infty$, into the longitudinal and transverse components, but now in $\mathcal{H}_q\otimes M_2(\mathds{C})$ and not in $\mathcal{H}_q$. This requires a slight generalization of the analysis presented in section \ref{Sec5}. This, however, is not very useful in this context and we do not pursue it here anymore.

It is finally clear from the above analysis that the upper bound \eqref{1.14} is saturated in the infinitesimal case through the sequence $\left\lbrace \mathds{P}_N \pi(a_s) \mathds{P}_N \right\rbrace$ in the limit $N\rightarrow\infty$, allowing one to identify
\begin{equation}
d(\rho_0, \rho_{dz}) = d(\dyad{0}{0}, \dyad{dz}{dz}) = \sqrt{2\theta} \, \abs{dz}.
\end{equation}
Invoking translational symmetry of the distance (See \eqref{trans_inv} in Appendix \ref{App1})
\begin{equation}\label{inf-dis}
d(\rho_0,\rho_{dz})=d(\rho_z, \rho_{z + dz}) = \sqrt{2\theta} \, \abs{dz} ~~\forall~~ z \in \left\lbrace zt, t \in [0,1] \right\rbrace
\end{equation}
and one concludes that for finitely separated states, one can write
\begin{equation}
d(\rho_0, \rho_{z}) = \sqrt{2\theta} \, \abs{z}, \label{fin-dis}
\end{equation}
reproducing the result \eqref{1.15.2} and identify the straight line joining $z = 0$ to $z$ to be geodesic of the Moyal plane enabling one to integrate the infinitesimal distance (\ref{inf-dis}) along this geodesic to compute finite distance. In fact, one can easily see at this stage that the distance \eqref{fin-dis} can be written as the sum of distances $d(\rho_0,\rho_{zt})$ and $d(\rho_0,\rho_{zt})$ as,
\begin{equation}
d(\rho_0,\rho_{z})=d(\rho_0,\rho_{zt})+d(\rho_{zt},\rho_{z}), \label{dis-zt}
\end{equation}
where $\rho_{zt}$ is an arbitrary intermediate pure state from the one-parameter family of pure states, introduced in \eqref{eqA1}, so that the respective triangle inequality becomes an equality.

As we shall subsequently see this feature will not persist for other generic non-commutative spaces and we will demonstrate this through the example of the fuzzy sphere later. Before that we, however, complete our study of the Moyal plane by computing the distance between the so-called discrete ``harmonic oscillator" states in the next section.


\section{Connes distance between discrete ``harmonic oscillator" states} \label{sec4}

\subsection{Distance between infinitesimally separated discrete ``harmonic oscillator" states $\ket{n}$ and $\ket{n+1}$ in the Moyal plane}

For the discrete case, we consider a pair of states, which are separated by an ``infinitesimal" distance. By this we mean the nearest states, which are eigenstates of $b^\dagger b$. To compute the distance between the states $\rho_{n+1} \equiv |n+1\rangle \langle n +1|$ and $\rho_{n} \equiv |n\rangle\langle n|$ we take a similar approach i.e. start with
\begin{equation}
d(\omega_{n+1}, \omega_n) = \sup_{a \in B} \abs{\text{tr}(\rho_{n+1}a) - \text{tr}(\rho_{n}a)}
\end{equation}
and re-express this as the difference in the expectation value of the transformed algebra element and that of itself in the same state $\ket{n}$ as
\begin{align*}
	d(\omega_{n+1}, \omega_n) & = \sup_{a \in B} \abs{\langle n | \frac{b}{\sqrt{n+1}} a \frac{b^\dagger}{\sqrt{n+1}}| n \rangle - \langle n | a | n \rangle}\\
	& = \sup_{a \in B} \frac{1}{n+1} \abs{\langle n | ([b,a] + ab)b^\dagger - (n+1)a| n \rangle}.\\
\end{align*}
On simplification this yields
\begin{equation}\label{2.1.1}
d(\omega_{n+1}, \omega_n) = \sup_{a \in B} \frac{1}{\sqrt{n+1}} \abs{\langle n |[b,a]| n+1 \rangle} = \sup_{a \in B} \frac{1}{\sqrt{n+1}} \abs{\langle n+1 |[b^{\dagger},a]| n \rangle}.
\end{equation}
Making use of the inequality \eqref{1.12} and the Bessel's inequality (written in terms of the matrix elements $A_{ij}$ of an operator $\hat{A}$ in some orthonormal bases)
\begin{equation}
\|A\|^2_{op} \geq \sum\limits_{i} |A_{ij}|^2 \geq |A_{ij}|^2,
\label{2.1.2}
\end{equation}
this reduces to
\begin{equation}
d(\omega_{n+1}, \omega_n) \leq \sqrt{\frac{\theta}{2(n+1)}}.
\label{2.1.4}
\end{equation}
Again the RHS will correspond to the required distance, provided that we can find at least one optimal element $a_s$ s.t. the above inequality is saturated. We try with the lower bound \eqref{3.1.9}:
\begin{equation}
a_s = \frac{d\rho}{\|[\mathcal{D}, \pi(d\rho)]\|_{op}},
\label{2.1.4.1}
\end{equation}
where $d\rho = \rho_{n+1} - \rho_{n} = \dyad{n+1}{n+1} - \dyad{n}{n}$. The operator norm $\|[\mathcal{D}, \pi(d\rho)] \|_{op}$ can now be computed using the eigen-spinor basis \eqref{Dir_bas_M}.  Here, for $d\rho=|n+1\rangle\langle n+1|-|n\rangle\langle n|$, we have
\begin{equation}
[\mathcal{D},\pi(d\rho)]=\sqrt{\frac{2}{\theta}}\left( \begin{tabular}{c|c}
0 & $A$\\
\hline
$-A^\dagger$ & 0\
\end{tabular} \right); ~~~\text{where}~~~ A=\begin{pmatrix}
  -\sqrt{n}&0&0  \\
 0&2\sqrt{n+1}&0 \\
 0&0&-\sqrt{n+2}
\end{pmatrix}
\end{equation}

with the rows and columns labeled from top to bottom and left to right respectively by $_{+} \langle\langle n | ,\, _{+}\langle\langle n+1 | ,\, _{+}\langle\langle n+2 |$ and $|n\rangle\rangle_{-}, \allowbreak |n+1\rangle\rangle_{-}, \allowbreak |n+2\rangle\rangle_{-}$. From this we get the operator norm as
\begin{equation}
\|[\mathcal{D}, \pi(d\rho)] \|_{op} =2\sqrt{\frac{2(n+1)}{\theta}}.
\end{equation}
Since  $\Tr(d\rho)^2=2$, we have
\begin{equation}
d(\omega_{n+1}, \omega_n) =|\text{tr}(d\rho~a_s)|= \frac{\text{tr}(d\rho)^2}{\|[\mathcal{D}, \pi(d\rho)] \|_{op}}=\sqrt{\frac{\theta}{2(n+1)}},
\label{2.1.5}
\end{equation}
demonstrating that the  result obtained in \cite{cag,FSBC} hold for the  ``harmonic oscillator" basis, unlike the coherent state of the previous section. Finally we would like to make an observation that this distance depends on $n$ and is a reflection of the fact that the ``translational invariance" is violated in the sense that $d(\omega_{n+1}, \omega_n) \neq d(\omega_n, \omega_{n-1})$ and this stems from the fact that $\left( \rho_{n+1} - \rho_{n} \right)$ and $\left( \rho_{n} - \rho_{n-1} \right)$ are not related by any unitary transformation despite their HS norms being equal: $\left[ \, \| \rho_{n+1} - \rho_{n} \|_{tr} = \| \rho_{n} - \rho_{n-1} \|_{tr} \right]$ (see below \eqref{N-trans}). This has to be contrasted with \eqref{inf-dis} in the coherent state basis.


\subsection{Distance between finitely separated discrete ``harmonic oscillator" states $\ket{n}$ and $\ket{m}$ in the Moyal plane} \label{sec5.2}

As this was already discussed in \cite{cag}, we just quote the essential results. Here we are concerned with the computation of the distance between $\rho_{n} \equiv \dyad{n}{n}$ and $\rho_{m} \equiv \dyad{m}{m}$ with the difference between the two integer labels $m$ and $n$ being $m-n \geq 2$. We start by writing,
\begin{align*}
d(\omega_{m}, \omega_{n}) & = \sup_{a \in B} \abs{\text{tr}(\rho_{n+k}a) - \text{tr}(\rho_{n}a)} ~~~;~~\mathrm{where}~ k = m - n \\
& = \sup_{a \in B} \abs{ \sum_{i=1}^{k} \left((\rho_{n+i} - \rho_{n+(i-1)}), a\right) }.\
\end{align*}
Making use of \eqref{2.1.1} along with \eqref{2.1.2} and \eqref{1.12}, one gets
\begin{equation} \label{finite_upper_bound}
d(\omega_{m}, \omega_{n}) \leq \sqrt{\frac{\theta}{2}}\sum_{i=1}^{k} \frac{1}{\sqrt{n+i}}.
\end{equation}

To find an optimal element $a_{s} \in B$ for which the above inequality is saturated, we must have
\begin{equation*}
\abs{\sum_{i=1}^{k} (a_{s})_{n+i,n+i} - (a_{s})_{n+(i-1),n+(i-1)}} = \sqrt{\frac{\theta}{2}} \sum_{i=1}^{k} \frac{1}{\sqrt{n+i}}.
\end{equation*}
Now if we let  $(a_{s})_{m,m} = 0 $ it implies $\abs{(a_{s})_{n,n}} = \sqrt{\frac{\theta}{2}} \sum_{i=1}^{k} \frac{1}{\sqrt{n+i}}$. Constructing such an $a_{s} \in B$ we get,
\begin{equation}
a_{s}^{(m-1)} = \sqrt{\frac{\theta}{2}} \sum_{p=0}^{m-1} \sum_{i=1}^{m-p} \frac{1}{\sqrt{p+i}} \dyad{p}{p}, 
\label{2.2.4}
\end{equation}
where $m = n + k$. This can be regarded as a member of the set $\left\lbrace a_{s}^{(N)} \right\rbrace$ with $N = 0, 1, 2, 3, \cdots$ with each member satisfying the ball condition : $\Big\| \left[ \mathcal{D}, \pi(a_{s}^{(N)}) \right] \Big\|_{op} = 1$ and all $a_{s}^{(N)}$'s with $N \geq m$ can be regarded as the optimal element to compute $d(\omega_m, \omega_n)$. To make it more transparent, let us rewrite it in an expanded form:
\begin{equation} \label{a_s_N}
a_{s}^{(N)} = \sqrt{\frac{\theta}{2}} \sum_{p=0}^{N} \sum_{i=1}^{N+1-p} \frac{1}{\sqrt{p+i}} \dyad{p}{p} = \sqrt{\frac{\theta}{2}} \left[ \sum_{i=1}^{N+1} \frac{1}{\sqrt{i}} \dyad{0}{0} +  \sum_{i=2}^{N+1} \frac{1}{\sqrt{i}} \dyad{1}{1} + \cdots + \frac{1}{\sqrt{N+1}} \dyad{N}{N} \right]
\end{equation}
The relevant distance can now be read-off from the difference of the coefficients of $\dyad{m}{m} = \rho_m$ and $\dyad{n}{n} = \rho_n$. Further this distance saturates i.e. reaches a plateau (whenever $N = m$), to the value
\begin{equation}
d(\omega_{m}, \omega_{n}) = \sqrt{\frac{\theta}{2}} \sum_{i=1}^{m-n} \frac{1}{\sqrt{n+i}}.
\label{2.2.5}
\end{equation}
and remains constant thereafter for all $a_s^{(N)}$ with $N \geq m$. From \eqref{2.2.5} it is also seen that the distances for the ``harmonic oscillator" states saturates the ``triangle inequality", 
\begin{equation}
d(\omega_{m}, \omega_{n}) = d(\omega_{m}, \omega_{l})+d(\omega_{l}, \omega_{n})  ~~~~\mathrm{for}~ n \leq l \leq m,
\label{2.2.6}
\end{equation}
reproducing the result of \cite{cag}.

In this paper we were concerned only with distances like \eqref{2.2.5}, where $n < m < \infty$. However note that, from the structure of $a_{s}^{(N)}$ \eqref{a_s_N}, it is clear that in the ``harmonic oscillator" basis it takes a diagonal form so that its operator norm can be read-off easily from the coefficients of $\dyad{0}{0} = \rho_0$, which is the largest and see that it diverges as $N \to \infty$.
\begin{equation} \label{mod_a_s_N}
\| a_{s}^{(N)} \|_{op} = \abs{Tr\left[ \left( \rho_0 a_{s}^{(N)} \right) \right]} = \abs{\langle 0 | a_{s}^{(N)} | 0 \rangle} = \sqrt{\frac{\theta}{2}} \sum_{i=1}^{N+1} \frac{1}{\sqrt{i}} \to \infty ~\mathrm{as}~ N \to \infty
\end{equation}
In other words, the unit ball in the Moyal plane is not norm-bounded, as observed in \cite{cag}. It should be mentioned in this context that there is no contradiction as the ball B \eqref{ConDis} defined through semi-norm $\|[\mathcal{D}, \pi(a)]\|_{op}$ has nothing to do with the operator norm $\| a \|_{op}$. In fact, formally defining $a_{s}^{(\infty)} := \lim_{N\rightarrow \infty} a_{s}^{(N)}$, we immediately see that  $a_{s}^{(\infty)}$ although belongs to B \eqref{ConDis}, is not an element of the algebra i.e. $a_{s}^{(\infty)} \not\in \mathcal{A} = \mathcal{H}_q$. Thus it does not fulfil the second condition, we have imposed in section \ref{sec2.5}, as it does not belong to the multiplier algebra either. We therefore leave out of our consideration the cases, where $a_{s}^{(\infty)}$ or operators like this can possibly play the role of optimal element. 

The fact that there are indeed such examples of pair of states, were shown already in \cite{cag}, where this operator or rather the entire families of such operators $\left\lbrace a_{s}^{(N)} \right\rbrace$ were used to compute a sequence of lower bounds $\left\lbrace L_{(N)} \right\rbrace$ for the distance between a pair of states $\rho(0) = \dyad{0}{0}$ and $\rho(s) = \frac{1}{\zeta(s)} \sum\limits_{m=1}^{\infty} \frac{1}{m^s} \dyad{m}{m}$. The latter state represents, in turn a family of density matrices parametrized by a real parameter $s$ through Riemann-Zeta function $\zeta(s)$. The sequence of lower bounds $\left\lbrace L_{(N)} \right\rbrace$ was obtained, in turn, by computing the inner products (as in \eqref{dis}) with the elements of the above sequence \eqref{a_s_N} $\left\lbrace a_{s}^{(N)} \right\rbrace$ as
\begin{equation}
d\left( \rho(s), \rho(0) \right) \geq \abs{\left( \left( \rho(s) - \rho(0) \right), a_{s}^{(N)} \right)} : = L_{(N)}
\end{equation}
It was further observed in \cite{cag} that this $L_{(N)}$'s also diverge as $N \to \infty$, for $s \leq \frac{3}{2}$. Note that since $\rho(s)$ has infinite number of components the above inner product yields progressively better estimates of the distance as $N$ increases, as $L_{(N)}$ is a monotonically increasing function of $N$. Not only that, it turns out that this sequence of \emph{lower bounds has no upper bound}, resulting in a divergent distance. This indicates that one cannot identify any optimal element, fulfilling \eqref{dis} for the pair of states and such pathological examples are not considered in this paper.

Finally note that $a_{s}$ \eqref{2.2.4} is no longer proportional to $\Delta \rho = \rho_{m} - \rho_{n} = \dyad{m}{m} - \dyad{n}{n}$ unlike (\ref{2.1.4.1}). Consequently, the distance computed here \eqref{2.2.5} will exceed the lower bound \eqref{3.1.9} and $\Delta\rho_\perp$ contributes non-trivially in (\ref{N_finite}).


\section{Fuzzy Sphere} \label{sec6}

We approach the problem of the fuzzy sphere in quite the same way as the Moyal plane. There are, however, some fundamental differences between these cases and we will comment on these as we proceed. To begin with, we shall first try to find the distance in the discrete basis and later we will look at the continuous coherent state basis. As will be seen in the subsequent discussion, it is convenient to adopt different techniques for these two cases, namely we need to use the Dirac eigen-spinors in the latter case.

\subsection{Distance Between Discrete States}

We begin with a particular fuzzy sphere of radius $r_n$, corresponding to a particular $n$. The discrete set of basis is indexed by $n_3$ as $\ket{n, n_3}$ or just $\ket{n_3}$ in an abbreviated from, where the index `$n$' is suppressed. We shall rather use the subscript `$n$' to denote the distance function $d_n (\omega, \omega^\prime)$ between a pair of states $\omega$ and $\omega^\prime$.

\subsubsection{Infinitesimal Distance}

We first compute the distance between the states $\rho_{n_{3}+1} \equiv |n_{3}+1\rangle \langle n_{3} +1|$ and $\rho_{n_{3}} \equiv |n_{3}\rangle\langle n_{3}|$  (this being the "infinitesimal separation" as far as discrete basis is concerned). Similar to the Moyal case, we start with
\begin{equation}
d_n(\omega_{n_3+1}, \omega_{n_3}) = \sup_{a \in B} \abs{\text{tr}(\rho_{n_3+1}a) - \text{tr}(\rho_{n_3}a)} = \sup_{a \in B} \frac{1}{\sqrt{n(n+1)-n_{3}(n_{3}+1)}} \abs{\langle n_{3} |[J_{-},a]| n_{3}+1 \rangle}.
\label{4.45}
\end{equation}
Again invoking Bessel's inequality \eqref{2.1.2}, we can write
\begin{equation}
d_n(\omega_{n_3+1}, \omega_{n_3})  \leq \frac{\|[J_{-},a]\|_{op}}{\sqrt{n(n+1)-n_{3}(n_{3}+1)}}  = \frac{\|[J_{+},a]\|_{op}}{\sqrt{n(n+1)-n_{3}(n_{3}+1)}}  \leq \frac{r_n}{\sqrt{n(n+1)-n_{3}(n_{3}+1)}},
\label{4.46}
\end{equation}
where we have made use of the inequality \eqref{A.7}, as proved in appendix \ref{ap_fuz}, where for $a \in B$, we have
\begin{equation}
\| [J_{-}, a] \|_{op} = \|[J_{+},a] \|_{op} \leq r_n  ;\ \ r_n = \lambda \sqrt{n(n+1)}.
\label{4.47}
\end{equation}
Like in the case of the Moyal plane, we try with the form corresponding to the lower bound (\ref{3.1.9}):
\begin{equation}
a_{s} = \frac{d\rho}{\|[\mathcal{D}, \pi\left(d\rho\right)]\|_{op}} ~~\mathrm{where}~~ d\rho = \dyad{n_{3}+1}{n_{3}+1} - \dyad{n_{3}}{n_{3}}.
\end{equation}
The operator norm $\|[\mathcal{D}, \pi(d\rho)] \|_{op}$ can now be computed using the Dirac operator eigen-spinor basis \eqref{Dir_bas_F} as
\begin{equation}
[\mathcal{D},\pi(d\rho)]=\frac{1}{r_n}\left( \begin{tabular}{c|c}
0 & $A$\\
\hline
$-A^\dagger$ & 0\
\end{tabular} \right),
\end{equation}
where
\begin{equation}
A = diag \left( \, -\sqrt{n(n+1)-n_3(n_3-1)}, ~ 2\sqrt{n(n+1)-n_3(n_3+1)}, ~ -\sqrt{n(n+1)-(n_3+1)(n_3+2)} ~ \right)
\end{equation}
with the rows and columns labeled from top to bottom and left to right respectively by $_+\langle\langle n,n_3-1 |, \allowbreak\,_+\langle\langle n,n_3 |_{+}, \allowbreak\,_+\langle\langle n,n_3+1 |$ and $| n,n_3-1 \rangle\rangle_{-}, \allowbreak | n,n_3 \rangle\rangle_{-}, \allowbreak n,n_3+1 \rangle\rangle_{-}$.  From this we get the operator norm as
\begin{equation}
\|[\mathcal{D}, \pi(d\rho)] \|_{op} =\frac{2}{r_n}\sqrt{n(n+1)-n_3(n_3+1)}.
\end{equation}

Again as ~tr$(d\rho)^2=2$, we can readily compute this infinitesimal distance to get,
\begin{equation}
d(\omega_{n_3+1}, \omega_{n_3}) =|\text{tr}(d\rho~a_s)|= \frac{\text{tr}(d\rho)^2}{\|[\mathcal{D}, \pi(d\rho)] \|_{op}}=\frac{r_n}{\sqrt{n(n+1)-n_3(n_3+1)}}.
\label{2.1.5.1}
\end{equation}
There is yet another optimal element given by
\begin{equation}
a_{s} = \frac{r_n}{\sqrt{n(n+1)-n_{3}(n_{3}+1)}} \dyad{n_{3}+1}{n_{3}+1}.
\end{equation}
For both these elements, the  "infinitesimal distance " is the upper bound itself. Furthermore it also coincides with the lower bound \eqref{3.1.9}. Thus,
\begin{equation}
d_n(\omega_{n_{3}+1}, \omega_{n_{3}}) = \frac{\lambda \sqrt{(n(n+1))}}{\sqrt{n(n+1)-n_{3}(n_{3}+1)}}.
\label{4.49}
\end{equation}
This reproduces the result of \cite{Var,Devi}.

\subsubsection{Finite Distance}

\paragraph{Discrete Basis}

For the finite case, we try to compute the distance between the states $\rho_{n_{3}} \equiv \dyad{n_{3}}{n_{3}}$ and $\rho_{m_{3}} \equiv \dyad{m_{3}}{m_{3}}$ with $\abs{m_{3}-n_{3}} \geq 2$ and $-n \le m_{3},n_{3} \le n$. Again, adopting the same technique as in the Moyal plane, we write
\begin{align*}
d_n(\omega_{m_{3}}, \omega_{n_{3}}) & = \sup_{a \in B} \abs{\text{tr}(\rho_{n_{3}+k}a) - \text{tr}(\rho_{n_{3}}a)} ~~~;~~\mathrm{where}~ k = m_{3} - n_{3} \\
& = \sup_{a \in B} \abs{ \sum_{i=1}^{k} \text{tr}\left((\rho_{n_{3}+i} - \rho_{n_{3}+(i-1)}), a\right) } \\
& \leq \sup_{a \in B}  \sum_{i=1}^{k} \frac{ \abs{\langle n_{3} +(i-1) | [J_{-},a]| n_{3}+i \rangle}}{\sqrt{n(n+1)- (n_{3} +i)(n_{3} +i-1)}}\\
& \leq \sum_{i=1}^{k} \frac{r_n}{\sqrt{n(n+1) - (n_{3}+ i)(n_{3}+ i -1) }}.
\end{align*}
It can be easily verified that the element $a_{s} \in B$, for which the above inequality is saturated is given by
\begin{equation}
a_{s} = \sum_{p=n_{3}}^{m_{3}-1} \left(\sum_{i=1}^{m_{3}-p} \frac{r_n}{\sqrt{n(n+1) - (p+i) (p+i -1)}} | p \rangle \langle p | \right),
\label{4.50}
\end{equation}
which gives
\begin{equation}
d_n(\omega_{m_{3}}, \omega_{n_{3}}) = \sum_{i=1}^{k} \frac{r_n}{\sqrt{n(n+1) - (n_{3}+ i)(n_{3}+ i -1) }}.
\label{4.51}
\end{equation}
From the above equation it is evident that for the discrete basis,
\begin{equation}
d_n(\omega_{m_{3}}, \omega_{n_{3}}) = d_n(\omega_{m_{3}}, \omega_{l_{3}}) + d_n(\omega_{l_{3}}, \omega_{n_{3}})  ~~~~\mathrm{for} \ n_{3} \leq l_{3} \leq m_{3}
\label{4.52}
\end{equation}
i.e. the distance between finitely separated states are additive and they saturate the triangle inequality for the fuzzy sphere, much like in the case of the Moyal plane \eqref{dis-zt}. In particular the distance between the north pole (N), denoted by $\omega_n = \dyad{n}{n}$ and south pole (S), denoted by $\omega_{-n} = \dyad{-n}{-n}$ is given by
\begin{equation} \label{DisNS}
d_n(N, S) = d_n(\omega_{n}, \omega_{-n}) = \sum_{k=1}^{2n} \frac{r_n}{\sqrt{k(2n + 1 - k)}}, ~~~\forall ~n.
\end{equation}
Since both the discrete states $\omega_n$ and $\omega_{-n}$, representing north and south poles are also coherent states, we should be able to reproduce the same results in our coherent state calculations, to be taken up in the next subsection. Furthermore, since the $n = 1/2$ case corresponds to maximal non-commutativity let us just write down the explicit distances between the north (N) and south poles (S) for the lowest three cases $n = 1/2$, $n=1$ and $n = 3/2$ respectively - as special cases. They are given by
\begin{equation} \label{dis_halfs}
d_{1/2} (\mathrm{N}, \mathrm{S}) =  r_{1/2} ~~;~~ d_1(\mathrm{N}, \mathrm{S}) = \sqrt{2} \, r_1~~;~~d_{3/2} (\mathrm{N}, \mathrm{S})=\Big(\frac{1}{2}+\frac{2\sqrt{3}}{3}\Big)r_{3/2}=1.6547~r_{3/2}.
\end{equation}
The coefficients in front of $r_{1/2}$,  $r_1$ and $r_{3/2}$ here indicates that they are highly deformed spaces and the corresponding distances are way below the corresponding commutative spheres $\pi r_{1/2}$,  $\pi r_1$  and  $\pi r_{3/2}$ respectively. This is just indicative of the fact that none of the fuzzy spheres with fixed $n$ allows one to define a geodesic in the conventional sense. We have more to say on this in the following subsections. However, one can expect the results to match with the commutative ones in the large $n$ limit so that the distance as given in \eqref{DisNS}, in the limit $n \to \infty$, should match asymptotically with $\pi r_n$, which is the geodesic distance between north and south pole for a commutative sphere.

To implement the commutative limit, we take $k/n = x_k$ and $\Delta x=x_k-x_{k-1}=1/n$ as the increment such that $\Delta x \to 0$ as $n \to \infty$. Then we can rewrite
\begin{equation}
\begin{split}
\lim\limits_{n \to \infty} \frac{d_n(N, S)}{r_n} & = \lim\limits_{n \to \infty} \sum_{k=1}^{2n} \frac{1}{\sqrt{k(2n + 1 - k)}} = \lim\limits_{n \to \infty} \sum_{k=1}^{2n} \frac{\frac{1}{n}}{\sqrt{\frac{k}{n}(2 + \frac{1}{n} - \frac{k}{n})}} \\
 & =\lim\limits_{\Delta x \to 0}\sum_{x=x_1}^{x_{2n}=2}\frac{\Delta x}{\sqrt{x_k(2+\Delta x-x_k)}}\to \int\limits_{0}^{2} \frac{dx}{\sqrt{x(2 - x)}} = 2 \int\limits_{0}^{1} \frac{ dt}{\sqrt{1 - t^2}}=\pi\
\end{split}
\end{equation}
where we have set $x=1-t$ to obtain the last integral.

\paragraph{Coherent States : Upper Bound of Finite Distance}

We begin by noting that the north and south poles are naturally associated with the extreme ends $(\theta = 0, \pi)$ of the one-parameter family of (pure) density matrices
\begin{equation} \label{Per_Coh_St}
\rho_\theta \equiv | \theta \rangle \langle \theta | = U_F (\theta) | n \rangle \langle n | U_F^\dagger (\theta) \in \mathcal{H}_n ~;~ U_F (\theta) = e^{-i\theta J_2}
\end{equation}
built out of Perelomov coherent states $| \theta \rangle = U_F (\theta) | n \rangle$ obtained by the action of $SU(2)$ in its $n$-th representation on $| n \rangle \equiv | n, n \rangle = | \theta = 0 \rangle$. Note also that we have taken without loss of generality the azimuthal angle $\phi = 0$. We can also parametrise the above family of (pure) density matrices by a suitable stereographic variable $z$ ($|z| = \tan \left( \frac{\theta}{2} \right)$) as $\rho_z \equiv \rho_\theta$, so that we can define as in the Moyal plane \eqref{1.4}, the action
\begin{equation}
\omega_z (a) = tr\left( \rho_z a \right) = \langle n | U_F^\dagger (z, \overline{z}) a U_F (z, \overline{z}) | n \rangle ~;~ a^\dagger = a \in \mathcal{H}_n \equiv \mathcal{A}
\end{equation}
In fact, we can emulate wholesale the analysis of the Moyal plane by defining the function $W(t) = \omega_{zt} (a) = tr\left( \rho_{zt} a \right)$, with $t \in [0,1]$ being a real parameter to rewrite $|\omega_z (a) - \omega_0 (a)|$, occurring in the distance formula
\begin{equation} \label{eq137}
d\left(\omega_z (a), \omega_0 (a) \right)= \sup\limits_{a \in B} |\omega_z (a) - \omega_0 (a)|
\end{equation}
as
\begin{equation}
\abs{\omega_z (a) - \omega_0 (a)} = | \int_{0}^{1} \derivative{W(t)}{t} dt |  \leq  \int_{0}^{1} | \derivative{W(t)}{t} | dt.
\label{4.56}
\end{equation}
A straightforward computation then yields
\begin{equation}
\abs{ \derivative{W(t)}{t} } = \left(\frac{|z| dt}{1 + |z|^{2} t^{2}}\right) \abs{\omega_{zt}\left(\left[ (J_+ - J_-), a \right] \right) }.
\label{4.58}
\end{equation}
An upper bound for the second factor can then be obtained by successive applications of inequalities of Cauchy-Schwartz and $\| \left[ J_\pm, a \right] \|_{op} \leq \lambda \sqrt{n(n+1)} = r_n$ \eqref{A.7} as
\begin{equation} \label{upp_bou_cohst}
\begin{split}
\abs{\omega_{zt}\left(\left[ \left( J_+ - J_- \right), a \right]\right)} & = \abs{\omega_{zt}\left([J_{+}, a] \right) - \omega_{zt}\left([J_{-}, a \right)}\\
 & \leq \sqrt{2} \sqrt{ \abs{\omega_{zt}\left( [J_{+}, a] \right) }^{2} + \abs{\omega_{zt}\left([J_{-} , a]\right) }^{2}}\\
 & \leq \sqrt{2} \sqrt{ \|[J_{+}, a]\|_{op}^{2} + \|[J_{-} , a] \|_{op}^{2}} ~~ \leq 2 r_n \
\end{split}
\end{equation}
Finally making use of all the equations (\ref{eq137} - \ref{upp_bou_cohst}), it is quite straightforward to show that
\begin{equation}
d\left( \omega_z, \omega_0 \right) = d \left( \omega_\theta, \omega_0 \right) \leq r_n \theta
\end{equation}
Therefore the upper bound of the Connes distance on the fuzzy sphere is actually the geodesic distance on the commutative sphere
\begin{equation} \label{Fuzz_up_bound}
d(\omega_{z} , \;\omega_{0}) = \sup_{a \in B} \abs{\omega_{z}(a) - \omega_{0}(a)} \le 2 r \tan^{-1} \abs{z} = r_n \theta.
\end{equation}

The stark difference with the Moyal plane is that we cannot find any algebra element saturating this inequality \eqref{Fuzz_up_bound}, not even through a sequence or through projections. This limit is actually the distance on a commutative sphere and is reached only for the commutative limit, $n \to \infty$ as we have shown earlier. This implies that for any finite $n$ representation, the distance between two points on the fuzzy sphere is less than the geodesic distance for a commutative sphere (see \eqref{dis_halfs}, for example) and the distance does not follow the conventional `geodesic' path as we know it. Indeed, we will show that for $n = 1/2$, the distance actually corresponds to half of the chordal distance between a pair of points on the surface of the sphere or more precisely between the associated pure states and interpolated by a one parameter family of mixed states. Only in the limit $n\rightarrow\infty$ this slowly deforms to become the great circle path on the surface.

\subsection{Ball condition in the eigen-spinor basis} \label{subsec_DirOp_fuz}

For the calculation of distances for coherent states or more precisely the operator-norm occurring in the ball condition \eqref{ConDis}, we now make use of the eigen-spinors \eqref{Dir_bas_F} of the Dirac operator \eqref{fuzz_dir}. We will first sketch the outline of the algorithm. One can calculate in a straightforward manner (at least in principle) the commutator $[\mathcal{D},\pi(a)]$ in the above-mentioned eigen-spinor basis \eqref{Dir_bas_F} to obtain
\begin{equation} \label{6.3.1}
[\mathcal{D},\pi(a)] = \frac{1}{r_n} \left( \begin{array}{c|c}
0_{(2n+2) \times (2n+2)} & A_{(2n+2) \times 2n} \\
\hline 
-~A^\dagger_{2n \times (2n+2)}&0_{(2n) \times (2n)}\
\end{array} \right),
\end{equation}
where we denote the non-vanishing rectangular matrices by $A$ and $A^\dagger$ as
\begin{equation} \label{6.3.2}
\begin{split}
A_{(2n+2) \times 2n}=(2n+1)_{+}\langle\langle n,n_3|\pi(a)| n, n_3^\prime \rangle\rangle_- ~~\Rightarrow~~_+\langle\langle n, n_3 | [\mathcal{D},\pi(a)] | n, n_3^\prime \rangle\rangle_- & \equiv \frac{1}{r_n}A_{(2n+2) \times 2n}\\
A^\dagger_{2n \times (2n+2)}=(2n+1)_{-}\langle\langle n,n'_3|\pi(a)| n, n_3 \rangle\rangle_+~~\Rightarrow ~~_-\langle\langle n, n_3^\prime | [\mathcal{D},\pi(a)] | n, n_3 \rangle\rangle_+ & \equiv -\frac{1}{r_n} A^\dagger_{2n \times (2n+2)}
\end{split}
\end{equation}
with the respective ranges for $n_3$ and $n'_3$ given by: $-n-1\le n_3\le n$ and $n-1\le n'_3\le n-1$.

The occurrences of rectangular null matrices in the diagonal blocks of \eqref{6.3.1} stems from the degeneracy of the spectrum of the Dirac operator (see comments below \eqref{eig_fuzz}). This yields a block diagonal form for
\begin{equation}\label{144}
[\mathcal{D},\pi(a)]^\dagger [\mathcal{D},\pi(a)] = \frac{1}{r_n^2} \left( \begin{tabular}{c|c}
$(A A^\dagger)_{(2n+2) \times (2n+2)}$ & $0_{(2n+2) \times 2n}$\\
\hline
$0_{2n \times (2n+2)}$ & $(A^\dagger A)_{2n \times 2n}$\
\end{tabular} \right)
\end{equation}
and, from the C$^*$-algebra, property we have
\begin{equation} \label{n1}  
\| [\mathcal{D},\pi(a)]\|_\text{op}^2=\| [\mathcal{D},\pi(a)]^\dagger [\mathcal{D},\pi(a)] \|_\text{op} = \frac{1}{r_n^2} \| AA^\dagger \|_\text{op} = \frac{1}{r_n^2} \| A^\dagger A \|_\text{op}.
\end{equation}

In what follows we shall make use use of \eqref{3.1.9} to first obtain the lower bound of the infinitesimal distance for a general $n$ representation. This will help us improve the estimate by introducing a suitable transverse part $\Delta\rho_\perp$ in \eqref{N_finite}. We consider the $n = 1/2$ and $n = 1$ representations separately in the subsequent subsections and try to calculate the finite Connes distance between the `north pole' and some other points represented by appropriate coherent states. We will see that for $n = 1/2$ the lower bound will itself correspond to the exact distance, leaving no room for $\Delta\rho_\perp$ to contribute. This is in contrast for $n \geq 1$, where $(\Delta \rho_\perp)$ will indeed play an important role in improving our estimation of the lower bound, except for $\theta = \pi$.

\subsection{Lower Bound for Infinitesimally Separated Coherent States}

Following the algorithm as explained in the previous subsection \ref{subsec_DirOp_fuz}, we easily find that $d\rho$ is in this case given by
\begin{equation}
d\rho = \dyad{dz}{dz} - \dyad{n}{n}.
\end{equation}
Using the infinitesimal version of \eqref{Per_Coh_St}, this yields
\begin{equation}
d\rho = d\theta \sqrt{\frac{n}{2}} \big[ \dyad{n}{n-1} + \dyad{n-1}{n} \big].
\end{equation}
Making use of \eqref{6.3.2}, we get the matrix form in the Dirac eigenbasis for $[\mathcal{D}, \pi(d\rho)]$ as
\begin{equation} 
[\mathcal{D}, \pi(d\rho)] \equiv \frac{1}{r_n} \left( \begin{tabular}{c|c}
0 & $A$\\
\hline
$-A^\dagger$ & 0\
\end{tabular} \right),
\end{equation}
where
\begin{equation}
A  =\begin{pmatrix}
d\theta \sqrt{n(n + \frac{1}{2})} & 0 & 0\\
0 & -d\theta \sqrt{n(n - \frac{1}{2})} & 0\\
-d\theta n \sqrt{2} & 0 & 0\
\end{pmatrix}.
\end{equation}
Here rows/columns are labeled from up to down/left to right respectively by $_{+} \langle\langle n | ,\, _{+}\langle\langle n-1 | ,\, _{+}\langle\langle n-2 |$ and $|n-1\rangle\rangle_-,|n-2\rangle\rangle_-,|n-3\rangle\rangle_-$. Subsequently, we get
\begin{equation}
[\mathcal{D}, \pi(d\rho)]^\dagger [\mathcal{D}, \pi(d\rho)] = \frac{1}{r_n^2} \left( \begin{tabular}{c|c}
$A A^\dagger$ & 0\\
\hline
0 & $A^\dagger A$\
\end{tabular} \right)
\end{equation}
with
\begin{equation}
A^\dagger A = n (d\theta)^2 \begin{pmatrix}
3n + \frac{1}{2} & 0 & 0\\
0 & n - \frac{1}{2} & 0\\
0 & 0 & 0\
\end{pmatrix}.
\end{equation}
Since $\|[\mathcal{D}, \pi(d\rho)]^\dagger [\mathcal{D}, \pi(d\rho)]\|_\text{op} = \frac{1}{r_n^2} \|A^\dagger A\|_\text{op}$, as follows from \eqref{144}, we have
\begin{equation}
\|[\mathcal{D}, \pi(d\rho)]\|_\text{op} = \frac{1}{r_n} \sqrt{\|A^\dagger A\|_\text{op}} = \frac{d\theta}{r_n} \sqrt{n (3n + \frac{1}{2})}.
\end{equation}
Using this, the lower bound \eqref{3.1.9} comes out to be
\begin{equation} \label{low_bound_fuz}
d(\omega_{z + dz}, \omega_z) \geq \frac{tr(d\rho^2)}{\| [\mathcal{D}, \pi(d\rho)] \|_\text{op}} = r_n d\theta \sqrt{\frac{2n}{6n + 1}}.
\end{equation}
Clearly, for $n = 1/2$ this comes out to be $r_{1/2} \frac{d\theta}{2}$. We show later that it coincides with the correct infinitesimal version of the finite distance for the $n = 1/2$ case, to be computed in the next subsection.

\subsection{The $n = 1/2$ fuzzy sphere} \label{section2}

Clearly here the algebra element can be taken to be a traceless hermitian $2 \times 2$ matrix without loss of generality. The traceless condition stems from the fact that the $2 \times 2$ identity matrix $\mathds{1}_2$ commutes with Dirac operator and therefore does not contribute to the ball condition. One can thus parametrise any generic algebra element `$a$' in terms of a $3$-vector $\vec{a} \in \mathds{R}^3$, by writing
\begin{equation}\label{al_element}
a = \vec{a}.\vec{\sigma}
\end{equation}
where $\vec{\sigma}$ stands for the three Pauli matrices.

This enables us to express all the matrix elements of $[\mathcal{D}, \pi(a)]$ in terms of the coefficients $a_i$. For any generic pair of states $|m\rangle\rangle_+$ and $|m'\rangle\rangle_-$ we have, by making use of the general framework sketched in section \eqref{subsec_DirOp_fuz}, the matrix element,
\begin{equation}
_+\langle\langle m | [\mathcal{D}, \pi(a)] | m^\prime \rangle\rangle_- = \frac{2}{r_{1/2}} ~_+\langle\langle m | \pi(a) | m^\prime \rangle\rangle_-,
\end{equation}
where $m$ takes values $-\frac{3}{2}, -\frac{1}{2}, \frac{1}{2}$ and $m^\prime = -\frac{1}{2}$. On explicit computation, one finds the following matrix elements for the commutator:
\begin{equation}
_+\Big\langle\Big\langle -\frac{3}{2} \Big| \pi(a) \Big| -\frac{1}{2} \Big\rangle\Big\rangle_-  = \frac{1}{\sqrt{2}} (a_1 + i a_2);~
_+\Big\langle\Big\langle -\frac{1}{2} \Big| \pi(a) \Big| -\frac{1}{2} \Big\rangle\Big\rangle_- = a_3;~
_+\Big\langle\Big\langle \frac{1}{2} \Big| \pi(a) \Big| -\frac{1}{2} \Big\rangle\Big\rangle_- = - \frac{1}{\sqrt{2}} (a_1 - i a_2)
\end{equation}
so that the final form of the commutator matrix (\ref{6.3.2}) in this case can be written as
\begin{equation}
\begin{split}
[\mathcal{D}, \pi(a)] & \equiv \frac{1}{r_{1/2}} \left( \begin{tabular}{c|c}
$0_{3 \times 3}$ & $\mathrm{A}_{3 \times 1}$\\
\hline
$- \mathrm{A}^\dagger_{1 \times 3}$ & 0\
\end{tabular} \right)
\end{split}, ~\text{where}~\mathrm{A}_{3 \times 1} =2~ _+\langle\langle m | [\mathcal{D}, \pi(a)] | m^\prime \rangle\rangle_-= \begin{pmatrix}
\sqrt{2} (a_1 + i a_2)\\
2a_3\\
- \sqrt{2} (a_1 - i a_2)
\end{pmatrix}.
\end{equation}
Finally,
\begin{equation}
[\mathcal{D}, \pi(a)]^\dagger [\mathcal{D}, \pi(a)] = \frac{1}{r^2_{1/2}} \left( \begin{tabular}{c|c}
$\mathrm{A}\mathrm{A}^\dagger$ & 0 \\
\hline
0 & $\mathrm{A}^\dagger \mathrm{A}$\
\end{tabular} \right).
\end{equation}

Since $\|\mathrm{A}^\dagger \mathrm{A}\|_\text{op} = \|\mathrm{A} \mathrm{A}^\dagger\|_{op} =r^2_{1/2} \| [\mathcal{D}, \pi(a)]^\dagger [\mathcal{D}, \pi(a)] \|_\text{op}$ and $\mathrm{A}^\dagger \mathrm{A}$ is just a number, we have
\begin{equation}
\| [\mathcal{D}, \pi(a)]^\dagger [\mathcal{D}, \pi(a)] \|_\text{op} = \frac{1}{r^2_{1/2}} \|\mathrm{A}^\dagger \mathrm{A}\|_\text{op} = \frac{4}{r^2_{1/2}} \left( a_1^2 + a_2^2 + a_3^2 \right) = \frac{4}{r^2_{1/2}} |\vec{a}|^2.
\end{equation}
Thus, the ball condition reduces to
\begin{equation}
\| [\mathcal{D}, \pi(a)] \|_\text{op} = \frac{2}{r_{1/2}} |\vec{a}| \leq 1.
\end{equation}
Equivalently, and interestingly, this yields a solid ball in $\mathbb{R}^3$:
\begin{equation} \label{n2}
| \vec{a} | \leq \frac{r_{1/2}}{2}.
\end{equation}
We have mentioned that from the symmetry of the space, we can choose any point as the North pole of the sphere. Here, for $n=\frac{1}{2}$, all pure states are coherent states such that we can parametrise a pair of generalized points (pure states) as
\begin{enumerate}
\item $\rho_N = \rho_{\theta = 0}=\begin{pmatrix}
1\\0
\end{pmatrix}\begin{pmatrix}
1&0
\end{pmatrix} = \begin{pmatrix}
1 & 0 \\
0 & 0 \
\end{pmatrix},$
\item $\rho_P = \rho_{\theta_0} =U(\theta_0)\begin{pmatrix}
1\\0
\end{pmatrix}\begin{pmatrix}
1&0
\end{pmatrix}U^\dagger(\theta_0)= \frac{1}{2}\begin{pmatrix}
1 + \cos \theta_0 & \sin \theta_0 \\
\sin \theta_0 & 1 - \cos \theta_0 
\end{pmatrix} $,\; $U(\theta_0) = \begin{pmatrix}
\cos \frac{\theta_0}{2} & - \sin \frac{\theta_0}{2}\\
\sin \frac{\theta_0}{2} & \cos \frac{\theta_0}{2}\
\end{pmatrix}
\in SU(2)$,
\end{enumerate} 
where in general $\omega_\theta(a)=\text{tr}(\rho_\theta~a)$. We define $\Delta \rho = \rho_{\theta_0} - \rho_0 $. Being a traceless hermitian matrix, this too can be expanded like (\ref{al_element}):
\begin{equation}
\Delta \rho = \vec{\Delta \rho}.\vec{\sigma}, ~\text{where}~(\Delta \rho)_1  = \frac{\sin \theta_0}{2} ;~(\Delta \rho)_2 = 0; ~(\Delta \rho)_3  = \frac{-1 +\cos \theta_0}{2}. \label{Delta-rho}
\end{equation}
With this, we have
\begin{equation} \label{B_condition}
|\omega_{\theta_0}(a)-\omega_0(a)|=|\text{tr}_{\mathcal{H}_n}(\Delta \rho ~a )| = |2 \vec{a}.\vec{\Delta \rho}|.
\end{equation}
Since  both $\vec{a}$ and $\vec{\Delta \rho}\in \mathds{R}^3$, the supremum of $ |\vec{a}.\vec{\Delta \rho}|$ will be attained when $\vec{a}$ and $\vec{\Delta\rho}$ are parallel or anti-parallel to each other. Thus, with the ball condition (\ref{n2}) and \eqref{Delta-rho}, we get the spectral distance between a pair of pure states $\rho_0$ and $\rho_{\theta_0}$ as
\begin{equation} \label{dist_n_half}
d_{\frac{1}{2}}(\omega_{\theta_0},\omega_0)=\sup_{| \vec{a} | \leq \frac{r_{1/2}}{2}} |\omega_{\theta_0}(a)-\omega_0(a)|= r_{\frac{1}{2}} \sqrt{(\Delta \rho)_1^2 + (\Delta \rho)^2_3}=r_{\frac{1}{2}} \, \sin {\frac{\theta_0}{2}}. 
\end{equation}
This is just half of the chordal distance connecting $NP$, reproducing the result of \cite{Var}. Since the supremum is attained here where $\vec{a}$ and $\vec{\Delta \rho}$ are parallel, the distance function just corresponds to the lower bound (\ref{3.1.9}). The corresponding infinitesimal distance $d(\omega_{d\theta},\omega_0)=\frac{1}{2}r_{\frac{1}{2}}d\theta$ can easily be seen to match exactly with (\ref{low_bound_fuz}), by setting $n=\frac{1}{2}$.

\begin{figure}[h]
\centering\includegraphics[width=7cm]{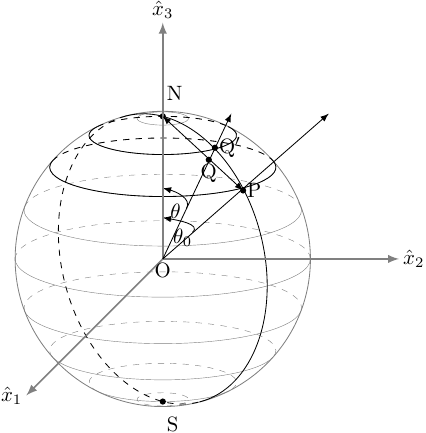}
\caption{Space of Perelomov's $SU(2)$ coherent states for $n=\frac{1}{2}$.}
\label{fig:1}
\end{figure}

%

Let us now consider a family of mixed states (as shown in the figure \ref{fig:1}). A generic mixed state $\rho_Q$, represented by the point $Q$ on the chord $NP$ inside the sphere, is obtained from the following convex sum:
\begin{equation} \label{mix-half}
\rho_t =(1-t)\rho_N + t\rho_P= \frac{1}{2}\begin{pmatrix}
2-t(1-\cos\theta_0)&t\sin\theta_0\\ t\sin\theta_0&t(1-\cos\theta_0)
\end{pmatrix}; ~~~0\leq t \leq 1.
\end{equation}
Clearly, $\rho_0=\rho_N$ and $\rho_1=\rho_P$. 
Introducing $(\Delta\rho)_{QN}=\rho_t-\rho_N$ and $(\Delta\rho)_{PN}=\rho_P-\rho_t$ such that the corresponding components (the counterparts of \eqref{Delta-rho}) are given by
\begin{eqnarray}
\nonumber((\Delta\rho)_{QN})_1=\frac{t~\sin\theta_0}{2},~((\Delta\rho)_{QN})_2=0, ~((\Delta\rho)_{QN})_3=-\frac{t(1-\cos\theta_0)}{2}\\\nonumber((\Delta\rho)_{PN})_1=\frac{(1-t)\sin\theta_0}{2},~((\Delta\rho)_{PN})_2=0, ~((\Delta\rho)_{PN})_3=-\frac{(1-t)(1-\cos\theta_0)}{2},
\end{eqnarray}
 we get the spectral distances between the mixed state $\omega_t$ and $\omega_N$ and $\omega
_P$ respectively as
\begin{equation} \label{dis-in-t}
d(\omega_{t},\omega_N)=t~r_{\frac{1}{2}}  \sin {\frac{\theta_0}{2}} ~~~\text{and}~~~d(\omega_{P},\omega_t)=(1-t)~r_{\frac{1}{2}} \sin {\frac{\theta_0}{2}}.
\end{equation}
The fact that the distance of this mixed state $\omega_t$ from the extremal, pure states $\omega_N$ and $\omega_P$ are proportional to the parameters $t$ and $(1-t)$ respectively indicates that we can identify a unique pure state $Q^\prime$ (see fig 1) nearest to a given mixed state $Q$ just by extending the straight line $OQ$ from the center $O$ to the surface of the sphere. This distance can therefore be used alternatively to characterize the `mixedness' of a spin-$1/2$ system. Further, we have
\begin{equation}
d(\omega_{\theta_0},\omega_0)=d(\omega_{P},\omega_N)=d(\omega_{P},\omega_t)+d(\omega_{t},\omega_N).
\end{equation}
This is just analogous to \eqref{dis-zt}, except that the intermediate state $\omega_t$ is not pure. This chord \underline{cannot}
therefore be identified as a conventional geodesic. This same family can be parametrised alternatively \cite{Var} as
\begin{equation} \label{rho_theta}
\rho_\theta = \frac{1}{2} (\mathds{1}_2 + \vec{\sigma}.\vec{n_\theta}) = \frac{1}{2} \left( \begin{array}{cc}
1 + |\vec{n_\theta}| \cos \theta & |\vec{n_\theta}| \sin \theta \\
|\vec{n_\theta}| \sin \theta & 1 - |\vec{n_\theta}| \cos \theta \
\end{array} \right),
\end{equation}
where $|\vec{n_\theta}|$ is the magnitude the vector $\vec{n_\theta}$ parametrising each of the mixed states between the two extremal pure states and is given by
\begin{equation}
\vec{n_\theta} = |\vec{n_\theta}| \left( \begin{array}{c}
\sin \theta , ~0 , ~\cos \theta \
\end{array} \right).
\end{equation}

Clearly, $|\vec{n_\theta}|$ is strictly less than $1$ : $|\vec{n_\theta}| < 1$ except for the extremal pure states at $\theta = 0$ and $\theta = \theta_0$. Further the mixed state $\rho_\theta$ for the open interval $(0, \theta_0)$ represents a point $Q$ in the chord connecting the north pole $N (\theta = 0)$ and point $P (\theta = \theta_0)$, and therefore lies in the interior of the sphere. Indeed, these two different parameters $t$ and $\theta$ for the same state can be related by setting $\rho_\theta = \rho_t$, to get
\begin{equation}
t=\frac{1-|\vec{n}_\theta|\cos\theta}{1-\cos\theta_0}=\frac{|\vec{n}_\theta|\sin\theta}{\sin\theta_0}~\Rightarrow~ \abs{\vec{n_\theta}} = \frac{\cos (\frac{\theta_0}{2})}{\cos (\theta - \frac{\theta_0}{2})} \equiv \frac{\mathrm{OQ}}{r_{1/2}},~~ON = OP = r_{1/2}.
\end{equation}
Thus, we can recast the spectral distance between a mixed state represented by $\rho_\theta$ and the pure states $\rho_N$ and $\rho_P$ (\ref{dis-in-t}) respectively as
\begin{equation} \label{183}
d(\omega_{\theta},\omega_N)= \frac{r_{\frac{1}{2}}\sin\theta}{2\cos(\theta-\frac{\theta_0}{2})} ~~~\text{and}~~~d(\omega_{P},\omega_\theta)=\frac{r_{\frac{1}{2}}\sin(\theta_0-\theta)}{2\cos(\theta-\frac{\theta_0}{2})} .
\end{equation}
For the case $\theta=\theta_0$, we get $d(\omega_{\theta_0},\omega_N)=r_{\frac{1}{2}}\sin\frac{\theta_0}{2}$ and $d(\omega_{P},\omega_{\theta_0})=0$. 

Finally, we can also obtain the distance between pure states represented by $\rho_\pi=\begin{pmatrix}
0&0\\0&1
\end{pmatrix}$ (South pole S) and $\rho_{\theta_0}$ (P) as
\begin{equation} \label{theta-dist}
d(\rho_{\pi},\rho_{\theta_0})=r_{\frac{1}{2}}\cos\frac{\theta_0}{2}.
\end{equation}
This implies that
\begin{equation}
[d(\rho_0,\rho_{\theta_0})]^2+[d(\rho_{\pi},\rho_{\theta_0})]^2=r^2_{1/2}. \label{Pytha}
\end{equation}
That is, the Pythagoras identity $(NP^2+SP^2=NS^2)$ is obeyed.   All these features, however, will not persist for higher `$n$', as we shall see.

\subsection{Analogy with $\mathds{C}P^1$ model and mixed states}
For the spectral triple \eqref{cp1}, the space of vector states is $CP^1$ \cite{Mart3}. We can parametrize a pair of $CP^1$-doublets, associated to the pair of points lying in the same latitude (i.e. the same polar angle $\theta$) \cite{Devi} as
\begin{equation}
\chi=\begin{pmatrix}
\cos\frac{\theta}{2}~~~~~\\\sin\frac{\theta}{2}e^{i\phi}
\end{pmatrix}\longrightarrow\rho=\chi\chi^\dagger;~~ \chi'=\begin{pmatrix}
\cos\frac{\theta}{2}~~~~~\\\sin\frac{\theta}{2}e^{i\phi'}
\end{pmatrix}\longrightarrow\rho'=\chi'\chi'^\dagger. \label{B}
\end{equation}
Here, the spectral distance between these two points is obtained as
\begin{equation} \label{cp1_dist}
d(\omega_{\rho'},\omega_\rho)=\frac{2\sin\theta}{|D_1-D_2|}\Big|\sin\Big(\frac{\phi-\phi'}{2}\Big)\Big|,
\end{equation}
which corresponds to the distance measured along the chord connecting the pair of points ($\theta, \phi$) and ($\theta, \phi'$), at the same latitude $\theta$.

Now, let us define a family of mixed states out of this pair of pure states $\rho$ and $\rho'$ in an analogous way as for the $n=1/2$ representation \eqref{mix-half}:
\begin{equation}
\rho_t= (1-t)\rho+t\rho'. 
\end{equation}
 Similarly, we obtain the distances between the mixed state $\rho_t$ and the corresponding pure states $\rho$ representing the point ($\theta, \phi$) and $\rho'$ representing ($\theta, \phi'$) as
 \begin{equation}
 d(\omega_{\rho_t},\omega_{\rho})=t\frac{2\sin\theta}{|D_1-D_2|}\Big|\sin\Big(\frac{\phi-\phi'}{2}\Big)\Big|;~~\text{and}~~d(\omega_{\rho'},\omega_{\rho_t})=(1-t)\frac{2\sin\theta}{|D_1-D_2|}\Big|\sin\Big(\frac{\phi-\phi'}{2}\Big)\Big|.
 \end{equation}
Clearly, we have
\begin{equation}
d(\omega_{\rho'},\omega_{\rho})= d(\omega_{\rho'},\omega_{\rho_t})+d(\omega_{\rho_t},\omega_{\rho}).
\end{equation}


\subsection{The n = 1 fuzzy sphere} \label{sec1}
In this final example, dealing with the computation of distance for $n=1$, $\mathbb{S}_\ast^2$ will turn out to be the only non-trivial example where the formulae (\ref{rev_formula}, \ref{N_finite}) can be applied to compute an \emph{almost exact} distance. This computation for $n=1$, however, presented in this section, is expected to be much more complicated than the $n=\frac{1}{2}$ case, simply because we expect the deviations from the straight line chord and the associated Pythagoras relation (\ref{Pytha}) to start showing up from here onwards as the analogue of \eqref{B_condition} for $n = 1/2$ case will be missing here. While in the extreme $n\rightarrow \infty$ limit this should merge with the great circle of the commutative sphere. It is obvious that one needs to consider the entire $su(3)$ algebra, which in the $3\times 3$ matrix representation are spanned by 8 traceless Gell-Mann matrices, as in \eqref{al_element}. Again the identity matrix is not considered, since it commutes with the Dirac operator and hence will make no contribution to the operator norm $\lVert[\mathcal{D},\pi(a)]\rVert_{op}$. We therefore consider the algebra element $a$ to be traceless and hermitian throughout this section. It is quite tempting to start directly by identifying the algebra element $a$ as a linear combination of the Gell-Mann matrices. The role of $\Delta\rho_\perp$ turns out to be very important in (\ref{rev_formula}) and (\ref{N_finite}) for the $n=1$ fuzzy sphere and we employ the most general form of $\Delta\rho_\perp$ possible to improve the estimate of the spectral distance as best as we can from the lower bound \eqref{3.1.9} i.e. $\kappa = 0$ case in \eqref{rev_formula}. The determination of an exact value, however, even with the help of $Mathematica$, remains a daunting task.

\subsubsection{Ball condition, general strategy to compute infimum and general form of $\Delta\rho$} \label{sec5}

To begin with the $n = 1$ case we first proceed in the same way as in section \ref{section2} to obtain, using \eqref{6.3.2}, 
\begin{equation}
_+\langle\langle n_3|[\mathcal{D},\pi(a)]|n_3'\rangle\rangle_- = \frac{3}{r_1}~_+\langle\langle n_3|\pi(a)|n_3'\rangle\rangle_-~~\text{and}~~_-\langle\langle n_3'|[\mathcal{D},\pi(a)]|n_3\rangle\rangle_+ = -\frac{3}{r_1}~_-\langle\langle n_3'|\pi(a)|n_3\rangle\rangle_+,
\end{equation} 
where the ranges of $n_3, n_3'$ are respectively given by $-2\le n_3\le 1$ and $-1\le n'_3\le 0$ so that with the diagonal representation the commutator $[\mathcal{D},\pi(a)]$ takes the following off-block diagonal form:
\begin{equation}
 [\mathcal{D},\pi(a)] = \frac{1}{r_1}\left( \begin{tabular}{c|c}
$0_{4\times 4}$ & $A_{4\times 2}$\\
\hline
$-A^{\dagger}_{2\times 4}$ & $0_{2\times 2}$\
\end{tabular} \right),
\end{equation}
where the rectangular matrices $A_{4 \times 2}$ and $A^{\dagger}_{2\times 4}$  are given by
\begin{equation}
A_{4\times 2} = \begin{pmatrix}
-\sqrt{3}a_{1,0} & -\sqrt{6}a_{1,-1}
\\
\sqrt{2}(a_{1,1} - a_{0,0}) & (a_{1,0} - 2a_{0,-1})
\\
(2a_{0,1} - a_{-1,0}) & \sqrt{2}(a_{0,0} - a_{-1,-1})
\\
\sqrt{6}a_{-1,1} & \sqrt{3}a_{-1,0}
\end{pmatrix},
\end{equation}
\begin{equation}
A^{\dagger}_{2 \times 4} = \begin{pmatrix}
-\sqrt{3}a_{0,1} & \sqrt{2}(a_{1,1} - a_{0,0}) & (2a_{1,0} - a_{0,-1}) & \sqrt{6}a_{1,-1}\\
-\sqrt{6}a_{-1,1} & (a_{0,1} - 2a_{-1,0}) & \sqrt{2}(a_{0,0} - a_{-1,-1}) & \sqrt{3}a_{0,-1}
\end{pmatrix}.
\end{equation}

Note that here $a_{m,n}$ are the usual matrix elements $\bra m a \ket n$, with $m,n \in \{1,0,-1\}$. Using the above result for $[\mathcal{D},\pi(a)]$ we readily obtain:
\begin{equation}
[\mathcal{D}, \pi(d\rho)]^\dagger [\mathcal{D}, \pi(d\rho)] = \frac{1}{r_1^2} \left( \begin{tabular}{c|c}
$(A A^\dagger)_{4\times 4}$ & $0_{4\times 2}$\\
\hline
$0_{2\times 4}$ & $(A^\dagger A)_{2\times 2}$\
\end{tabular} \right).
\end{equation}
By exploiting the properties of the operator norm one has the freedom to choose between the two block-diagonal square matrices as $\frac{1}{r_1^2}\lVert AA^{\dagger}\rVert _{op} = \frac{1}{r_1^2}\lVert A^{\dagger}A\rVert_{op} = \lVert[\mathcal{D},\pi(a)]^{\dagger}[\mathcal{D},\pi(a)]\rVert_{op} = \lVert[\mathcal{D},\pi(a)]\rVert_{op}^2$. We choose to work with the more convenient one i.e. the $2\times 2$ matrix $(A^\dagger A)_{2\times 2}$ which turns out to be
\begin{equation} \label{mat1}
M := (A^\dagger A)_{2\times 2} = \begin{pmatrix}
M_{11} & M_{12}\\
M^*_{12} & M_{22} 
\end{pmatrix},
\end{equation}
where
\begin{equation*}
\begin{split}
M_{11} & = 3|a_{0,1}|^2 + 2(a_{0,0} - a_{1,1})^2 +|a_{0,-1} - 2a_{1,0}|^2 + 6|a_{1,-1}|^2, \\
M_{22} & = 3|a_{0,-1}|^2 + 2(a_{0,0} - a_{-1,-1})^2 +|a_{1,0} - 2a_{0,-1}|^2 + 6|a_{1,-1}|^2, \\
M_{12} & = \sqrt{2} \left\lbrace 3a_{1,-1}(a_{0,1} + a_{-1,0}) + (a_{0,0} - a_{1,1})(2a_{0,-1}-a_{1,0}) + (a_{0,0} - a_{-1,-1})(2a_{1,0} - a_{0,-1}) \right\rbrace. \
\end{split}
\end{equation*}
This matrix has two eigenvalues $E_{\pm}$:
\begin{equation} \label{eig_gen} 
E_{\pm} := \frac{1}{2}\left(P \pm \sqrt{Q}\right) = \frac{1}{2}\left( (M_{11} + M_{22}) \pm \sqrt{(M_{11} - M_{22})^2 + 4|M_{12}|^2} \right).
\end{equation}
Here both $P$ and $Q$ can be written as a sum of several whole square terms and thus they are both positive definite for any algebra elements $a$. Clearly,
\begin{equation}
E_+ \geq E_- ~~~~\forall~~ a \in B
\end{equation}
yielding, for a particular $a\in B$,
\begin{equation} \label{op-norm}
\lVert [\mathcal{D},\pi(a)] \rVert_{\text{op}} = \frac{1}{r_1} \sqrt{E_+}.
\end{equation}
The corresponding infimum $\inf_{a\in B}\lVert [\mathcal{D},\pi(a)] \rVert_{\text{op}}$ is computed by varying the entries in the algebra elements, within the admissible ranges and obtaining the global minimum of $E_+$. This gives
\begin{equation}
\inf_{a\in B}\lVert [\mathcal{D},\pi(a)] \rVert_{\text{op}}=\frac{1}{r_1}\text{min}\big(\sqrt{E_+}\big)= \frac{1}{r_1}\sqrt{\text{min}(E_+)}.
\end{equation}
The eigenvalue $E_+$ will always have a ``concave-up" structure in the parametric space as it can be written as the sum of square terms only, (\ref{mat1}) and (\ref{eig_gen}). There can be points in the parametric space where $E_+$ and $E_-$ are equal, namely points where $Q$ becomes $0$, but since $E_+$ can not become less than $E_-$, determining the minimum of $E_+$ will alone suffice in calculating the infimum of the operator norm as is clear from \eqref{op-norm}. So we work with $E_+$ alone and use \emph{Mathematica} to  get the desired result.

The pure states corresponding to points on the fuzzy sphere can be obtained by the action of the SU(2) group element
\begin{equation} \label{mat3}
\hat{U} = e^{i\theta \hat{J_2}} = \begin{pmatrix}
\cos^2\frac{\theta}{2} & \frac{1}{\sqrt{2}}\sin\theta & \sin^2\frac{\theta}{2} \\
-\frac{1}{\sqrt{2}}\sin\theta & \cos\theta & \frac{1}{\sqrt{2}}\sin\theta \\
\sin^2\frac{\theta}{2} & -\frac{1}{\sqrt{2}}\sin\theta & \cos^2\frac{\theta}{2}
\end{pmatrix}
\end{equation}
on the pure state $\dyad{1}{1}$ corresponding to the north pole ($N$) of $\mathds{S}_*^2$ (with $n = 1$) i.e. $\rho_\theta = \hat{U} \dyad{1}{1} \hat{U}^\dagger$. Note that we have taken for convenience the azimuthal angle $\phi = 0$. This can be done without loss of generality.

Correspondingly,
\begin{equation}
\Delta\rho = \rho_{\theta} - \rho_0 = \hat{U}|1\rangle\langle 1|\hat{U}^{\dagger} - |1\rangle\langle 1| = \begin{pmatrix}
\cos^4\frac{\theta}{2} - 1 & -\frac{1}{\sqrt{2}}\sin\theta\cos^2\frac{\theta}{2} & \sin^2\frac{\theta}{2}\cos^2\frac{\theta}{2}
\\
-\frac{1}{\sqrt{2}}\sin\theta\cos^2\frac{\theta}{2} & \frac{1}{2}\sin^2\theta & -\frac{1}{\sqrt{2}}\sin\theta\sin^2\frac{\theta}{2}
\\
\sin^2\frac{\theta}{2}\cos^2\frac{\theta}{2} & -\frac{1}{\sqrt{2}}\sin\theta\sin^2\frac{\theta}{2} & \sin^4\frac{\theta}{2}
\end{pmatrix} \label{mat2}
\end{equation}
like the $n = 1/2$ case \eqref{Delta-rho}, all entries are real here; indeed by writing $\Delta \rho = (\Delta \rho)_i \lambda_i$ ($\lambda_i$'s are the Gell-Mann matrices) the coefficients of $\lambda_2$, $\lambda_5$ and $\lambda_7$ vanishes. This $\Delta\rho$ however only provides us with a lower bound \eqref{3.1.9} of the distance in Connes' formula and the actual distance is reached by some optimal algebra element ($a_s$) of the form
\begin{equation} \label{new1}
a_s = \Delta\rho + \kappa\Delta\rho_{\perp}~; ~~~tr(\Delta\rho\Delta\rho_{\perp}) = 0, 
\end{equation}
for which the infimum is reached (say in \eqref{N_finite}). This should be contrasted with the optimal element, for which the supremum is reached in \eqref{Fuzz_up_bound}.

\subsubsection{Spectral distance using $\Delta\rho_\perp$} \label{section4}

We had calculated the lower bound of the infinitesimal distance \eqref{low_bound_fuz} for the general $\mathbb{S}_\ast^2$ with radius $r_n = \lambda \sqrt{n(n + 1)}$. We now first undertake the task of computing the \emph{true} infinitesimal distance for $n=1$ with radius $r_1 = \lambda \sqrt{2}$. As is clear from the analysis presented in section \ref{subsec_DirOp_fuz} that corresponding computation for $n \geq 3/2$ is virtually intractible because of the presence of too many independent parameters to vary. Anyway, this infinitesimal distance for $n=1$ fuzzy sphere will help us to guess a general form of the finite distances, the computation of which will be taken up in the next subsection. To this end let us employ the modified distance formula \eqref{rev_formula} by constructing the most general form of $\Delta\rho_\perp$ for both finite as well as infinitesimal distances.

\paragraph{Infinitesimal distance}

In this case $d\rho$ takes a simpler form, as can be seen by expanding (\ref{mat2}) and keeping only the leading order terms in $d\theta$:
\begin{equation}
d\rho = \rho_{d\theta} - \rho_0 = -\frac{d\theta}{\sqrt{2}}\big(|1\rangle\langle 0| + |0\rangle\langle 1|\big).
\label{new4}
\end{equation}
The most general structure of the transverse part  $d\rho_{\perp}$ here is obtained by taking all possible linear combinations of the generic states $|i\rangle\langle j|$ i.e.  
\begin{equation}
d\rho_{\perp} = \sum_{i,j} C_{ij}|i\rangle\langle j|~~; ~~i,j\in\{-1,0,+1\}, \label{new2}
\end{equation}
where $C_{ij} = C_{ji}^*$ because of the hermiticity of $d\rho_\perp$. Clearly, the complex parameters $C_{ij}$ are exact analogues of suitable combinations of $x_i$'s in \eqref{a_gell}. The orthogonality condition (\ref{new1}) here requires the coefficient $C_{10}$ to be purely imaginary. Moreover we can demand that the matrix representation of $\Delta\rho_\perp$ should be traceless as discussed in section \ref{sec1}. We thus impose $C_{11}+C_{22}+C_{33} = 0$. To better understand the significance of each term we write (\ref{new2}) in matrix form as follows:
\begin{equation}
d\rho_{\perp} = \begin{pmatrix}
\mu_1 & i\alpha_1 & \gamma
\\
-i\alpha_1 & \mu_0 & \beta
\\
\gamma^* & \beta^* & -(\mu_1 + \mu_0) 
\end{pmatrix}~;~~\mu_1,\mu_0,\mu_{-1}, \alpha_1\in\mathds{R} ~~\text{and}~~\beta,\gamma\in \mathds{C}.   \label{new3}
\end{equation}
With this our optimal algebra element $a_s$ \eqref{new1} becomes:
\begin{equation}
a_s = \begin{pmatrix}
\mu_1 & -\frac{d\theta}{\sqrt{2}} + i\alpha_1 & \gamma
\\
-\frac{d\theta}{\sqrt{2}}-i\alpha_1 & \mu_0 & \beta
\\
\gamma^* & \beta^* & -(\mu_1 + \mu_0) 
\end{pmatrix},
\end{equation}
where we have absorbed $\kappa$ inside the coeffiecients of $d\rho_\perp$ \eqref{new3}. With $7$ independent parameters, it is extremely difficult to vary all the parameters simultaneously to compute the infimum analytically. However, as far as infinitesimal distances are concerned, it may be quite adequate to take each parameter to be non-vanishing one at a time. Thus, by keeping one of these diagonal/complex conjugate pairs like $\beta$ and $\gamma$ to be non-zero one at a time and computing the eigenvalues of the $2\times 2$ matrix (\ref{mat1}) using (\ref{new1}), (\ref{new4}) and (\ref{new3}), it is found that only the real part of $\beta$ contributes non-trivially to the infimum of the operator norm $\lVert[\mathcal{D},\pi(a)]\rVert_\text{op}$ in the sense that the operator norm of this object with $a_s = \begin{pmatrix}
0&-\frac{d\theta}{\sqrt{2}}&0\\-\frac{d\theta}{\sqrt{2}}&0&\beta\\0&\beta^*&0
\end{pmatrix}$ in \eqref{new1} is a monotonically increasing function of $Im(\beta)$ but yields a non-trivial value for the infimum which is less than the one with vanishing $\beta$ i.e. $d\rho$ itself. Like-wise, both real and imaginary parts of $\gamma$ and $\alpha_1$ in \eqref{new3} does not contribute to the infimum.

\begin{figure}[h]
\centering\includegraphics[width=8cm]{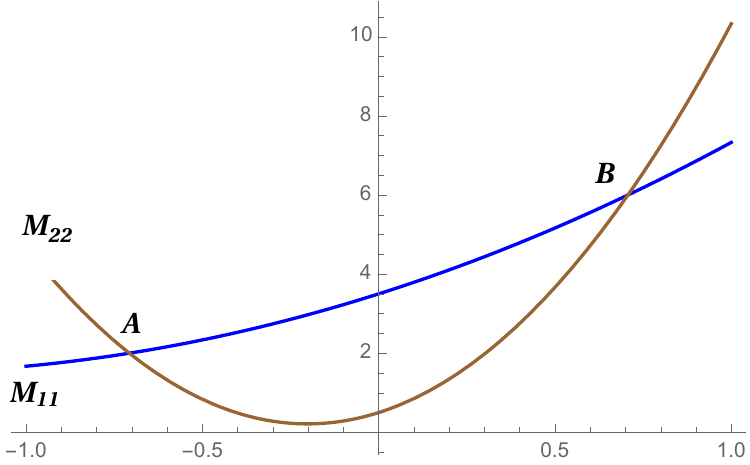}
\caption{Infimum corresponding to the plots of $M_{11}$ and $M_{22}$}\label{fig:2}
\end{figure}

On computation, we get
\begin{equation} \label{new5}
\lVert [\mathcal{D},\pi(a)]\rVert_{\text{op}}=\frac{1}{r_1}\sqrt{\text{max}\{M_{11},M_{22}\}}, ~~\text{where}~~M_{11} = \beta_1^2 + 2\sqrt{2}\beta_1d\theta + \frac{7}{2}d\theta^2 ~~\text{and}~~  
M_{22} = 7\beta_1^2 + 2\sqrt{2}\beta_1d\theta + \frac{1}{2}d\theta^2; 
\end{equation}
where $\beta_1 = Re(\beta)$ and the $2\times 2$ matrix \eqref{mat1} takes a diagonal form $diag\{M_{11},M_{22}\}$ thus trivially yielding eigenvalues \eqref{new5}. From the plot of these eigenvalues(see Fig.2) it is clear that the infimum of the operator norm over the full range of $\beta_1$ is given by the minimum value of the two intersections at $A$ ($\beta_1=-\frac{d\theta}{\sqrt{2}}$) and $B$ ($\beta_1=+\frac{d\theta}{\sqrt{2}}$) which comes out to be $\frac{\sqrt{2} d\theta}{r_1}$ i.e. $\lVert[\mathcal{D},\pi(a)]\rVert_\text{op}=\frac{\sqrt{2} d\theta}{r_1}$. Note in this context that $$E_+=\begin{cases}
M_{22}, ~~\text{for}~~\beta_1< -\frac{d\theta}{\sqrt{2}} ~~\text{and}~~\beta_1>+\frac{d\theta}{\sqrt{2}}~~i.e. ~\text{left of A and right of B}\\
M_{11}, ~~\text{for}~~-\frac{d\theta}{\sqrt{2}}<\beta_1< +\frac{d\theta}{\sqrt{2}} ~~i.e. ~\text{in between A and B}.
\end{cases}$$
Also, since $\lVert d\rho \rVert_{tr}^2 = d\theta^2$, as follows from \eqref{new4}, the infinitesimal spectral distance \eqref{rev_formula} is given by
\begin{equation} \label{eq1}
d_1(\rho_{d\theta}, \rho_{0}) = r_1\frac{d\theta}{\sqrt{2}}.
\end{equation}

We now corroborate the same result by varying all the $7$ parameters simultaneously. Of course we shall have to employ \emph{Mathematica} now. To that end, first note that $P,Q$ (\ref{mat1}) and ( \ref{eig_gen}) are now given as:
\begin{eqnarray}\label{P1}
P &=& 2\{ 1 + \mu_1^2 + 2\beta_1^2 + 3\alpha^2 + 4\mu_0^2 + 6|\gamma|^2 +  (\mu_1 + \mu_0)^2 + (\sqrt{2}\beta_1 + 1)^2 + (2\beta_2 - \alpha)^2  \}, \\ \nonumber
 \label{Q1}
Q &=&9\{ -1 + 4\mu_0\mu_1 + 2(|\beta|^2 + \mu_0^2 - \alpha^2 )  \}^2+ 4\{ 3(\mu_0 + \mu_1 + \gamma_1) + 4.242(\beta_1\mu_1 - \beta_2\gamma_2 - \alpha\gamma_2 - \beta_1\gamma_1)   \}^2  \\
&&+ 4\{ 3\gamma_2 + 4.242(\beta_2\mu_1 + \beta_2\gamma_1 + \alpha\gamma_1 - \beta_1\gamma_2 - \alpha\mu_0 - \alpha\mu_1) \}^2, 
\end{eqnarray}
where $\beta=\beta_1+i\beta_2$ and $\gamma=\gamma_1+i\gamma_2$. Interestingly enough we get the same infimum i.e. $2d\theta^2$ by finding the minimum of the eigenvalue $E_+$ as discussed in section \ref{sec5} with the $7$ parameter eigenvalue \eqref{eig_gen} where P and Q are given by (\ref{P1}) and (\ref{Q1}). We therefore recover the distance \eqref{eq1} which also matches with \eqref{better-dis-1} for $\theta \rightarrow d\theta$.

\paragraph{Finite distance} 

Before we embark into the computation of finite distance for $n=1$ fuzzy sphere let us recall the simple example of $H_2$-atom, where the energy gap between the ground state $(n=1)$ and first excited state $(n=2)$ is the largest one and the corresponding gaps in the successive energy levels go on decreasing and virtually become continuous for very large $n$ ($n\gg 1$). One can therefore expect a similar situation here too. Indeed, a preliminary look into the distance between north and south poles \eqref{dis_halfs} already support this in the sense that $\frac{\big(d_{3/2}(\mathrm{N},\mathrm{S})/r_{3/2}\big)}{\big(d_1(\mathrm{N},\mathrm{S})/r_1\big)}<\frac{\big(d_1(\mathrm{N},\mathrm{S})/r_1\big)}{\big(d_{1/2}(\mathrm{N},\mathrm{S})/r_{1/2}\big)} $. One therefore expects the distance function $d_1(\rho_0,\rho_\theta)$ to be essentially of the same form as that of $d_{\frac{1}{2}}(\rho_0,\rho_\theta)$, except to be scaled up by a $\sqrt{2}$-factor and a miniscule deformation in the functional form i.e. we can make the following ansatz for an approximate distance
\begin{equation} \label{d_guess}
d_1^* \left( \rho_0, \rho_\theta \right) \approx \sqrt{2} r_1 \sin \left( \frac{\theta}{2} \right) .
\end{equation}
This is also consistent with the infinitesimal form of the distance \eqref{eq1} (for another heuristic derivation of the same, see appendix \ref{sec6.6.2}). For large-values of $n$, the corresponding ratios $\frac{\big(d_{n}(\mathrm{N},\mathrm{S})/r_n\big)}{\big(d_{n-1}(\mathrm{N},\mathrm{S})/r_{n-1}\big)}\rightarrow 1$, and the functional deformations in the distance only for large variations in $n$, i.e. for $\Delta n = n_1 - n_2 \gg 1$ are expected to be pronounced.

We now begin our computation of the distance for finite angle $\theta$ by observing that for any finite angle $\theta$, the $\Delta\rho$ matrix (\ref{mat2}) can be directly used to compute the square of the HS norm $\lVert\Delta\rho\rVert_{\text{HS}}^2$. Moreover, this $\Delta\rho$ can be used as the algebra element $a$ to compute the eigenvalues of \eqref{eig_gen} and then the operator norm $\lVert[\mathcal{D},\pi(a)]\rVert_{op}$, yielding the lower bound for the spectral distance using \eqref{3.1.9}. More specifically for $\theta = \frac{\pi}{2}$ (i.e. the distance between the north pole $N$ and any point $E$ on the equator) the lower bound is found to be
\begin{equation} \label{dis_NE}
d_1(N, E) \ge 0.699r_1.
\end{equation}
Contrasting with the corresponding value \eqref{d_guess} of $d_1^* (N, E) = \sqrt{2} r_1 \sin \left(\frac{\pi}{4}\right) = r_1$, we see that the transverse component $\Delta \rho_\perp$ must play a role here. On the other hand for $\theta = \pi$ (i.e. the distance between north and south pole) we find $d(N, S) = \sqrt{2} \, r_1$ which matches exactly with the result of \eqref{dis_halfs} for the distance between discrete states $|1\rangle\langle 1|$ and $|-1\rangle\langle -1|$. This means that there is no contribution of $\Delta\rho_\perp$ to this distance for $\theta = \pi$. There are, however, non-trivial contributions from $\Delta\rho_{\perp}$ to the distance for any general value of the angle $\theta < \pi$ as we have illustrated above through the example of $\theta = \frac{\pi}{2}$. Now the most general $\Delta\rho_{\perp}$ in this case can be constructed as follows:
\begin{equation} \label{new9}
\begin{split}
\Delta\rho_{\perp} =  \left[ \mu_1 \dyad{1}{1} + \mu_0 \dyad{0}{0} \right. & + \mu_{-1} \dyad{-1}{-1} + \alpha \dyad{1}{0} + \alpha^* \dyad{0}{1} \\
 & \left. + \gamma \dyad{1}{-1} + \gamma^* \dyad{-1}{1} + \beta \dyad{0}{-1}| + \beta^* \dyad{-1}{0}  \right],
\end{split}
\end{equation}
where $\mu_1,\mu_0,\mu_{-1} \in \mathds{R}$. Again we can take $a_s$ \eqref{new1} to be traceless as before, which implies $\mu_1 + \mu_0 + \mu_{-1} = 0$ and we can eliminate one of them, say $\mu_{-1}$. Furthermore, imposing the orthogonality condition \eqref{new1} we have a relation between all the remaining 8 parameters and one of them can be eliminated from that. In this case, using the $\Delta\rho$ \eqref{mat2} and $\Delta\rho_\perp$ \eqref{new9} in \eqref{new1} while absorbing $\kappa$ inside the coefficients of $\Delta\rho_\perp$, we have
\begin{equation}
\text{tr}\Big(\Delta\rho~\Delta\rho_\perp\Big)=0~\Longrightarrow~ \mu_1 = - \frac{1}{2} \mu_0 \sin^2 \left(\frac{\theta}{2}\right) - \frac{1}{\sqrt{2}} \beta_1 \sin\theta + \cos^2 \left(\frac{\theta}{2}\right) \left( \gamma_1 + \mu_0 - \sqrt{2} \alpha_1 \cot \frac{\theta}{2} \right), 
\end{equation}
where $\alpha_1, \beta_1$ and $\gamma_1$ are the real components and $\alpha_2, \beta_2$ and $\gamma_2$ are the imaginary components of $\alpha, \beta$ and $\gamma$ respectively. With these substitutions the eigenvalues of the matrix \eqref{mat1} become a function of $7$ parameters. We now calculate P,Q \eqref{eig_gen} for the $\theta = \frac{\pi}{2}$ case, as an illustration, to get
\begin{eqnarray}
 \nonumber P &=& 5 + \frac{45}{4}\mu_0^2+8(|\alpha|^2+|\beta|^2)+2(\alpha_1^2 + \beta_1^2) + \gamma_1^2+12|\gamma|^2 -4\gamma_1+6\mu_0 \\
& & -3\sqrt{2}\mu_0(\beta_1+\alpha_1)-4\alpha_1\beta_1+3\gamma_1\mu_0-2\sqrt{2}\gamma_1(\beta_1+\alpha_1)-8\alpha_2\beta_2,\\
\nonumber Q &=& 9\Big[\big\{1+2|\alpha|^2-3\mu_0^2-2|\beta|^2-\gamma_1+\frac{1}{2}\mu_0+2\sqrt{2}\beta_1-2\mu_0\gamma_1+2\sqrt{2}\mu_0(\alpha_1+\beta_1)\big\}^2+\frac{1}{2}\Big\{\big(\sqrt{2}+2\sqrt{2}\alpha_1^2-2\sqrt{2}\beta_1^2 \\\nonumber 
& &    +2\sqrt{2}\gamma_1-4\beta_1+\sqrt{2}\mu_0 +\mu_0\beta_1-4\gamma_2\beta_2-6\gamma_1\alpha_1-4\gamma_2\alpha_2-5\alpha_1\mu_0\big)^2+ \big(2\beta_2-2\sqrt{2}\gamma_2-2\alpha_2-6\gamma_1\beta_2\\ 
& &-2\sqrt{2}\alpha_1\alpha_2 +4\alpha_1\gamma_2+2\sqrt{2}\alpha_1\beta_2+\beta_2\mu_0+4\gamma_2\beta_1+2\sqrt{2}\beta_1\beta_2-2\gamma_1\alpha_2+5\alpha_2\mu_0-2\sqrt{2}\beta_1\alpha_2 \big)^2 \Big\}\Big].
\end{eqnarray}
Minimizing $E_+$ as before, we get
\begin{equation}
d(N, E) = r_1,
\end{equation}
which, quite remarkably, matches exactly with \eqref{d_guess}. As for $\theta = \pi$, the terms P,Q \eqref{eig_gen} of the eigenvalues $E_{\pm}$ come out to be
\begin{eqnarray}
P &=& 4+ 9\mu_0^2+8(|\alpha|^2+|\beta|^2)+12|\gamma|^2-8(\alpha_1\beta_1+\alpha_2\beta_2), \\
\nonumber Q &=& 36\big(|\beta|^2 - |\alpha|^2 - 2\mu_0 \big)^2 + 18\Big[\Big\{(\beta_1+\alpha_1)(2\gamma_1+\mu_0)+2(\beta_1-\alpha_1)+ 2\gamma_2(\alpha_2+\beta_2)\Big\}^2   \\
& & + \Big\{(\beta_2+\alpha_2)(2\gamma_1-\mu_0)-2(\beta_2-\alpha_2)- 2\gamma_2(\alpha_1+\beta_1)\Big\}^2 \Big].
\end{eqnarray}

Again we need to calculate the spectral distance by minimizing $E_+$ which gives $d_1(\rho_{\theta = \pi},\rho_0 ) = \sqrt{2}r_1$. This is precisely the lower bound result $d_1(N,S)$ of \eqref{dis_halfs} as discussed previously and hence support our claim that $\Delta\rho_\perp$ will not contribute here at all. We now present in Table 1 both the distances i.e. the one calculated using the formula \eqref{d_guess} and the other calculated using the same approach i.e. by finding the global minima of the eigenvalue $E_+$ (say $d_1$) for various angles between $0$ and $\pi$.
\begin{center}

\begin{tabular}{|c|c|c|}
\hline
Angle (degree)  &  $d_1^*/r_1$ & $d_1/r_1$\\
\hline
10 & 0.1232568334 & 0.1232518539 \\
20 & 0.2455756079 & 0.2455736891 \\
30 & 0.3660254038 & 0.3660254011 \\
40 & 0.4836895253 & 0.4836894308 \\
50 & 0.5976724775 & 0.5976724773 \\
60 & 0.7071067812 & 0.7071067811 \\
70 & 0.8111595753 & 0.8111595752 \\
80 & 0.9090389553 & 0.9090389553 \\
90 & 1 & 0.9999999998 \\
100 & 1.0833504408 & 1.0833504407 \\
110 & 1.1584559307 & 1.1584559306 \\
120 & 1.2247448714 & 1.2247448713 \\
130 & 1.2817127641 & 1.2817127640 \\
140 & 1.3289260488 & 1.3289260487 \\
150 & 1.3660254038 & 1.3660254037 \\
160 & 1.3927284806 & 1.3927284806 \\
170 & 1.4088320528 & 1.4088320527 \\
\hline
\end{tabular}
\captionof{table}{Data set for various distances corresponding to different angles}

\end{center}
It is very striking that the distance $d_1^*$ \eqref{d_guess} matches almost exactly with $d_1$ for all these angles as one sees from table 1. This strongly suggests that \eqref{better-dis-1} is indeed very very close to the exact distance! In fact, for larger angles like $50\degree$ and above the results agree upto $9$ decimal places, whereas for smaller angles ($<50\degree$) they agree upto $5$ decimal places and show some miniscule deviations from $6$ decimals onwards. One can expect to see more pronounced deformations away from \eqref{d_guess} in the functional form when the overall ratio scale of magnification starts reducing monotonically with $n\rightarrow\infty$ and eventually merge with the commutative results.


\section{Conclusions} \label{sec_con}
We have provided here a general algorithm to compute the finite spectral distance on non-commutative spaces, in our Hilbert-Schmidt operator formulation, and we find here that the formula quoted in \cite{FSBC} actually corresponds to the lower bound. However, as far as the computation of infinitesimal distances is concerned, this formula is quite adequate. This is because it can reproduce the local infinitesimal distance up to an overall numerical factor as shown in \cite{FSBC, Devi}. We should, however, keep in mind that the knowledge of infinitesimal distances may not be adequate to capture the geometry of a generic non-commutative space as such. These spaces, unlike a commutative differentiable manifold, may not allow a geodesic to be defined in the conventional sense. By the word ``conventional", we mean that the geodesic passes through a one parameter family of pure states. In this situation, one cannot simply integrate the infinitesimal distance to compute the finite distance.

Here, we have studied extensively the contrasting features of the geometry of Moyal plane ($\mathds{R}^2_*$) and that of the Fuzzy sphere ($\mathds{S}^2_*$). For this, we made use of both the above mentioned revised algorithm and also emulated the method of \cite{Mart} to compute the upper bound and then look for an optimal element saturating this upper bound. In the case of the Moyal plane, we succeed in identifying such optimal element `$a_s$' belonging to the multiplier algebra.  We then constructed a sequence of projection operators $\pi^N(a_s)$ in the finite dimensional subspace spanned by eigen-spinors of the Dirac operator that converge to $\pi(a_s)$ as $N \to \infty$ and saturates the upper bound, allowing us to identify the upper bound itself with the distance. Eventually, this enables us to relate the one parameter family of pure states to the geodesic of the Moyal plane which is nothing but the straight line. In contrast, on the fuzzy sphere, although an analogous upper bound can be constructed for any finite $n$-representation of $su(2)$, there simply does not exist an optimal element $a_s$ saturating the inequality. Indeed, for the case of extremal non-commutativity $n=\frac{1}{2}$, the finite distance turns out to be half the chordal distance. Here, except for the extremal points, the interpolating ``points" correspond to mixed states. This in turn helps us to find the distance between a given mixed state and a uniquely defined nearest pure state lying on the ``surface" of $\mathds{S}^2_*$. The corresponding distance can then be taken as an alternative characterization of the ``mixedness" of a state. This exercise shows that in Connes' framework no discrimination is made between pure and mixed states; it scans through the entire set of pure and mixed states to compute the supremum in \eqref{ConDis}.

All these calculations are enormously simplified by working in the eigen-spinor basis of the respective Dirac operator, so much so that we are able to compute the distance in the `$n=1$' fuzzy sphere, using this revised algorithm. Since this algorithm involves also the transverse $\Delta\rho_\perp$ components in addition to the longitudinal $\Delta\rho$ component, this becomes somewhat less user-friendly. For the `$n=1$' case, for example, it involves a minimization in seven parameters. Needless to say that we have to make use of \textit{Mathematica} after solving the quadratic characteristic equation. For higher $n$'s, the corresponding characteristic equations will not only involve higher degree polynomials, but will also involve a large number of independent parameters to be varied. Consequently, the computation for the  $n>1$ fuzzy sphere, even with the help of \textit{Mathematica}, remains virtually intractable and for the Moyal plane the number of parameters is simply infinite! To put our findings in a nutshell, we observe that the finite distance for $n=1$ and that of $n=\frac{1}{2}$ have almost the same functional form except for an overall scaling by a factor of $\sqrt{2}$ and a miniscule deformation at small `$\theta$' and that too only from the sixth decimal onwards.

Finally we would like to mention that this method, which involves working in the eigen-spinor basis of the Dirac operator, makes the computation quite transparent and convenient to execute if the matrices involved are not too big. In fact, this can be used to compute distances on double Moyal plane also \cite{KK}.


\appendix
\setcounter{equation}{0}
\renewcommand{\theequation}{A.\arabic{equation}}

\section{Appendix}

\subsection{Dirac Operator in Moyal plane, its response to $ISO(2)$ symmetry and translation as an isometry transformation} \label{App1}

In order to construct a Dirac Operator for the Moyal plane, we need to consider the momentum operator satisfying the non-commutative Heisenberg algebra (\ref{MP}, \ref{HA}). Now since only $\mathcal{H}_q$ can furnish a complete representation of the entire Heisenberg algebra we need to construct our Dirac operator on this space. Thus we consider the operators $\hat{P}_\alpha$, which acts adjointly only on $\mathcal{H}_q$. The Dirac operator is then constructed in terms of Pauli matrices $\sigma_1 = \begin{pmatrix}
0 & 1\\
1 & 0\end{pmatrix}
$ and $\sigma_2 = \begin{pmatrix}
0 & -i\\
i & 0\
\end{pmatrix}$ as

\begin{equation} \label{parent_Dirac_op}
\mathcal{D} \equiv \sigma_i \hat{P}_i = \sigma_1 \hat{P}_1 + \sigma_2 \hat{P}_2.
\end{equation}

Note that in this construction the Dirac operator has a natural action on $\mathcal{H}_q \otimes \mathds{C}^2$ through the adjoint action of $\hat{P}_i$, i.e. on a generic element $\Phi = \begin{pmatrix}
| \phi_1 ) \\
| \phi_2 ) \
\end{pmatrix} \in \mathcal{H}_q \otimes \mathds{C}^2$ it acts as

\begin{equation}
\mathcal{D} \Phi = \sqrt{\frac{2}{\theta}} \begin{pmatrix}
[ i\hat{b}^\dagger,| \phi_2 ) ]\\
[-i \hat{b},| \phi_1 ) ]
\end{pmatrix} .
\end{equation}

This, in turn implies that the action of the commutator $[\mathcal{D},\pi(a)]$ on $\Phi$, on using \eqref{corr4A1}, gives 

\begin{equation} \label{D-com}
[\mathcal{D},\pi(a)]\Phi =\sqrt{\frac{2}{\theta}}\begin{pmatrix}
0& [i \hat{b}^\dagger,a]\\
[-i \hat{b},a]
\end{pmatrix}\Phi.
\end{equation}

Now regarding $\Phi$ as a test function, we can identify the Dirac operator $\mathcal{D}$ as 

\begin{equation}
\mathcal{D}=\sqrt{\frac{2}{\theta}}\begin{pmatrix}
0&i\hat{b}^\dagger\\-i\hat{b}&0
\end{pmatrix}.
\end{equation}

This is further simplified by considering the transformation $\hat{b} \to i \hat{b}$ and $\hat{b}^\dagger \to -i \hat{b}^\dagger$, which just corresponds to a $SO(2)$ rotation by an angle $\pi/2$ in $\hat{x}_1, \hat{x}_2$ space. With this transformation the Dirac operator takes the following hermitian form:

\begin{equation} \label{Dir2}
\mathcal{D} = \sqrt{\frac{2}{\theta}} \begin{pmatrix}
0 & \hat{b}^\dagger\\
\hat{b} & 0 \
\end{pmatrix},
\end{equation}

which precisely has the same form as \eqref{DirOp}, used throughout the paper. The most important point to note is that this very structure of the Dirac operator allows us to make it also act directly from the left on $\Psi =\begin{pmatrix}
\ket{\psi_1}\\
\ket{\psi_2}\
\end{pmatrix} \in \mathcal{H}_c \otimes \mathds{C}^2$ \eqref{corr4A2}. In this context, we would like to mention that this Dirac operator $\mathcal{D}$ responds to a $SO(2)$ rotation in the $\hat{x}_1,\hat{x}_2$ plane by an arbitrary angle $\alpha$ as

\begin{equation}
\mathcal{D}=\sqrt{\frac{2}{\theta}}\begin{pmatrix}
0&\hat{b}^\dagger\\
\hat{b}&0
\end{pmatrix}\longrightarrow \mathcal{D}^{(\alpha)}=\sqrt{\frac{2}{\theta}}\begin{pmatrix}
0&\hat{b}^{(\alpha)\dagger}\\
\hat{b}^{(\alpha)}&0
\end{pmatrix}=\sqrt{\frac{2}{\theta}}\begin{pmatrix}
0&\hat{b}^\dagger e^{-i\alpha}\\
\hat{b}e^{i\alpha}&0
\end{pmatrix}. \label{Dirac-alpha}
\end{equation}

Since $[\hat{b}^{(\alpha)},\hat{b}^{(\alpha)\dagger}]=[\hat{b},\hat{b}^\dagger]=1$, one can build the tower of states, parametrized by $\alpha$, analogous to (\ref{Hc}) and is related to the ones in (\ref{Hc}) by a phase factor:

\begin{equation} \label{eq108}
\ket{n}^{(\alpha)}=\frac{(\hat{b}^{(\alpha)\dagger})^n}{\sqrt{n!}}\ket{0}=e^{-in\alpha}\ket{n}.
\end{equation}

On the other hand, under translation (\ref{CohSt}), the Dirac operator transforms as

\begin{equation} \label{Dirac-z}
\mathcal{D}\rightarrow \mathcal{D}^{(z)}=U(z,\bar{z})\mathcal{D}U^\dagger(z,\bar{z})=\sqrt{\frac{2}{\theta}}\begin{pmatrix}
0 & \hat{b}^\dagger-\bar{z}\\
\hat{b}-z & 0
\end{pmatrix} \equiv \sqrt{\frac{2}{\theta}} \begin{pmatrix}
0 & \hat{b}_z^\dagger\\
\hat{b}_z & 0\
\end{pmatrix}; ~\hat{b}_z = \hat{b} - z,
\end{equation}

with $\hat{b}_{z=0} = \hat{b}$. It is trivial to see that $\left[ \hat{b}_z, \hat{b}_z^\dagger \right] = \left[ \hat{b}, \hat{b}^\dagger \right] = 1$ and $\hat{b}_z | z \rangle = 0$ so that $| z \rangle$ can be associated with the new origin : $ | z \rangle \equiv | \tilde{0} \rangle$ \cite{FSBC}.  Thus the computation of infinitesimal distance between $\rho_z = | z ) = | z \rangle \langle z |$ and $\rho_{z+dz}$ can first be computed by conveniently choosing an equivalent spectral triple $(\pi^{(z)}, \mathcal{A}, \mathcal{D}^{(z)}, \mathcal{H}_c)$\footnote{By unitary equivalence of two spectral triples $(\pi_1(\mathcal{A}), \mathcal{H}_1, \mathcal{D}_1)$ and $(\pi_2(\mathcal{A}), \mathcal{H}_2, \mathcal{D}_2)$ we mean that there should exist a unitary operator $U: \mathcal{H}_1 \rightarrow \mathcal{H}_2$ such that $\mathcal{D}_2 = U\mathcal{D}_1 U^*$ and $\pi_2(\cdot) = U \pi_1(\cdot)U^*$ where $\pi_i$ $(i = 1,2)$ are representations of $\mathcal{A}$ in $\mathcal{H}_i$, respectively (See, for example the definition 2.2.2 in \cite{R2}). In the present example, of course, we are dealing with the same Hilbert space $\mathcal{H}_1 = \mathcal{H}_2 = \mathcal{H}_c$ \eqref{Hc}. }  with the Dirac operator $\mathcal{D}^{(z)}$ \eqref{Dirac-z}. But in this case both pairs of Dirac operators ($\mathcal{D}^{(z)}$ and $\mathcal{D}$) and $\left( \Delta \rho(z) \equiv \rho_{z+dz} - \rho_z,~\Delta \rho(0) \equiv \rho_{dz} - \rho_0 \right)$ are related by the same unitary operators $U(z, \overline{z})$ (\ref{CohSt}, \ref{Dirac-z}): 

\begin{equation}
\Delta\rho(z) = U(z,\bar{z}) 
\Delta\rho(0)U^\dagger(z,\bar{z}).
\end{equation}

Consequently, the ball condition remains unaffected under the simultaneous transformation in $\mathcal{D}$ and the optimal element $a_s$:

\begin{equation} \label{ball_inv}
\|[\mathcal{D}^{(z)}, \pi(a_s^{(z)})]\|_{op} = \|[\mathcal{D}, \pi(a_s)]\|_{op}; ~a_s^{(z)} = U(z,\bar{z})a_s U^\dagger(z,\bar{z}),
\end{equation}

so that the same value of the supremum is now attained by the new optimal element $a_s^{(z)}$, as this also ensures the invariance of the inner product:

\begin{equation}
(\Delta\rho(z), a_s^{(z)}) = (\Delta\rho(0), a_s).
\end{equation}

This, in turn, can be viewed as the unitary transformation $\pi(a) \rightarrow \pi^{(z)}(a) = U(z,\bar{z})\pi(a)U^\dagger (z,\bar{z})$ of the original representation $\pi$ itself.  This indicates that the value of the infinitesimal distances $d(\rho_0, \rho_{dz})$ and $d(\rho_z, \rho_{z+dz})$ computed through spectral triples $(\pi, \mathcal{A}, \mathcal{D}, \mathcal{H}_c)$ and $(\pi^{(z)}, \mathcal{A}, \mathcal{D}^{(z)}, \mathcal{H}_c)$ respectively are identical

\begin{equation}
d_{\mathcal{D}}(\rho_0, \rho_{dz}) = d_{\mathcal{D}^{(z)}}(\rho_z, \rho_{z+dz}).
\end{equation}

As illustrated through the spectral triple \eqref{cp1} below (See appendix A.2), this still does not imply that these distances are indeed equal. In fact, as it seems apparently that $\mathcal{D}$ \eqref{Dir2} does not commute with $U(z,\bar{z})$ \eqref{CohSt}: $[\mathcal{D}, U(z,\bar{z})] \not= 0$ violating the necessary condition for isometric transformation which is $[\mathcal{D}, U(z,\bar{z})] = 0$ in this case (See lemma 2.3 in \cite{R1}). However, since the difference between the Dirac operator $\mathcal{D}^{(z)}$ and $\mathcal{D}$ is such that their difference commutes with the diagonal representation of any algebra element: $[\mathcal{D}^{(z)}-\mathcal{D}, \pi(a)] = 0~\forall~ a\in \mathcal{A}$, it follows that the same ball condition $\|[\mathcal{D}, \pi(a)]\|_{op} \leq 1$ is obtained irrespective of whether one makes use of $\mathcal{D}$ or $\mathcal{D}^{(z)}$. And this has to be so, as after all the parent expression of the Dirac operator \eqref{parent_Dirac_op} is manifestly translationally invariant, and relates to the fact that the action of $\hat{P}_i$ on $|\psi)$ can as well be defined by replacing $\hat{x}_j$ with $\hat{x}_j + \alpha_j$ in \eqref{URep} for some constant $\alpha_j$. One therefore need not invoke any equivalent spectral triple and work with the same spectral triple $(\mathcal{A} = \mathcal{H}_q, \mathcal{H} = \mathcal{H}_c\otimes \mathds{C}^2, \mathcal{D})$ as introduced in section (2.3) throughout the Moyal plane.  Translational invariance of the distance then follows trivially

\begin{equation} \label{trans_inv}
d\left( \rho_z, \rho_{z+dz} \right) = d\left( \rho_0, \rho_{dz} \right).
\end{equation}

This can then be generalized for finite distance also i.e. $d(\rho_z, \rho_{z+w}) = d(\rho_0, \rho_w)$ for any finite complex number $w \in \mathds{C}$. Finally it can be shown to be rotationally $SO(2)$ symmetric also by making use of the identity $\|[\mathcal{D}^{(\alpha)}, \pi(a)]\|_{op} = \|[\mathcal{D}, \pi(a)]\|_{op}$ for the rotated $\mathcal{D}^{(\alpha)}$ \eqref{Dirac-alpha}, so that 

\begin{equation}
d(\rho_0, \rho_z) = d(\rho_0, \rho_{ze^{i\alpha}}),
\end{equation}

for any angle $\alpha$, proving that the distance is invariant under the entire Euclidean group $ISO(2)$.

\subsection{Contrasting features between spectral triples of $\mathds{R}^2_*$ and \eqref{cp1} }

To illustrate this point more clearly with a simpler example, let us reconsider the spectral triple \eqref{cp1}. In a suitable neighbourhood of $\mathds{C}P^1$ manifold - the space of pure states, the $\mathds{C}P^1$ doublet can be written as $\chi = \begin{pmatrix}
\cos\left(\frac{\theta}{2}\right) \\ \sin\left(\frac{\theta}{2}\right) e^{i\phi}
\end{pmatrix}$ \eqref{B} in terms of polar coordinate $\theta, \phi$ satisfying $\chi^\dagger \chi = 1$. The corresponding unit 3-vector $\vec{n}~ (\vec{n}^2 =1)$, on $\mathds{C}P^1 \approx  S^2$, are obtained by Hopf map $\vec{n} = \chi^\dagger \vec{\sigma} \chi$, as $\vec{n} = (\sin\theta\cos\phi, \sin\theta\sin\phi, \cos\theta)$. We can now associate, in the spirit of Gelfand and Naimark, the pure state

\begin{equation} \label{cp1_state}
\rho(\theta, \phi) = \chi \chi^\dagger = \frac{1}{2}\begin{pmatrix}
1 + \cos\theta & \sin\theta e^{-i\phi} \\ \sin\theta e^{i\phi} & 1 - \cos\theta
\end{pmatrix},
\end{equation}

to the point $\vec{n}(\theta, \phi)$. In particular, the pair of equatorial points $\vec{n}_1 = \vec{n}(\theta = \frac{\pi}{2}, \phi = 0) = (1,0,0)$ and $\vec{n}_2 = \vec{n}(\theta = \frac{\pi}{2}, \phi = \frac{\pi}{2}) = (0,1,0)$ are associated to the density matrices $\rho_1 = \rho (\frac{\pi}{2},0) = \begin{pmatrix}
1 & 1 \\ 1 & 1
\end{pmatrix}$ and $\rho_2 = \rho (\frac{\pi}{2},\frac{\pi}{2}) = \begin{pmatrix}
1 & -i \\ i & 1
\end{pmatrix}$ respectively. Being in the same latitude, the distance between these states is a finite one and is given in terms of chordal distance \eqref{cp1_dist}. However, if both these density matrices are subjected to the same unitary transformation $U_1(\theta) = exp(\frac{i}{2}\theta\sigma_1) \in SU(2)$ as $\rho_i \rightarrow U^\dagger_1(\theta) \rho_i U_1(\theta);$ $(i = 1,2)$, the state  $\rho_1$ stays at the equator as $\rho_1$ is invariant under $U_1(\theta)$ action: $\rho_1 \rightarrow U^\dagger_1(\theta) \rho_1 U_1(\theta) = \rho_1$, whereas $\rho_2$ is not invariant and moved to the latitude $\theta$. In particular, for $\theta = \frac{\pi}{2}$, the density matrix $\rho_2$ is transformed to that of north pole N with associated density matrix $\rho_N = \begin{pmatrix}
1 & 0 \\ 0 & 0
\end{pmatrix} = U^\dagger_1(\frac{\pi}{2}) \rho_2 U_1(\frac{\pi}{2})$. Being a state belonging to different latitude, the distance between pair of states diverges: $d(\rho_1, \rho_N) \rightarrow \infty$. This stems from the fact that this unitary transformation $U_1(\theta)$ does not commute with the Dirac operator $\mathcal{D}$ \eqref{cp1}: $[\mathcal{D}, U_1(\theta)] \not= 0$ and therefore is not an isometric transformation. One can however, restore the distance if we consider an equivalent spectral triple, where Dirac operator is also subjected to same unitary transformation: $\mathcal{D} \rightarrow U^\dagger_1(\frac{\pi}{2}) \mathcal{D} U_1(\frac{\pi}{2}) = \frac{1}{2}\begin{pmatrix}
D_1 + D_2 & i(D_1 - D_2) \\ -i(D_1 - D_2) & D_1 + D_2
\end{pmatrix} \equiv \mathcal{D}'$. And in contrast with the Moyal plane, the respective ball condition is not invariant under $SU(2)$ transformation.

\subsection{On the irreducibility condition}

The third condition actually corresponds to some irreducibility condition that we essentially reproduce here from \cite{FSBC}. To begin with, recall that there is a close relationship between normal state and density matrices so that the action of a state $\omega$ on an algebra element $a \in \mathcal{A} = \mathcal{H}_q$ can be uniquely written as \cite{Rob},
\begin{equation} \label{C1}
\omega (a) = \Tr_{\mathcal{H}_c} (\rho_{\omega} a),
\end{equation}
where $\rho_{\omega}$ is a hermitian, non-negative operator on $\mathcal{H}_c$ with $\Tr_{\mathcal{H}_c} (\rho_{\omega}) = 1$. This follows from the fact about normal states on von-Neumann algebra. Moreover, the representation of the algebra $\mathcal{A}$ on $\mathcal{H}_c$ \eqref{Hc} is irreducible i.e. $\mathcal{H}_c$ carries an irreducible representation of the oscillator algebra. Then it readily follows that von Neuman algebra generated by $\mathcal{A}$ in $\mathcal{B}(\mathcal{H}_c)$ is the whole of $\mathcal{B}(\mathcal{H}_c)$ itself. 

Thus for any pure normal state $\omega ~ \exists$ a pure density matrix $\rho_{\omega}$ satisfying $\rho_{\omega} = |\psi\rangle\langle \psi | = \rho_{\omega}^* = \rho_{\omega}^2$ for some unit vector $|\psi\rangle \in \mathcal{H}_c$ with $\langle \psi | \psi \rangle = 1$. Clearly, the rank one projection operator $\rho_\omega$ is in $\mathcal{H}_q$: $\rho_\omega \in \mathcal{H}_q$ and therefore it follows trivially that the difference $\Delta\rho := \rho_\omega - \rho_{\omega'}$ between any such pair of pure density matrices $\rho_\omega$ and $\rho_{\omega'}$ also belongs to $\mathcal{H}_q$ i.e. $\Delta\rho \in \mathcal{H}_q$.
Now using Schwartz inequality, we can write 
\begin{equation} \label{C2}
(\rho_{\omega}, a) = \Tr_{\mathcal{H}_c} (\rho_{\omega} a) \leq \|\rho_{\omega}\| \|a\| < \infty,
\end{equation}
and likewise for $\rho_{\omega'}$. Further the equality $\omega (a) = \omega' (a) = 0$ occurring in the third condition below \eqref{ball_sat} can be recast in the form:
\begin{equation} \label{C3}
\Tr_{\mathcal{H}_c} (\Delta\rho a) = (\Delta\rho, a) = 0 ~\forall ~a \in V_0
\end{equation}
Also the fact that $\Delta\rho \in \mathcal{H}_q$ implies that $\Delta\rho$ should be a compact HS operator. Indeed, this also follows trivially from the fact that the HS norm of $\Delta\rho$ is bounded above, as follows from
\begin{equation} \label{C4}
\|\Delta\rho \|_{\text HS}^2 =   \Tr_{\mathcal{H}_c} (\Delta\rho^2) \leq 2\left(1 - \Tr_{\mathcal{H}_c}(\rho_\omega\rho_{\omega'}) \right) \leq 2,
\end{equation}
where we have made use of the non-negativeness of both $\rho_\omega$ and $\rho_{\omega'}$. Finally coming to the equality 

\begin{equation} \label{C5}
\| [\mathcal{D}, \pi(a)] \|_{op} = 0
\end{equation}

occurring again in this third condition, we see that it implies, on making use of the form of the Dirac operator \eqref{DirOp} for the Moyal plane $\mathds{R}^2_*$,

\begin{equation}\label{C6}
\|[b,a]\|_{op} = \|[b^\dagger,a]\|_{op} = 0 \Rightarrow [b,a] = [b^\dagger,a] = 0 \Rightarrow a \propto \mathds{1}_c.
\end{equation}

But since $\mathds{1}_c \not\in \mathcal{H}_q$, as the algebra $\mathcal{A} = \mathcal{H}_q$ for the Moyal plane is non-unital, the set $V_0$ contains only the null element: $V_0 = \{ 0 \}$, so that $(\Delta\rho, a) = 0~\forall ~ a \in V_0$ \eqref{C3}  follows trivially. \\
For the fuzzy sphere $\mathds{S}^2_*$, the logic proceeds exactly in a similar fashion \cite{Devi}, except that $\mathds{1}_c$ is now an element of the corresponding unital algebra so that $V_0 = \{ \mathds{1}_c \}$. So in this case too we get $\omega (a) - \omega' (a) = (\Delta\rho, a) = 0$. Here $\mathcal{H}_c = \mathcal{F}_n$ \eqref{F-n} is a finite $(2n+1)$-dimensional Hilbert space and is associated with the fuzzy sphere corresponding to the n-th representation of $SU(2)$. Further, it carries an irreducible representation of $SU(2)$ algebra.

\subsection{Fuzzy Sphere}  \label{ap_fuz}

A similar kind of identity also follows for the case of the fuzzy sphere.  The Dirac operator for the fuzzy sphere is \eqref{fuzz_dir}
\begin{equation*}
\mathcal{D}\equiv \frac{1}{r_n} \hat{\vec{J}} \otimes \vec{\sigma} = \frac{1}{r_n}\begin{pmatrix}
\hat{J}_3 & \hat{J}_- \\ \hat{J}_+ & -\hat{J}_3
\end{pmatrix}.
\end{equation*}

Then for a hermitian algebra element $a=a^{\dagger} \in \mathcal{A}$,
\begin{equation}
\begin{split}
\|\comm{\mathcal{D}}{\pi(a)}\|^{2}_\text{op} & = \|\comm{\mathcal{D}}{\pi(a)}^\dagger\comm{\mathcal{D}}{\pi(a)}\|_\text{op}\\
& = \frac{1}{r_n^{2}}\left\|\left[\begin{array}{cc}
\comm{J_{+}}{a}^{\dagger}\comm{J_{+}}{a} + \comm{J_{3}}{a}^{\dagger}\comm{J_{3}}{a} & \comm{\comm{J_{-}}{a}}{\comm{J_{3}}{a}}\\
-\comm{\comm{J_{+}}{a}}{\comm{J_{3}}{a}}  & \comm{J_{-}}{a}^{\dagger}\comm{J_{-}}{a} +\comm{J_{3}}{a}^{\dagger}\comm{J_{3}}{a}
\end{array}\right]  \right\|_{op} \\
& \ge   \frac{1}{r_n^{2}}\sup_{{\stackrel{\psi \in \mathcal{H}}{\|\psi\| = 1}}} \bra{\psi_{1}} \comm{J_{+}}{a}^{\dagger}\comm{J_{+}}{a} + \comm{J_{3}}{a}^{\dagger}\comm{J_{3}}{a} \ket{\psi_{1}} \ \ \Bigg(\psi =\begin{pmatrix}
\ket{\psi_{1}}\\
0
\end{pmatrix}
\Bigg)\\
&  \ge  \frac{1}{r_n^{2}} \sup_{{\stackrel{\ket{\psi_{1}} \in \mathcal{H}_{c}}{\braket{\psi_{1}}{\psi_{1}} = 1}}} \bra{\psi_{1}} \comm{J_{+}}{a}^{\dagger}\comm{J_{+}}{a}\ket{\psi_{1}} +    \frac{1}{r_n^{2}}\sup_{{\stackrel{\ket{\psi_{1}} \in \mathcal{H}_{c}}{\braket{\psi_{1}}{\psi_{1}} = 1}}} \bra{\psi_{1}} \comm{J_{3}}{a}^{\dagger}\comm{J_{3}}{a} \ket{\psi_{1}}\\
& \ge \frac{1}{r_n^2} \left[ \|\comm{J_{+}}{a}\|^{2}_\text{op} + \|\comm{J_{3}}{a}\|^{2}_\text{op} \right].
\end{split}
\label{A.4}
\end{equation}

Therefore we get
\begin{equation}
\frac{1}{r_n} \|\comm{J_{+}}{a}\|_\text{op} \le \|\comm{\mathcal{D}}{\pi(a)}\|_\text{op}  \ \ \ \frac{1}{r_n} \|\comm{J_{3}}{a}\|_\text{op} \le \|\comm{\mathcal{D}}{\pi(a)}\|_\text{op}. \label{A.5}
\end{equation}
 
For $ a = a^{\dagger} \in \mathcal{A}, \comm{J_{+}}{a}^{\dagger} = - \comm{J_{-}}{a}$ and using $\|A\|_\text{op} = \|A^{\dagger}\|_\text{op}$, we get
\begin{equation}
\frac{1}{r_n} \|\comm{J_{-}}{a}\|_\text{op} \le \|\comm{\mathcal{D}}{\pi(a)}\|_\text{op} .
\label{A.6} 
\end{equation}
 For $a = a^{\dagger} \in B$ i.e $ \|\comm{\mathcal{D}}{\pi(a)}\|_\text{op}  \le 1$ it thus follows that
\begin{equation}
\| \left[J_{\pm}, a \right] \|_\text{op} \leq r_n = \lambda \sqrt{n(n+1)}.
\label{A.7}
\end{equation} 

\hfill $\square$

\subsection{An improved but realistic estimate of spectral distance $d_1 \left( \rho_0, \rho_\theta \right)$ : A heuristic analysis} \label{sec6.6.2}

The upper bound for the spectral distance, obtained previously in \eqref{Fuzz_up_bound} corresponded to that of the geodesic of a commutative sphere $\mathds{S}^2$, but that lies much above the realistic distance for any fuzzy sphere $\mathds{S}^2_*$ associated to the $n$-representation of $SU(2)$, as discussed in section $6.2$. It is therefore quite imperative that we try to have a more realistic estimate of this where this upper bound will be lowered considerably. Since the determination of the exact form of the spectral distance is extremely difficult and we will have to be content with a somewhat heuristic analysis here. To that end, we start here with the most general form of an algebra element $a$, as a linear combination of all the $8$ Gell-Mann matrices ($\lambda_i$) and look for an optimal element from $a \in B$ giving $\sup\abs{(\Delta\rho, a)}$, with some additional restrictions which are to be discussed later. We write

\begin{equation} \label{a_gell}
a = x_i\lambda_i = \begin{pmatrix}
x_3 + \frac{x_8}{\sqrt{3}} & x_1 - ix_2 & x_4 - ix_5 \\
x_1 + ix_2 & -x_3 + \frac{x_8}{\sqrt{3}} & x_6 - ix_7 \\
x_4 + ix_5 & x_6 + ix_7 & -\frac{2x_8}{\sqrt{3}}
\end{pmatrix}
\end{equation}
in analogy with \eqref{al_element} for $n = 1/2$ $\mathds{S}^2_*$. Again the rows/columns are labeled from top to bottom/left to right by $\big(\langle 1|,\langle 0|,\langle -1|\big)/\big(|1\rangle, |0\rangle,|-1\rangle\big)$. Now we calculate tr$(\Delta\rho a)$ using the $\Delta\rho$ matrix \eqref{mat2} and the above algebra element \eqref{a_gell} to get
\begin{equation} \label{tr}
tr(\Delta\rho a) = \left(x_3 + \frac{x_8}{\sqrt{3}} \right)\left(\cos^4\frac{\theta}{2} - 1\right) - \frac{2x_8}{\sqrt{3}}\sin^4\frac{\theta}{2} + \frac{x_4}{2}\sin^2\theta + \frac{1}{2}\sin^2\theta\left(\frac{x_8}{\sqrt{3}} - x_3\right) - \sqrt{2}\sin\theta\left(x_1\cos^2\frac{\theta}{2} + x_6\sin^2\frac{\theta}{2} \right).
\end{equation}

This clearly demonstrates the expected independence of imaginary components viz. $x_2,x_5$ and $x_7$. We therefore set $x_2 = x_5 = x_7 = 0$ to begin with. This simplifies the matrix elements of $M$ \eqref{mat1}, using \eqref{a_gell}, as
\begin{eqnarray} \label{mat4}
M_{11} &=& 3x_1^2 + 6x_4^2 + 8x_3^2 + (x_6 - 2x_1)^2, \\
M_{22} &=& 3x_6^2 + 6x_4^2 + (x_1-2x_6)^2 + 2\left( \sqrt{3}x_8 - x_3 \right)^2 \label{mat4-1}, \\ 
M_{12} &=& \sqrt{2} \left( 3x_4(x_1 + x_6) + 2x_3(x_1 - 2x_6) - (2x_1 - x_6)\left(x_3 - \sqrt{3}x_8\right)  \right) \label{mat4-}.
\end{eqnarray}
Like-wise the above expression \eqref{tr} simplifies as
\begin{eqnarray}
 \nonumber |tr(\Delta\rho a)|&=&\Bigg|\Bigg[\sin\Big(\frac{\theta}{2}\Big)\Bigg\{\sin\Big(\frac{\theta}{2}\Big)\Big\{x_3+3x_3\cos^2\Big(\frac{\theta}{2}\Big)+\sqrt{3}x_8\sin^2\Big(\frac{\theta}{2}\Big)\Big\}\\
 & & +\cos\Big(\frac{\theta}{2}\Big)\Big\{2\sqrt{2}x_1\cos^2\Big(\frac{\theta}{2}\Big)+2\sqrt{2}x_6\sin^2\Big(\frac{\theta}{2}\Big) -2x_4\sin\Big(\frac{\theta}{2}\Big)\cos\Big(\frac{\theta}{2}\Big)\Big\}\Bigg\}\Bigg]\Bigg|. \label{guess}
\end{eqnarray} 
Our aim is to obtain a simple form of the ``Ball" condition and eventually of Connes spectral distance \eqref{ConDis} so that we might obtain an improved estimate for the spectral distance over the lower bound \eqref{3.1.9}, obtained by making use of $\Delta\rho$ \eqref{mat2}, which is more realistic than \eqref{Fuzz_up_bound}. We shall see shortly that with few more additional restrictions, apart from the previous ones (like vanishing of $x_2, x_5$ and $x_7$ imposed already) it is possible to simplify the analysis to a great extent, which in turn yields a distance estimate which has the same mathematical structure as that of the exact distance \eqref{dist_n_half} for the $n = 1/2$ case upto an overall factor. To that end, we impose the following new constraints: 
\begin{equation}\label{constraints}
x_1=x_6; ~~x_3=\frac{x_8}{\sqrt{3}};~~x_4=0,
\end{equation}
as a simple observation of \eqref{guess} suggests that it simplifies even further to the following form:
\begin{equation}\label{guess2}
 |tr(\Delta\rho a)| =\Big|\Big[\sin\Big(\frac{\theta}{2}\Big)\Big\{4x_3\sin\Big(\frac{\theta}{2}\Big)+2\sqrt{2}x_1\cos\Big(\frac{\theta}{2}\Big)\Big\}\Big]\Big| =2\sqrt{2}\sqrt{x_1^2+2x_3^2}\Big|\sin\Big(\frac{\theta}{2}\Big) \cos\Big(\zeta-\frac{\theta}{2}\Big)\Big|,
\end{equation}
where ~$\cos\zeta=\frac{x_1}{\sqrt{x_1^2+2x_3^2}}$~ and ~$\sin\zeta=\frac{\sqrt{2}x_3}{\sqrt{x_1^2+2x_3^2}}$~. Moreover, putting the above constraints \eqref{constraints} in the equation \eqref{mat4}- \eqref{mat4-}, we get 
\begin{equation}
M_{11}=M_{22}=4x_1^2+8x_3^2~~;~~~~M_{12}=0.
\end{equation}
With this, the eigenvalue $E_+$ \eqref{eig_gen} and the corresponding ball condition \eqref{op-norm} can be obtained as
\begin{equation} \label{E-plus}
E_+=4x_1^2+8x_3^2 ~~\Rightarrow ~~\lVert [\mathcal{D},\pi(a)] \rVert_{\text{op}} = \frac{1}{r_1}2\sqrt{x_1^2+2x_3^2}\le 1.
\end{equation}
Using this ball condition in \eqref{guess2}, we get
\begin{equation}
|\text{tr}(\Delta\rho a)| \le \sqrt{2}r_1\sin\Big(\frac{\theta}{2}\Big) \cos\Big(\zeta-\frac{\theta}{2}\Big).
\end{equation}
Hence, a suggestive form of the spectral distance between a pair of pure states $\rho_0=|1\rangle\langle 1|$ and $\rho_\theta=U|1\rangle\langle 1|U^\dagger$ for the $n=1$ representation can be easily obtained by identifying the optimal value of the last free parameter $\zeta$ to be given by $\zeta=\frac{\theta}{2}$. This yields
\begin{equation}
d^a_1=\sup_{a\in B}\big\{|tr(\Delta\rho a)|\big\} =\sqrt{2}r_1\sin\Big(\frac{\theta}{2}\Big).\label{better-dis-1}
\end{equation}
The corresponding form of the optimal algebra element $a_s$ is obtained after a straightforward computation to get  
\begin{equation}
\hat{a}_s=\frac{r_1}{2}\begin{pmatrix}
\sqrt{2}\sin\big(\frac{\theta}{2}\big)&\cos\big(\frac{\theta}{2}\big)&0\\\cos\big(\frac{\theta}{2}\big)&0&\cos\big(\frac{\theta}{2}\big)\\0&\cos\big(\frac{\theta}{2}\big)&-\sqrt{2}\sin\big(\frac{\theta}{2}\big)
\end{pmatrix}. 
\end{equation} 

When $\theta=\pi$, the distance is exactly the same between the two pure coherent states $|1\rangle\langle 1|$ and $|-1\rangle\langle -1|$ and the above distance \eqref{better-dis-1} gives $d_1^* (\rho_0,\rho_\pi)=\sqrt{2}r_1$ which exactly matches with the one computed in \eqref{dis_halfs} using the discrete formula \eqref{DisNS}.

Note that we have made use of all the restrictions $x_2=x_4=x_5=x_1-x_6=x_3-\frac{x_8}{\sqrt{3}}=0$ and $\zeta=\frac{\theta}{2}$, imposed at various stages. Finally, we would like to mention that this simple form \eqref{better-dis-1} was obtained by imposing the above ad-hoc constraints resulting in $M_{11}-M_{22}=M_{12}=0$. Consequently, one cannot a priori expect this to reflect the realistic distance either. At best, this can be expected to be closer to the realistic one, compared to \eqref{Fuzz_up_bound}. The only merit in \eqref{better-dis-1} is that it has essentially the same structure as that of \eqref{dist_n_half} for the $n=\frac{1}{2}$ case, which turns out indeed to be the case with a greater degree of accuracy.

\section*{Acknowledgement}
We thank Prof. A. P. Balachandran for his valuable suggestions. We also thank Alpesh Patil and Aritra Narayan Bose for collaborating during the initial states of the paper. Biswajit Chakraborty thanks Prof. Debashish Goswami for valuable discussions. Kaushlendra Kumar thanks SNBNCBS authorities for the hospitality and studentship provided during his stay where parts of this work were completed. Kaushlendra Kumar also acknowledges the financial support from CSIR in the form of JRF.



\begin{thebibliography}{25}

\bibitem{FSBC}
Scholtz F G and Chakraborty B, Spectral triplets, statistical mechanics and emergent geometry in non-commutative quantum mechanics, \emph{J. Phys. A: Math. Theor.} \textbf{46}, 085204 (2013).

\bibitem{Mart}
Martinetti P and Tomassini L, Noncommutative geometry of the Moyal plane: translation isometries, Connes' distance on coherent states, Pythagoras equality, \emph{Commun. Math. Phys.} \textbf{323}, 107-141 (2013).

\bibitem{Dop}
Doplicher S, Fredenhagen K, and Roberts J E, The quantum structure of spacetime at the Planck scale and quantum fields, \emph{Commun. Math. Phys.} \textbf{172}, 187--220 (1995).

\bibitem{Mad}
Madore J, The fuzzy sphere, \emph{Class. Quant. Grav.} \textbf{9}, 69-87 (1992).

\bibitem{Jab}
Hammou A B, Lagraa M, and Sheikh-Jabbari M M, Coherent state induced star product on $R^3_\lambda$ and the fuzzy sphere, \emph{Phys. Rev. D} \textbf{66}, 025025 (2002).

\bibitem{Devi}
Devi Y C, Prajapat S, Mukhopadhyay A K, Chakraborty B and Scholtz F G, Connes distance function on fuzzy sphere and the connection between geometry and statistics, \emph{J. Math. Phys.} \textbf{56}, 041707 (2015).

\bibitem{Con}
Connes A, \emph{Non-Commutative Geometry} (Academic Press, New York, 1994).

\bibitem{van}
Van den Dungen K and van Suijlekom W D, Particle physics from almost-commutative spacetimes, \emph{Rev. Math. Phys.} \textbf{24} 1230004 (2012).

\bibitem{cag}
Cagnache E, D’Andrea F, Martinetti P and Wallet J-C, The spectral distance in the Moyal plane, \emph{J. Geom. Phys.} \textbf{61} 1881--1897 (2011).

\bibitem{Mart2}
Martinetti P, Mercati F and Tomassini L, Minimal length in quantum space and integrations of the line element in noncommutative geometry, \emph{Rev. Math. Phys.} \textbf{24} 1250010 (2012).

\bibitem{Lat}
Latremoliere F, Quantum locally compact metric spaces, \emph{J. Funct. Anal.} \textbf{264} 362-402 (2013); Wallet J-C, Connes distance by examples: homothetic spectral metric spaces, \emph{Rev. Math. Phys.} \textbf{24} 1250027 (2012).

\bibitem{Perelomov} A M Perelomov, Generalized coherent states and some of their applications, \textit{Sov.Phys.Usp.} \textbf{20} 703 (1977).

\bibitem{Grosse2} H. Grosse , P. Prešnajder, The construction of Non-commutative manifolds using coherent states, \textit{Lett. in Math. Phys.}  \textbf{28} 239-250 ,1993.

\bibitem{Var}
D'Andrea F, Lizzi F, and Varilly J C, Metric properties of the fuzzy sphere, \emph{Lett. Math. Phys.} \textbf{103} 183--205 (2013).

\bibitem{SCGV}
Scholtz F G, Chakraborty B, Govaerts J and Vaidya S, Spectrum of the non-commutative spherical well, \emph{J. Phys. A: Math. Theor.} \textbf{40} 14581 (2007).

\bibitem{Roh}
Scholtz F G, Gouba L, Hafver A and Rohwer C M, Formulation, interpretation and application of non-commutative quantum mechanics, \emph{J. Phys. A: Math. Theor.} \textbf{42} 175303 (2009).

\bibitem{Liz}
Galluccio S, Lizzi F and Vitale P, Twisted noncommutative field theory with the Wick-Voros and Moyal products, \emph{Phys. Rev. D} \textbf{78} 085007 (2008); Balachandran A P and Martone M, Twisted quantum fields on Moyal and Wick--Voros planes are inequivalent, \emph{Mod. Phys. Lett. A} \textbf{24} 1721 (2009); Balachandran A P, Ibort A, Marmo G and Martone M, Inequivalence of quantum field theories on noncommutative spacetimes: Moyal versus Wick-Voros planes, \emph{Phys. Rev. D} \textbf{81} 085017 (2010).

\bibitem{Val}
Provost J P and Vallee G, Riemannian structure on manifolds of quantum states, \emph{Commun. Math. Phys.} \textbf{76} 289--301 (1980).

\bibitem{Asht}
Ashtekar A and Schilling T A, Geometrical Formulation of Quantum Mechanics. In: Harvey A. (eds) On Einstein's Path (Springer, New York, 1999).


\bibitem{FA} D'Andrea F, Pythagoras Theorem in Noncommutative Geometry. In: Noncommutative geometry and optical transport \textbf{676} (Contemporary Mathematics, 2016).

\bibitem{Basu}
Basu P, Chakraborty B and Scholtz F G, A unifying perspective on the Moyal and Voros products and their physical meanings, \emph{J. Phys. A: Math. Theor.} \textbf{44} 285204 (2011).


\bibitem{Wat}
Carow-Watamura U and Watamura S, Chirality and Dirac operator on noncommutative sphere, \emph{Commun. Math. Phys.} \textbf{183} 365-382 (1997).

\bibitem{Gay}
Gayral V, Gracia-Bond\'ia J M, Iochum B, Sch\"ucker T and Varilly J C, Moyal planes are spectral triple, \emph{Commun. Math. Phys.} \textbf{246}, 569–623 (2004).

\bibitem{Rob}
Bratelli O and Robinson D W, \emph{Operator Algebras and Quantum Statistical Mechanics} \textbf{1} 2nd edn. (Springer, Berlin, 1987) p:76.

\bibitem{R1}
Martinetti P, From Monge to Higgs: a survey of distance computation in noncommutative geometry, \emph{Contemp. Maths.} \textbf{676} 1-46 (2016).

\bibitem{Mart3}
Iochum B, Krajewski T and Martinetti P, Distances in finite spaces from noncommutative geometry, \emph{J. Geom. Phys.} \textbf{37} 100-125 (2001).

\bibitem{KK}
Kumar K and Chakraborty B, Spectral distances on doubled Moyal plane using Dirac eigen-spinors, \textit{Phys. Rev. D} \textbf{97}, 086019 (2018).

\bibitem{R2}
Goswami D and Bhowmik J, \emph{Quantum Isometry Groups}, (Springer, New Delhi, 2016).

\bibitem{Conway}
Conway J B, \emph{A Course in Functional Analysis} 2nd edn (Springer-Verlag, New York, 1990) p 32.


\end{thebibliography}
\end{document}